\documentclass[trackchanges]{aastex701}

\usepackage{indentfirst}
\usepackage{xcolor,graphicx,float}
\usepackage{subfigure}
\usepackage{color}
\usepackage{multirow}
\usepackage{rotating} 
\usepackage{booktabs}
\usepackage{threeparttable}
\usepackage[section]{placeins} 

\begin{document}

\title{Eight Gamma-Ray Bursts Observed with the Fermi Gamma-ray Burst Monitor: Significant Quasi-Thermal Components and Conditional Jet Constraints}

\author[orcid=0009-0008-4550-4081]{Ze-lian Du}
\affiliation{College of Physics and Electronics information, Yunnan Normal University, Kunming 650500, People's Republic of China}
\email[show]{duzelian@163.com}

\author[orcid=0000-0003-3846-0988]{Zhao-Yang Peng} 
\affiliation{College of Physics and Electronics information, Yunnan Normal University, Kunming 650500, People's Republic of China}
\email[show]{pengzhaoyang412@163.com}

\author[orcid=0000-0001-5681-6939]{Jia-Ming Chen}
\affiliation{Department of Astronomy, School of Physics and Astronomy, Key Laboratory of Astroparticle Physics of Yunnan Province, Yunnan University, Kunming 650091, People’s Republic of China}
\email{chenjiaming5821@163.com}

\author{Yue Yin}
\affiliation{Department of Physics, Liupanshui Normal College, Liupanshui 553004, People's Republic of China}
\email{564089310@qq.com}

\author{Ting Li}
\affiliation{State-owned Assets and Laboratory Management Office, Yunnan Normal University, Kunming 650500, People’s Republic of China}
\email{369271054@qq.com}

\begin{abstract}
 
The composition of gamma-ray burst (GRB) jets remains uncertain. We analyze eight Fermi Gamma-ray Burst Monitor (GBM) GRBs with reported redshift information, selected from 126 events by prespecified fluence and spectral criteria. After excluding compound-model DIC evaluations classified as unstable by the posterior-stability tests, a stable added-blackbody (BB) comparison gives $\Delta\mathrm{DIC}>10$ in 203 of 215 accepted bins (60--100\% per burst); independently, a multicolor-blackbody plus power-law (mBB+PL) decomposition gives $F_{\rm mBB}/F_{\rm tot}>0.5$ in 214 of 215 bins. The cutoff-power-law (CPL)--mBB correlations ($r=0.83$ and 0.78) exceed the Band--mBB correlations ($r=0.41$ and 0.48), demonstrating model dependence. Time-integrated BB fractions span 0.005--0.311, with six of eight below 0.12. We then apply the non-dissipative photosphere prescriptions of \citet{2013A&A...551A.124H} (H2013) and \citet{2015ApJ...801..103G} (G2015). After excluding boundary or physically inadmissible inversions, the H2013 numbers with $\epsilon_{\rm Th}>0.5$ are 1/2, 1/4, and 2/8 among valid solutions at $R_0=21$, 100, and 1000~km, respectively; the corresponding G2015 numbers with $\sigma_0<1$ are 0/1, 1/4, and 1/8. The large number of invalid solutions at small $R_0$ (six and seven at 21~km for H2013 and G2015) is itself evidence that those assumed radii are incompatible with the adopted inversion for most events. Among bursts having valid solutions at more than one radius, increasing $R_0$ generally lowers $\epsilon_{\rm Th}$ and raises $\sigma_0$. Thus launch radius, efficiency, and spectral decomposition dominate the conditional jet constraints; no universal jet class or unique non-thermal mechanism is inferred.

\end{abstract}


\section{Introduction\label{Sections1}}
Gamma-ray bursts (GRBs) are among the most energetic explosive phenomena in the universe, releasing enormous amounts of energy on timescales from milliseconds to thousands of seconds. Photons with energies well above 10~GeV have been detected from some GRBs, and their instantaneous luminosities can exceed the total luminosity of the Milky Way. The prompt emission is the earliest detected signal from a GRB. Despite more than half a century of observations, the origin of the prompt emission, jet composition, energy dissipation, particle acceleration, and radiation mechanisms remain debated \citep{2015PhR...561....1K,2018pgrb.book.....Z,2019MmSAI..90...57M}. Spectral and jet-composition analyses can therefore clarify the radiation and energy-dissipation processes operating during this phase.

Early theories proposed that GRB outflows originate as pure thermal fireballs composed of photons and electron--positron pairs \citep{1978MNRAS.183..359C,1986ApJ...308L..43P,1986ApJ...308L..47G}. A modified blackbody from a simple fireball photosphere, however, does not reproduce the typical observed spectrum, which is commonly described by the Band function \citep{1993ApJ...413..281B}. Baryon-dominated fireball models were subsequently developed to account for these observations \citep{1990ApJ...363..218P,1990ApJ...365L..55S,1993ApJ...405..278M,1993MNRAS.263..861P}. In these models, the prompt spectrum may contain a quasi-thermal photospheric component and a non-thermal component from an optically thin region \citep{2000ApJ...530..292M,2004ApJ...614..827R,2006ApJ...642..995P}. Internal shocks can dissipate outflow kinetic energy and accelerate relativistic electrons, whose synchrotron radiation may produce the observed non-thermal gamma rays \citep{1994ApJ...430L..93R,1997ApJ...490...92K,1998MNRAS.296..275D,1996ApJ...466..768T,2000ApJ...543..722L,2020NatAs...4..210Z}. Poynting-flux-dominated outflow models have also been extensively explored \citep{2001A&A...369..694S,2002ApJ...581.1236Z,2006A&A...450..887G,2006A&A...457..763G,2014MNRAS.445.3892B}. In this limit, the thermal contribution is weak, and the prompt emission may be generated through internal-collision-induced magnetic reconnection and turbulence (ICMART; \citealt{2011ApJ...726...90Z,2014ApJ...782...92Z,2015ApJ...805..163D,2022ApJ...927..173S}).

Most observed GRB spectra have a non-thermal shape that is well described by the Band function \citep{1993ApJ...413..281B}. Such spectra are often attributed to emission generated by internal shocks or magnetic reconnection above the photosphere. The Band function is a smoothly joined broken power law with low- and high-energy photon indices $\alpha$ and $\beta$, whose typical values are approximately $-1.0$ and $-2.2$, respectively \citep{2000ApJS..126...19P}. Spectra with these typical indices are broadly compatible with synchrotron emission \citep{2016ApJ...816...72Z}. In the $\nu F_{\nu}$ representation, the spectrum peaks at $E_p$, typically near 300~keV.

Although non-thermal empirical models describe most GRB spectra, quasi-thermal components can provide important constraints on the outflow \citep{2002A&A...393..409G,2003A&A...406..879G,2005ApJ...625L..95R,2009ApJ...702.1211R,2010ApJ...709L.172R,2014ApJ...785..112D,2017IJMPD..2630018P}. Analyses with the Compton Gamma Ray Observatory Burst and Transient Source Experiment (CGRO/BATSE) and with Fermi have identified such components in a number of bursts \citep{2004ApJ...614..827R,2013ApJ...770...32G,2013MNRAS.432.3237G,2017ApJ...846..138G,2016MNRAS.456.2157I,2022ApJ...932...25C}. Adding a blackbody (BB) component to a non-thermal baseline can improve the fit \citep{2019ApJS..245....7L}, but statistical preference for the additional component does not by itself establish a photospheric origin. The fitted thermal contribution is often subdominant, as in GRB~100724B \citep{2011ApJ...727L..33G}, GRB~110721A \citep{2012ApJ...757L..31A}, and the short GRB~120323A \citep{2013ApJ...770...32G}. The coexistence of thermal-like and non-thermal spectral structure therefore motivates jet models ranging from hot fireballs to Poynting-flux-dominated outflows and hybrid configurations containing both components \citep{2013A&A...551A.124H,2015ApJ...801..103G,2020ApJ...894..100L}.

Several GRBs, most notably GRB~090902B, show strong thermal-like spectral structure \citep{2009ApJ...706L.138A,2012MNRAS.420..468P,2022ApJ...934L..22D,2022ApJ...940..142W,2022MNRAS.517.2088S,2024ApJ...961..137S,2024ApJ...972..132C}. Systematic comparisons of the associated jet constraints remain limited. The present sample comprises eight Fermi/GBM GRBs with reported redshift information and significant quasi-thermal components. Statistical support for spectral curvature, the fitted component flux fraction, and the additional physical assumption of a photospheric origin are treated as distinct levels of inference. The time-integrated fits are then interpreted with the non-dissipative photosphere prescription of \citet{2013A&A...551A.124H} (H2013) and the hybrid-jet prescription of \citet{2015ApJ...801..103G} (G2015), evaluated over assumed launch radii and efficiencies. This design provides a linked test from spectral phenomenology to engine-scale interpretation: it determines which conclusions are fixed by the data, which depend on the photospheric prescription, and which change with the launch radius or efficiency. Comparing the directional behavior of two physically distinct frameworks also isolates model-dependent degeneracies that cannot be recognized from a single best-fit jet classification. The resulting parameter-space map, rather than a categorical label for each burst, is the central contribution of the study.

The paper is organized as follows. Section~\ref{Sections2} describes the sample selection and spectral fitting. Section~\ref{Sections3} presents the time-resolved spectral results. Section~\ref{Sections4} uses the H2013 and G2015 prescriptions to infer conditional jet parameters from the time-integrated fits. Section~\ref{Sections5} discusses the results and summarizes the conclusions. A standard flat cosmology is adopted throughout, with $H_0 = 70 \, \text{km} \, \text{s}^{-1} \, \text{Mpc}^{-1}$, $\Omega_m = 0.3$, $\Omega_r = 0$, and $\Omega_\Lambda = 0.7$.

\section{Data and Analysis Methods\label{Sections2}}

\subsection{Sample Selection}

Public Time-Tagged Event (TTE) data from the Fermi Gamma-Ray Burst Monitor (GBM) were used. GBM comprises 12 sodium iodide (NaI) detectors covering approximately 8~keV--1~MeV and two bismuth germanate (BGO) detectors covering approximately 200~keV--40~MeV \citep{2009ApJ...702..791M}. NaI detectors with favorable viewing angles and without spacecraft blockage were selected together with the BGO detector on the same side of the spacecraft. The detectors used for each event are listed in Table~\ref{table2}. The TTE data were downloaded from the Fermi Science Support Center\footnote{\url{https://heasarc.gsfc.nasa.gov/W3Browse/fermi/fermigbrst.html}}.

The GBM archive was searched from 2008 July, when routine science operations began, through 2025 May. The search returned 126 bursts with reported redshifts and public TTE data; these events constitute the parent sample for the quantitative screening below.

(1) Events were retained when their fluence (time-integrated energy) for the source interval, as provided by the MPE GRB catalog, exceeded $10^{-5}\,\mathrm{erg\,cm^{-2}}$. Visual inspection was used only as a data-quality check: events with telemetry gaps, severe instrumental artifacts, or photon statistics inadequate for time-resolved spectroscopy were excluded. No light-curve morphology was used as a selection criterion. This step left 39 events (87 rejected).

(2) For each candidate, the Band and cutoff-power-law (CPL) baselines were compared with models containing an additional blackbody (BB) or power-law (PL) component. Using the deviance information criterion (DIC), we define $\Delta\mathrm{DIC}=\mathrm{DIC}_{\rm simple}-\mathrm{DIC}_{\rm complex}$, so positive values favor the complex model. Requiring $\Delta\mathrm{DIC}>10$ for the relevant nested comparisons left 16 events (23 rejected). This criterion tests whether an additional spectral component is statistically supported; it does not measure that component's energy fraction and, by itself, does not establish a photospheric origin.

(3) The time-resolved spectra were then fitted with the multicolor-blackbody (mBB) and mBB+PL models. The event-level criterion required more than half of the accepted time bins to have an mBB contribution above 50\% of the mBB+PL 10--1000~keV energy flux, with a central fitted value of $-1.5<m<1$. Eight events satisfied these prespecified requirements (eight rejected). In the final sample, 214 of 215 accepted bins satisfy the flux criterion and all 215 have central $m$ values in the adopted range (Section~\ref{Sections3}). The parameter-correlation tests presented in Section~\ref{Sections3} were not used in this selection and are therefore descriptive results rather than selection criteria.

The first selection cut is based exclusively on catalog fluence over the source interval, not on peak or time-averaged flux. It favors energetic, well-measured events and is not a completeness criterion. Requiring both redshift information and high fluence additionally favors optically well-followed, relatively bright events. The resulting sample is therefore not representative of the full GBM population, and fractions measured within these eight events are not population occurrence rates.

The adopted redshifts and their provenance are summarized in Table~\ref{table2}. GRB~211211A was retained because it satisfies the same spectral criteria, but it is physically distinct from most long-duration collapsar GRBs: its association with a $z=0.076$ galaxy and a kilonova supports a compact-binary-merger origin. It is therefore identified explicitly in all sample-level interpretations.

For GRB~231129C, we adopt the estimated value $z=0.5$ reported by \citet{2024ApJ...972..132C}. This value is not a spectroscopic redshift, and the source does not quote a quantitative uncertainty; it is therefore treated as an estimate rather than as a probabilistic redshift measurement.

\subsection{Spectral Fitting Method}

Time-resolved and time-integrated spectral analyses were performed with the Multi-Mission Maximum Likelihood framework (3ML; \citealt{2015arXiv150708343V}) using the MultiNest nested sampler with 500 live points. Related Bayesian GRB applications are described by \citet{2019MNRAS.490..927B} and \citet{2021ApJS..254...35L}. For each detector separately, source-free intervals before and after the burst were fitted with polynomials of order 0--4; the order was selected by likelihood-ratio testing, and the resulting background estimate and uncertainty were propagated into the source fit.

The priors reproduce those in the analysis scripts. For Band, $\alpha\sim\mathrm{TN}(-1,0.5;-1.5,1)$, $\beta\sim\mathrm{TN}(-2,0.5;-5,-1.6)$, $E_p\sim\mathrm{LN}(\ln300,1)$, and $K\sim\mathrm{LN}(\ln0.1,1)$. For CPL, $\alpha\sim\mathrm{TN}(-1,0.5;-5,1)$, $E_p\sim\mathrm{LN}(\ln300,1)$, and $K\sim\mathrm{LN}(\ln0.1,1)$. The BB component in compound fits has $kT\sim\mathrm{LN}(\ln200,1)$ and $K\sim\mathrm{LN}(\ln10^{-6},1)$; the PL component in Band/CPL compound fits has $\alpha_{\rm PL}\sim\mathrm{TN}(-1,1;-5,1)$ and $K\sim\mathrm{LogU}(10^{-4},10^2)$. For standalone mBB, $m\sim\mathrm{TN}(0.5,1;-10,10)$, $kT_{\min}\sim\mathrm{LN}(\ln2,1)$, $kT_{\max}\sim\mathrm{LN}(\ln100,1)$, and $K\sim\mathrm{LogU}(10^{-5},10^5)$. The mBB+PL fits use the same $kT_{\min}$, $kT_{\max}$, and mBB-normalization priors but $m\sim\mathrm{TN}(3,1;-10,10)$, $\alpha_{\rm PL}\sim\mathrm{TN}(-3,1;-5,1)$, and $K_{\rm PL}\sim\mathrm{LogU}(10^{-4},10^6)$. Here TN, LN, and LogU denote truncated normal, log-normal, and log-uniform distributions; energies are in keV and normalizations use the model units. Because MultiNest is a nested sampler, MCMC chain length, burn-in, and $\hat R$ do not apply. Each comparison used identical intervals, detectors, channels, and backgrounds. Fits with a component accumulated at a prior boundary were flagged rather than interpreted as detections. The complete prior profile and sampling configuration accompany the machine-readable results.

Bayesian Blocks were used to segment the TTE light curve from the brightest unblocked NaI detector, with false-alarm probability $p=0.01$ \citep{2013ApJ...764..167S}; the resulting boundaries were applied to the other selected detectors. Only blocks satisfying $S\geq30$ were retained as independent spectra \citep{2018ApJS..236...17V}; lower-significance blocks were not used in the time-resolved fits. The significance was calculated from the selected-detector count spectra over the same active energy channels used in the corresponding spectral fit.

\subsection{Spectral Models}
Most GRB spectra are well described by non-thermal empirical models. Two standard forms were adopted: the smoothly broken power law (Band function) and the cutoff power law (CPL).

The Band function is

\begin{equation}
{N(E)_{B{\rm{and}}}} = A\left\{ \begin{array}{l}
\left( {\frac{E}{{100keV}}} \right)^\alpha\exp \left( { - \frac{E}{{{E_0}}}} \right),E < \left( {\alpha  - \beta } \right){E_0}\\
{\left( {\frac{{\left( {\alpha  - \beta } \right){E_0}}}{{100keV}}} \right)^{\alpha  - \beta }}\exp \left( {\beta  - \alpha } \right){\left( {\frac{E}{{100keV}}} \right)^\beta },E > \left( {\alpha  - \beta } \right){E_0}
\end{array} \right.
\end{equation}

where $A$ is the normalization factor with units of photons $\mathrm{s^{-1}\,cm^{-2}\,keV^{-1}}$. The parameters $\alpha$ and $\beta$ are the low- and high-energy photon indices, respectively; $E_0$ is the break energy at which the two power-law segments join smoothly. In the $\nu F_{\nu}$ spectrum, the peak energy is $E_p=(2+\alpha)E_0$ (in keV).

The CPL function is
\begin{equation}
{N(E)_{\rm CPL}} = A{\left( {\frac{E}{{100\,\mathrm{keV}}}} \right)^\alpha }\exp \left( { - \frac{\left( {\alpha + 2} \right)E}{{E_p}}} \right)
\end{equation}
where $A$ is the normalization constant with units of photons $\mathrm{s^{-1}\,cm^{-2}\,keV^{-1}}$. Here, $\alpha$ is the low-energy spectral index and $E_p$ is the peak energy in keV.

To test for additional thermal and non-thermal components, the Planck blackbody (BB) function and power-law (PL) model were added separately to the Band and CPL baselines, and the resulting changes in goodness of fit were evaluated. The Planck blackbody and power-law functions are defined as follows:

The Planck function is
\begin{equation}
{N(E)_{\rm BB}} = A\frac{{{E^2}}}{{\exp \left[ {E/kT_{\rm BB}} \right] - 1}}
\end{equation}
where $A$ is the normalization constant, $k$ is the Boltzmann constant, and $kT_{\rm BB}$ is the blackbody temperature.

The power-law (PL) model is
\begin{equation}
{N(E)_{\rm PL}} = A{\left( {\frac{E}{{100\,\mathrm{keV}}}} \right)^\lambda }
\end{equation}

Thermal emission is commonly represented by the Planck function. In a relativistic GRB jet, however, angular and Doppler effects produce a distribution of observed temperatures, and the photospheric radius increases with viewing angle \citep{2008ApJ...682..463P}. The multicolor BB (mBB) model describes this broadened photospheric spectrum by summing Planck functions over a range of temperatures \citep{2010ApJ...709L.172R}. It has been widely used to fit GRBs with thermal-like spectra (e.g., \citealt{2021ApJS..255...25I,2022MNRAS.517.2088S,2024ApJ...961..137S,2024ApJ...972..132C}); the form adopted here follows \citet{2018ApJ...866...13H}:
\begin{equation}
N(E)_{\rm mBB} = \frac{8.0525(m+1)A_{\rm mBB}}{\left[\left(\frac{T_{\max}}{T_{\min}}\right)^{m+1} - 1\right]} \left(\frac{kT_{\min}}{\mathrm{keV}}\right)^{-2} I(E),
\end{equation}
\begin{equation}
I(E) = \left(\frac{E}{kT_{\min}}\right)^{m-1} \int_{\frac{E}{kT_{\max}}}^{\frac{E}{kT_{\min}}} \frac{x^{2-m}}{e^x - 1} \, dx.
\end{equation}
Here, $x=E/kT$, $K=L_{39}/D_{{\rm L},10\,{\rm kpc}}^2$, and \(N(E)\) has units of \(\mathrm{ph\,cm^{-2}\,keV^{-1}\,s^{-1}}\). Within the range \(kT_{\min}\)--\(kT_{\max}\), the temperature index $m$ determines the spectral shape. When \(m=2\), the mBB model approximates a pure blackbody spectrum.

Model quality is evaluated with the Deviance Information Criterion (DIC; \citealt{2019ApJS..245....7L,2021ApJ...920...53C}), defined as
\begin{equation}
{DIC} = -2 \log \left[ {P(\text{data}\mid \vec{\theta})}\right] + 2{P_{DIC}}
\end{equation}
where $\vec{\theta}$ is the posterior mean and $P_{\rm DIC}$ is the effective number of parameters. A smaller DIC indicates a better expected fit after model complexity is penalized. Only models fitted to the same data, detectors, energy channels, and background treatment are compared.

The DIC difference, $\Delta \text{DIC} = \text{DIC}_j - \text{DIC}_i$, is used to compare two competing spectral models. Following the classification of \citet{2018ApJ...866...13H}, $\Delta \text{DIC}$ is interpreted as follows:
(a) $0 \le \Delta \text{DIC} \le 2$: The fits obtained with models $j$ and $i$ are statistically indistinguishable;
(b) $2 < \Delta \text{DIC} \le 6$: Positive evidence favors model $i$;
(c) $6 < \Delta \text{DIC} \le 10$: Strong evidence favors model $i$;
(d) $\Delta \text{DIC} > 10$: Very strong relative support for model $i$ under the adopted priors and likelihood.

DIC is used here as a screening statistic rather than a stand-alone component detection. The added component must also be constrained away from its null boundary, with acceptable residual and PG-statistic-per-degree-of-freedom (PG-stat/dof) diagnostics. Posterior-stability tests identified 20 grossly discordant negative DIC evaluations from compound thermal models in 14 bins. These include the two entries shown in Table~\ref{table1} ($-828.6$ for CPL+BB+PL at 0.21--0.30~s and $-1169.2$ for CPL+BB at 0.30--0.40~s), whose parameters extend to prior edges. Refitting the two displayed cases with the same detectors, intervals, model definitions, and 500 live points repeatedly proposed $E_p$ outside its hard 10--10000~keV domain because the adopted log-normal prior has nonzero support beyond those bounds; the refits produced out-of-bounds exceptions and no stable posterior. The apparently acceptable PG-stat/dof values therefore do not establish posterior stability. A uniform fit-quality rule is applied to all bins: these DIC evaluations are reported in the machine-readable table but are excluded from preference counts and physical interpretation. Affected time bins remain in the analysis when stable Band, CPL, mBB, and mBB+PL fits are available; an added-BB detection is counted only when a stable Band+BB or CPL+BB comparison gives $\Delta\mathrm{DIC}>10$.

\section{Time-resolved Spectral Analysis\label{Sections3}}
\subsection{Thermal Component Analysis}

GRB~210619B provides the highest-statistics illustration. Bayesian Blocks yielded 75 intervals, of which 41 satisfied $S\geq30$. Band is preferred to CPL in 40 of these intervals. After the two unstable compound-model DIC entries described above are excluded, each bin still has a stable alternative baseline comparison; adding BB or PL yields $\Delta\mathrm{DIC}>10$ in all 41 intervals. These results show that additional curvature or a tail is favored over the adopted empirical baselines. They do not, by themselves, establish the energetic dominance or unique physical origin of either component.

Thermal-flux dominance is assessed separately from DIC using the mBB+PL decomposition. Here $F_{\rm th}$ in Table~\ref{table1} denotes $F_{\rm BB}$ for models containing a single-temperature BB and $F_{\rm mBB}$ for models containing mBB; it is blank for models without a thermal component. $F_{\rm tot}$ is the total model energy flux. The component flux fraction is defined as $f_{\rm mBB}=F_{\rm mBB}/(F_{\rm mBB}+F_{\rm PL})$, with all component fluxes integrated over 10--1000~keV. A time bin is labeled mBB-flux dominated when $f_{\rm mBB}>0.5$; no additional posterior-probability cut was imposed. This operational flux criterion, rather than the relative DIC improvement produced by BB and PL, defines the quasi-thermal subset. The label is therefore treated as model dependent rather than as a population probability.

Applying this definition to the complete time-resolved output gives 214 mBB-flux-dominated bins among 215 accepted bins (99.5\%). The per-burst counts are 15/15, 13/13, 41/41, 102/103, 11/11, 20/20, 7/7, and 5/5 for GRBs~130518A, 210610B, 210619B, 211211A, 231129C, 240825A, 241228B, and 250407A, respectively (Table~\ref{table2}). The only exception is the 15.26--16.07~s interval of GRB~211211A, for which $F_{\rm mBB}=0.97\times10^{-6}\,\mathrm{erg\,cm^{-2}\,s^{-1}}$, $F_{\rm tot}=2.75\times10^{-6}\,\mathrm{erg\,cm^{-2}\,s^{-1}}$, and $f_{\rm mBB}=0.353$. Its mBB+PL parameters are also weakly constrained, including $m=0.77^{+0.05}_{-1.89}$. Because no secondary posterior-probability threshold is imposed, this interval remains part of the operational flux-fraction count. All 215 central $m$ values lie within $-1.5<m<1$.

\begin{figure}[htbp]
\centering
\includegraphics[width=0.32\textwidth]{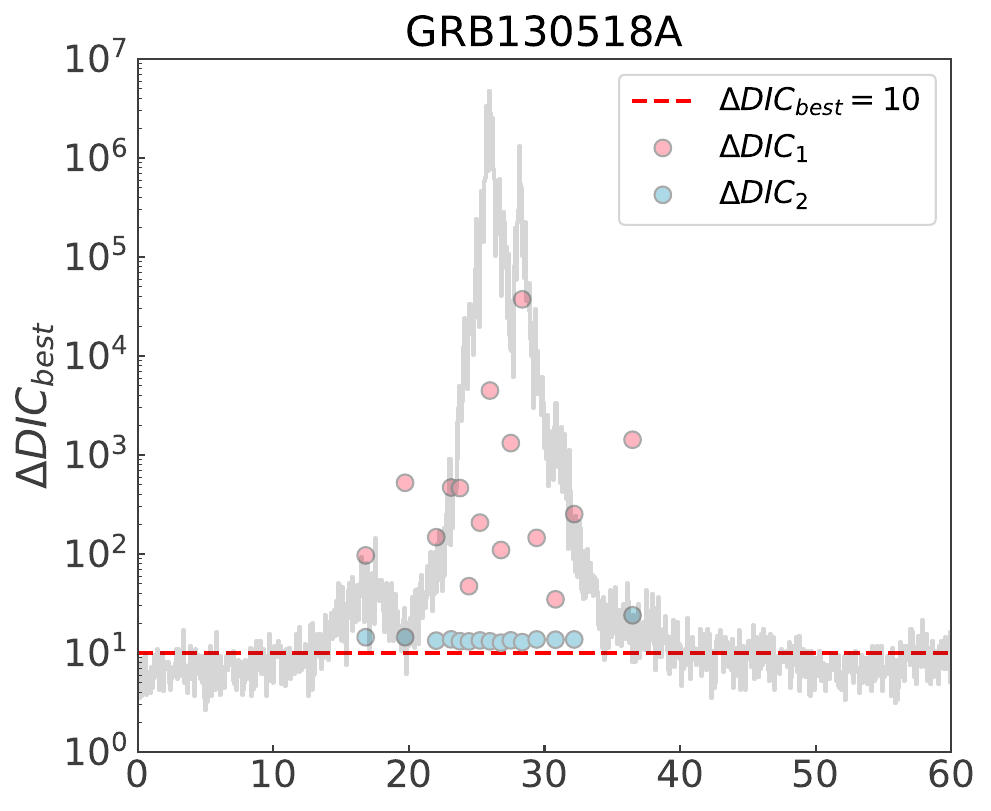}
\includegraphics[width=0.32\textwidth]{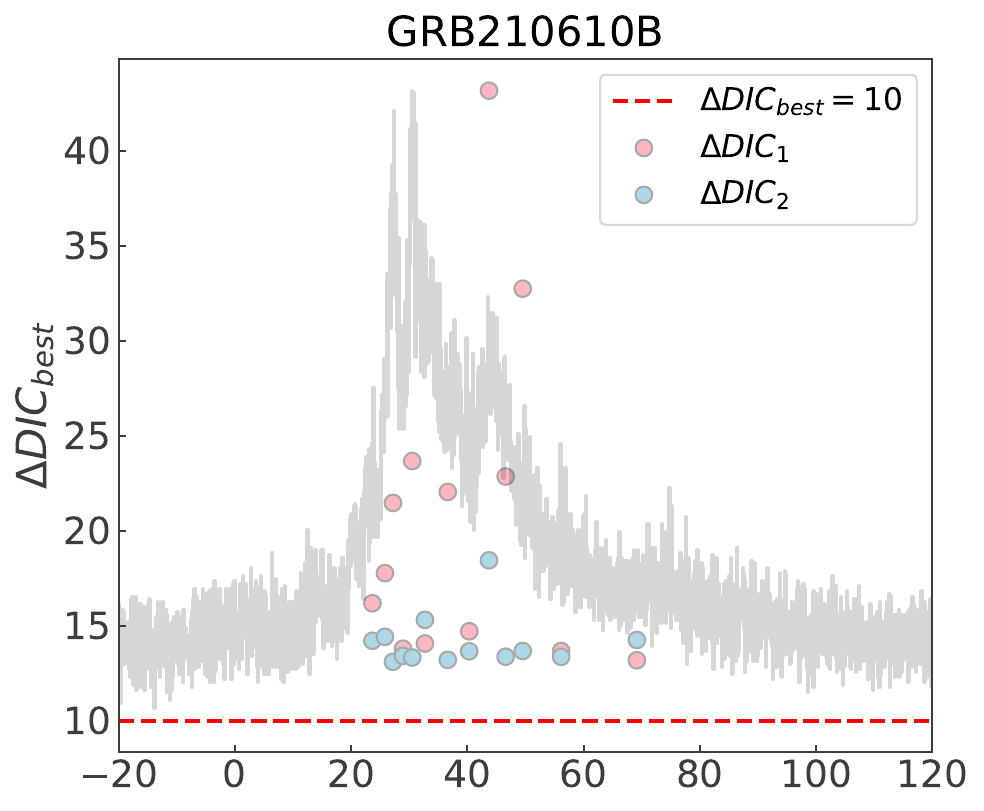}
\includegraphics[width=0.32\textwidth]{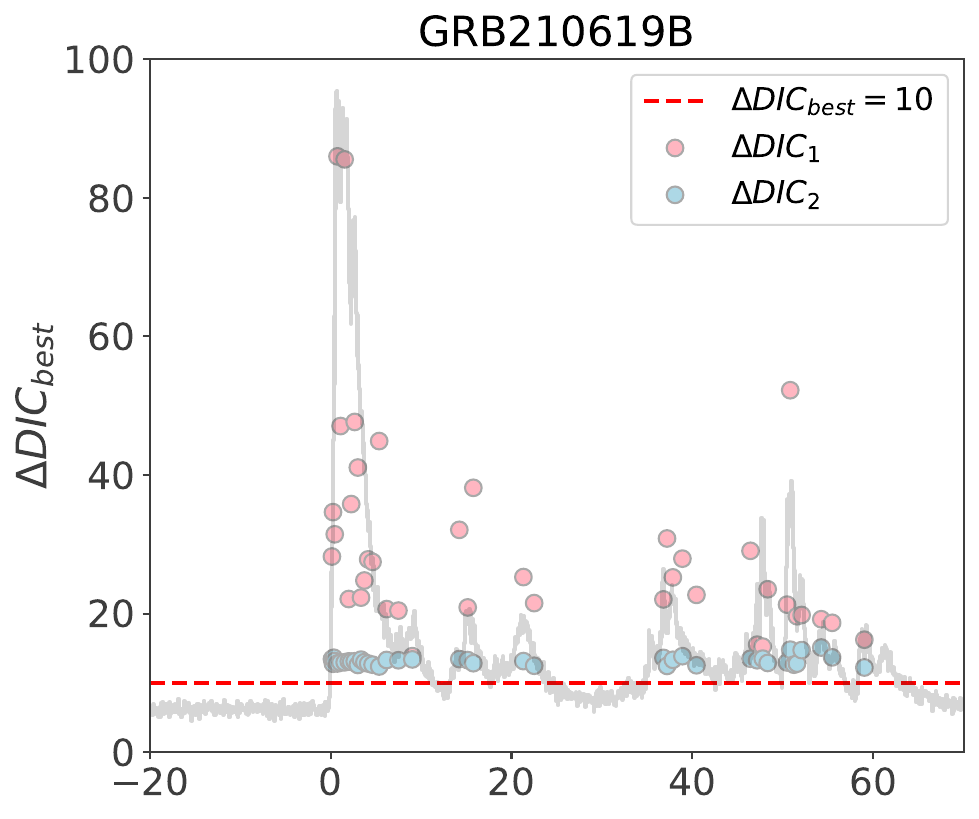}
\includegraphics[width=0.32\textwidth]{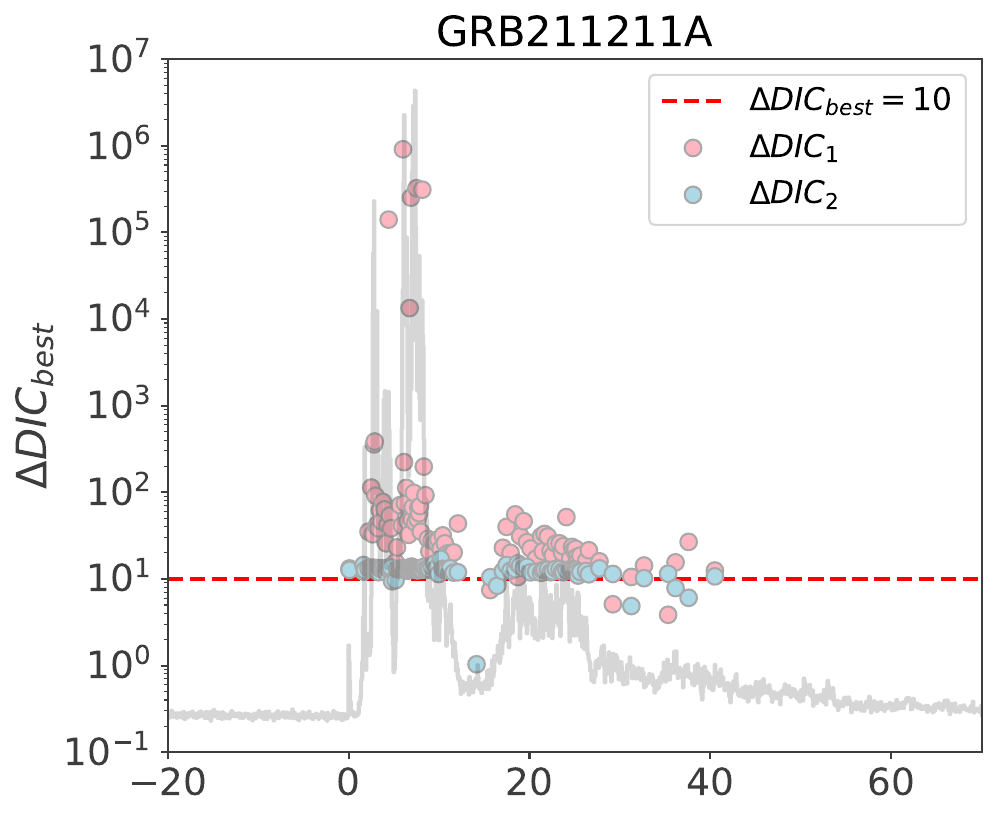}
\includegraphics[width=0.32\textwidth]{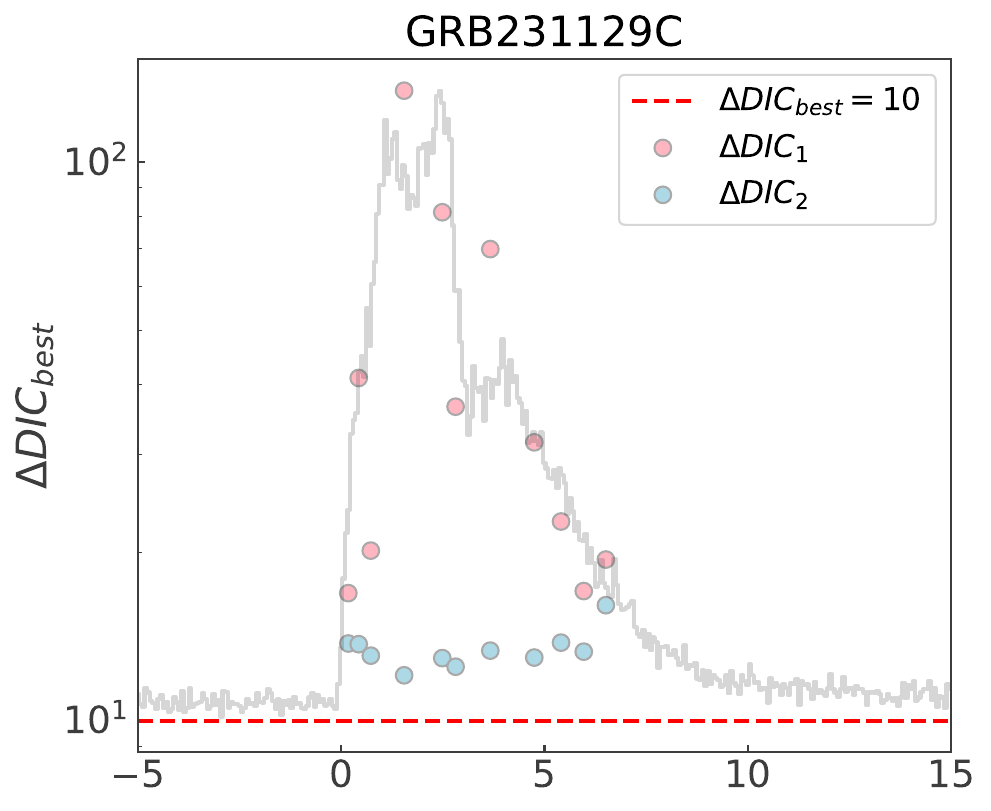}
\includegraphics[width=0.32\textwidth]{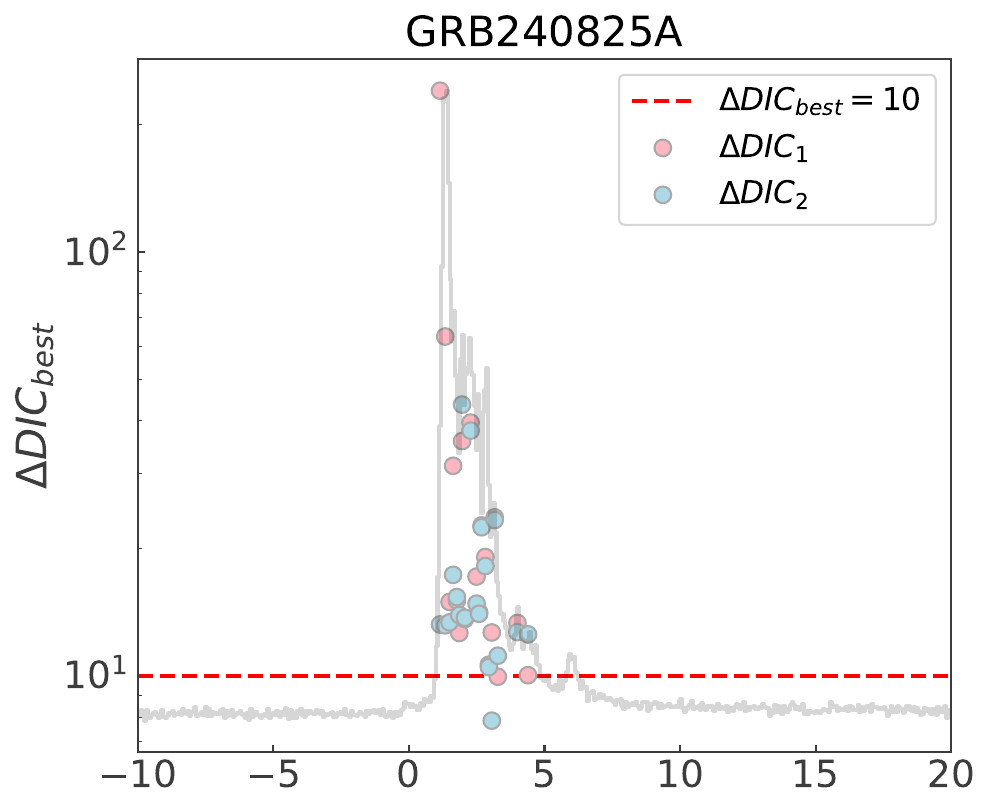}
\includegraphics[width=0.32\textwidth]{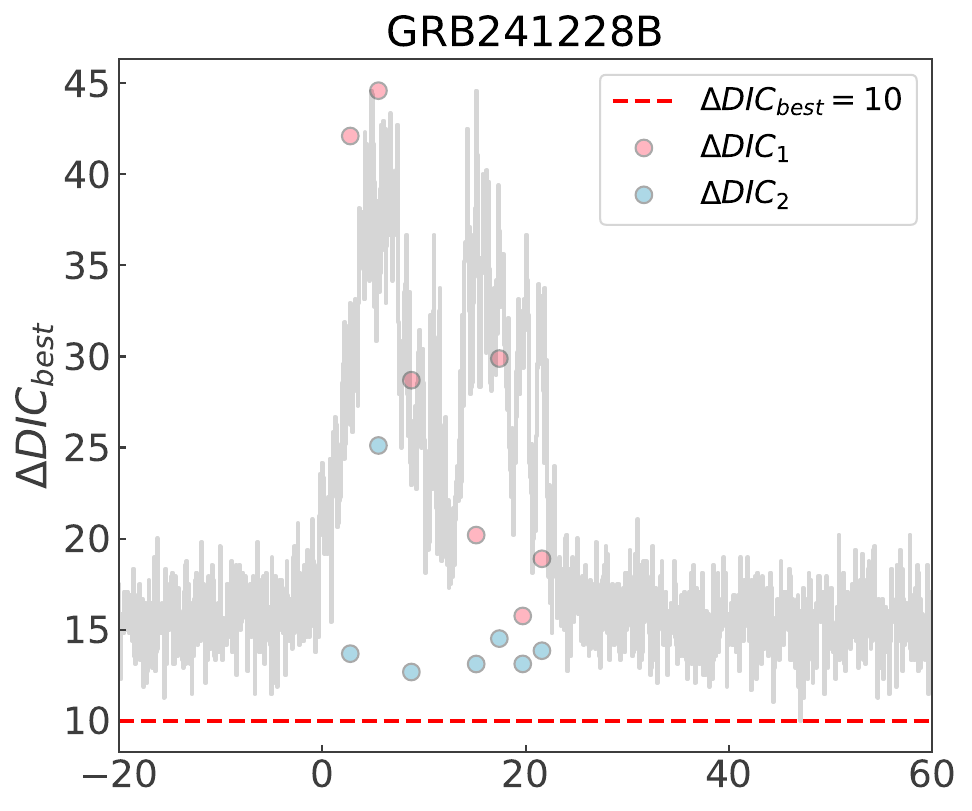}
\includegraphics[width=0.32\textwidth]{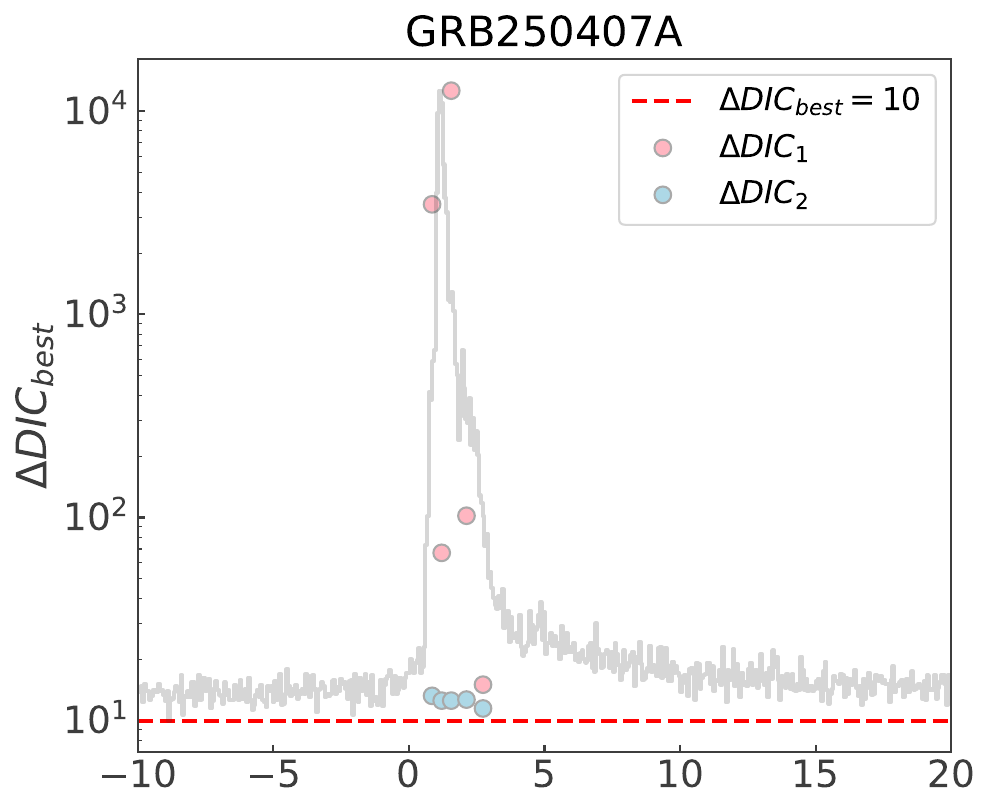}
   \figcaption{Time evolution of $\Delta\mathrm{DIC}_{\rm best}$. The red dashed line marks $\Delta\mathrm{DIC}_{\rm best}=10$; values above this line indicate strong relative support for an additional thermal-like or non-thermal component under the adopted model comparison.} \label{fig 1}
\end{figure}

Table~\ref{table1} gives four illustrative bins for GRB~210619B. Table~\ref{table:mrt} shows a five-row excerpt of the 215-row machine-readable table, which provides the accepted-bin mBB+PL posterior summaries, component fluxes, flux fractions, prior identifier, fit-quality flag, and DIC and PG-stat/dof values for all ten fitted models. Figure~\ref{fig 1} summarizes the DIC comparisons; it is interpreted as evidence for additional spectral structure, not as a plot of component energy fractions. The direct flux counts in Table~\ref{table2} and the bin-by-bin machine-readable table provide the separate evidence for mBB-flux dominance. Because the sample was selected using this event-level property, the 214/215 fraction characterizes the selected sample and is not an occurrence rate for the GBM population.

\begin{table*}[p]
\centering
\caption{Fits to the First Four Time-resolved Spectra of GRB~210619B}
\label{table1}
\resizebox{\textwidth}{!}{%
\begin{tabular}{cccccccccc}
\hline
\hline

Time Bin& \multirow{2}*{$S$} & \multirow{2}*{Model} & \multirow{2}*{$\alpha/m$} & $kT/kT_{\max}$ &\textbf{$F_{\rm th}\times10^{-6}$} & $E_p$ & \textbf{$F_{\rm tot}\times10^{-6}$}& \multirow{2}*{DIC} & \multirow{2}*{\textbf{PG-stat/dof}}\\
$(s)$& & & &(keV)&$(ergs^{-1}cm^{-2})$ &(keV)&$(ergs^{-1}cm^{-2})$& & \\

\hline

\multirow{10}{*}{0.07 - 0.21} & \multirow{10}{*}{38} & Band & $-0.34^{+0.07}_{-0.13}$ &  &  & $607^{+155}_{-64}$ & $21.1^{+0.9}_{-0.8}$ & 941.1 & 1.92 \\
             &  & CPL & $-0.69^{+0.06}_{-0.05}$ &  &  & $1421^{+208}_{-196}$ & $22.6^{+0.7}_{-0.6}$ & 992.5 & 2.02 \\
             &  & Band+BB & $-0.22^{+0.01}_{-0.27}$ & $234^{+41}_{-190}$ & $0.4^{+3.5}_{-0.4}$ & $391^{+396}_{-46}$ & $21.2^{+0.8}_{-0.8}$ & 912.9 & 1.92 \\
             &  & CPL+BB & $-0.81^{+0.04}_{-0.06}$ & $71^{+11}_{-4}$ & $4.7^{+0.9}_{-0.8}$ & $2567^{+299}_{-382}$ & $20.5^{+0.6}_{-0.5}$ & 949.0 & 1.93 \\
             &  & Band+PL & $-0.32^{+0.07}_{-0.13}$ &  &  & $577^{+159}_{-46}$ & $21.1^{+0.9}_{-0.8}$ & 927.7 & 1.93 \\
             &  & CPL+PL & $-0.47^{+0.10}_{-0.08}$ &  &  & $813^{+147}_{-74}$ & $22.3^{+0.7}_{-0.7}$ & 957.4 & 1.97 \\
             &  & Band+BB+PL & $-0.16^{+0.05}_{-0.31}$ & $223^{+25}_{-181}$ & $0.4^{+3.6}_{-0.3}$ & $355^{+414}_{-77}$ & $21.3^{+0.8}_{-0.8}$ & 899.8 & 1.93 \\
             &  & CPL+BB+PL & $-0.79^{+0.08}_{-0.04}$ & $70^{+8}_{-7}$ & $4.4^{+0.9}_{-0.9}$ & $2338^{+205}_{-584}$ & $20.3^{+0.8}_{-0.6}$ & 923.5 & 1.93 \\
             &  & mBB & $-0.27^{+0.06}_{-0.06}$ & $1334^{+211}_{-168}$ &  &  & $20.4^{+0.6}_{-0.5}$ & 992.4 & 1.96 \\
             &  & mBB+PL & $-0.37^{+0.10}_{-0.08}$ & $1046^{+375}_{-153}$ & $18.1^{+0.8}_{-0.8}$ &  & $20.5^{+0.6}_{-0.6}$ & 967.1 & 1.92 \\
            \hline
            \multirow{10}{*}{0.21 - 0.30} & \multirow{10}{*}{40} & Band & $-0.41^{+0.10}_{-0.10}$ &  &  & $580^{+118}_{-86}$ & $28.2^{+1.1}_{-1.1}$ & 572.2 & 1.15 \\
             &  & CPL & $-0.73^{+0.04}_{-0.05}$ &  &  & $1389^{+196}_{-142}$ & $30.7^{+0.9}_{-0.9}$ & 610.0 & 1.22 \\
             &  & Band+BB & $-0.25^{+0.00}_{-0.26}$ & $330^{+54}_{-286}$ & $0.6^{+5.6}_{-0.6}$ & $356^{+336}_{-21}$ & $28.4^{+1.2}_{-1.1}$ & 537.6 & 1.14 \\
             &  & CPL+BB & $-0.84^{+0.04}_{-0.04}$ & $67^{+9}_{-4}$ & $5.4^{+1.0}_{-1.1}$ & $2270^{+243}_{-344}$ & $28.1^{+0.9}_{-0.8}$ & 577.7 & 1.15 \\
             &  & Band+PL & $-0.44^{+0.15}_{-0.06}$ &  &  & $592^{+108}_{-110}$ & $28.2^{+1.2}_{-1.1}$ & 559.4 & 1.15 \\
             &  & CPL+PL & $-0.70^{+0.14}_{-0.01}$ &  &  & $1206^{+18}_{-317}$ & $30.4^{+0.9}_{-0.9}$ & 579.6 & 1.20 \\
             &  & Band+BB+PL & $-0.21^{+0.03}_{-0.27}$ & $315^{+20}_{-269}$ & $1.0^{+5.8}_{-0.9}$ & $349^{+320}_{-5}$ & $28.4^{+1.1}_{-1.1}$ & 509.8 & 1.14 \\
             &  & CPL+BB+PL & $-0.25^{+0.40}_{-0.59}$ & $465^{+379}_{-404}$ & $5.2^{+1.9}_{-1.3}$ & $361^{+1734}_{-876}$ & $28.1^{+0.9}_{-0.9}$ & $-828.6^{\dagger}$ & 1.14 \\
             &  & mBB & $-0.18^{+0.06}_{-0.08}$ & $1007^{+175}_{-129}$ &  &  & $28.4^{+1.0}_{-0.9}$ & 619.7 & 1.18 \\
             &  & mBB+PL & $-0.27^{+0.15}_{-0.01}$ & $831^{+131}_{-175}$ & $25.2^{+1.3}_{-1.1}$ &  & $28.6^{+0.9}_{-0.8}$ & 594.4 & 1.14 \\
            \hline
            \multirow{10}{*}{0.30 - 0.40} & \multirow{10}{*}{54} & Band & $-0.48^{+0.06}_{-0.06}$ &  &  & $602^{+68}_{-66}$ & $41.1^{+1.2}_{-1.2}$ & 661.8 & 1.33 \\
             &  & CPL & $-0.68^{+0.04}_{-0.05}$ &  &  & $1010^{+110}_{-82}$ & $44.3^{+1.0}_{-1.0}$ & 718.4 & 1.45 \\
             &  & Band+BB & $-0.45^{+0.03}_{-0.10}$ & $56^{+73}_{-15}$ & $0.2^{+1.5}_{-0.2}$ & $563^{+109}_{-28}$ & $41.2^{+1.2}_{-1.2}$ & 648.5 & 1.33 \\
             &  & CPL+BB & $-0.36^{+0.43}_{-0.57}$ & $466^{+369}_{-391}$ & $9.6^{+1.7}_{-1.7}$ & $395^{+1745}_{-942}$ & $40.9^{+1.0}_{-1.0}$ & $-1169.2^{\dagger}$ & 1.39 \\
             &  & Band+PL & $-0.49^{+0.08}_{-0.04}$ &  &  & $589^{+78}_{-47}$ & $41.2^{+1.2}_{-1.2}$ & 648.2 & 1.34 \\
             &  & CPL+PL & $-0.59^{+0.09}_{-0.04}$ &  &  & $807^{+64}_{-96}$ & $43.4^{+1.1}_{-1.1}$ & 685.1 & 1.39 \\
             &  & Band+BB+PL & $-0.47^{+0.06}_{-0.07}$ & $41^{+86}_{-0}$ & $0.3^{+1.5}_{-0.3}$ & $580^{+88}_{-49}$ & $41.2^{+1.2}_{-1.2}$ & 635.1 & 1.34 \\
             &  & CPL+BB+PL & $-0.35^{+0.13}_{-0.04}$ & $464^{+6}_{-97}$ & $9.6^{+2.4}_{-1.6}$ & $427^{+14}_{-67}$ & $40.7^{+1.2}_{-1.2}$ & 334.9 & 1.35 \\
             &  & mBB & $-0.18^{+0.06}_{-0.08}$ & $788^{+200}_{-101}$ &  &  & $41.2^{+1.3}_{-1.3}$ & 728.7 & 1.41 \\
             &  & mBB+PL & $-0.14^{+0.14}_{-0.11}$ & $538^{+121}_{-110}$ & $37.1^{+1.9}_{-1.8}$ &  & $41.9^{+1.2}_{-1.2}$ & 697.2 & 1.36 \\
            \hline
            \multirow{10}{*}{0.40 - 0.50} & \multirow{10}{*}{70} & Band & $-0.38^{+0.06}_{-0.07}$ &  &  & $471^{+57}_{-39}$ & $50.9^{+1.3}_{-1.3}$ & 800.5 & 1.62 \\
             &  & CPL & $-0.70^{+0.03}_{-0.04}$ &  &  & $984^{+92}_{-66}$ & $56.8^{+1.1}_{-1.1}$ & 899.7 & 1.82 \\
             &  & Band+BB & $-0.23^{+0.06}_{-0.09}$ & $341^{+33}_{-66}$ & $9.2^{+2.5}_{-2.1}$ & $329^{+43}_{-26}$ & $51.5^{+1.3}_{-1.2}$ & 769.1 & 1.58 \\
             &  & CPL+BB & $-0.89^{+0.04}_{-0.01}$ & $62^{+4}_{-2}$ & $10.3^{+1.1}_{-1.1}$ & $2027^{+50}_{-327}$ & $51.4^{+0.9}_{-0.9}$ & 824.1 & 1.64 \\
             &  & Band+PL & $-0.35^{+0.04}_{-0.09}$ &  &  & $454^{+69}_{-25}$ & $50.9^{+1.3}_{-1.2}$ & 787.4 & 1.62 \\
             &  & CPL+PL & $-0.58^{+0.05}_{-0.05}$ &  &  & $745^{+69}_{-53}$ & $55.5^{+1.1}_{-1.1}$ & 855.7 & 1.75 \\
             &  & Band+BB+PL & $-0.20^{+0.05}_{-0.10}$ & $347^{+29}_{-68}$ & $9.4^{+2.3}_{-2.0}$ & $321^{+43}_{-21}$ & $51.4^{+1.2}_{-1.2}$ & 757.1 & 1.59 \\
             &  & CPL+BB+PL & $-0.28^{+0.17}_{-0.01}$ & $450^{+12}_{-77}$ & $12.9^{+2.7}_{-2.3}$ & $356^{+14}_{-48}$ & $49.9^{+1.2}_{-1.2}$ & 780.4 & 1.62 \\
             &  & mBB & $-0.32^{+0.04}_{-0.06}$ & $1010^{+205}_{-97}$ &  &  & $51.0^{+1.2}_{-1.2}$ & 858.5 & 1.68 \\
             &  & mBB+PL & $-0.42^{+0.08}_{-0.07}$ & $898^{+175}_{-167}$ & $45.8^{+1.7}_{-1.5}$ &  & $51.2^{+1.2}_{-1.1}$ & 826.9 & 1.62 \\

\hline
\hline
\end{tabular}%
}

\noindent
\footnotesize Note. $F_\mathrm{th}$ is $F_\mathrm{BB}$ for single-blackbody models and $F_\mathrm{mBB}$ for multicolor-blackbody models; $F_\mathrm{tot}$ is the total model flux in 10--1000~keV. PG-stat/dof is the PG-statistic divided by the number of degrees of freedom. Values marked $\dagger$ are unstable, prior-edge DIC evaluations excluded from model-preference counts and physical interpretation.

\end{table*}

\begin{table*}[htbp]
    \centering
    \caption{Sample and model-comparison summary}
    \label{table2}
    \resizebox{\textwidth}{!}{%
    \begin{tabular}{ccccccccc}
        \hline
        \hline
        \multirow{2}{*}{GRB} & Time Period & \multirow{2}{*}{Selected Detectors} & \multirow{2}{*}{$z$} & \textbf{Fluence}$\times10^{-5}$ & $T_{90}$ & \multicolumn{2}{c}{Bins with $\Delta\mathrm{DIC}>10$} & \multirow{2}{*}{$N(f_{\rm mBB}>0.5)/N_{\rm acc}$} \\
        \cline{7-8}
        & $(s)$ & & & \textbf{($\mathrm{erg\,cm^{-2}}$)} & $(s)$ & added BB & added PL & \\
        \hline
        GRB 130518A & \(0 \sim 60\) & $n3,n6,n7,b0$ & 2.488 & 9.459 & 48.577 & $100\%(15/15)$ & $100\%(15/15)$ & \textbf{$15/15$} \\
        GRB 210610B & \(10 \sim 95\) & $n9,na,nb,b1$ & 1.1345 & 11.042 & 50.041 & $100\%(13/13)$ & $100\%(13/13)$ & \textbf{$13/13$} \\
        GRB 210619B & \(0 \sim 70\) & $n4,n7,n8,b1$ & 1.937 & 30.248 & 54.785 & $100\%(41/41)$ & $100\%(41/41)$ & \textbf{$41/41$} \\
        GRB 211211A & \(0 \sim 55\) & $n1,n2,na,b0$ & 0.076 & 50.118 & 34.305 & \textbf{$92\%(95/103)$} & $91\%(94/103)$ & \textbf{$102/103$}\\
        GRB 231129C & \(0.1 \sim 8\) & $n3,n6,n7,b0$ & 0.5 & 7.8699 & 5.824 & $100\%(11/11)$ & $100\%(11/11)$ & \textbf{$11/11$} \\
        GRB 240825A & \(0.1 \sim 8\) & $n6,n7,nb,b1$ & 0.659 & 10.017 & 3.968 & $90\%(18/20)$ & $90\%(18/20)$ & \textbf{$20/20$}\\
        GRB 241228B & \(0 \sim 25\) & $n6,n7,n8,b1$ & 2.674 & 4.2677 & 19.456 & $100\%(7/7)$ & $100\%(7/7)$ & \textbf{$7/7$} \\
        GRB 250407A & \(0.1 \sim 10\) & $n3,n4,n5,b0$ & 1.36 & 4.5797 & 6.272 & \textbf{$60\%(3/5)$} & $100\%(5/5)$ & \textbf{$5/5$} \\
        \hline
        \hline
    \end{tabular}%
    }

\noindent
\footnotesize Note. The Fluence values are taken directly from the MPE GRB catalog and represent time-integrated energy used for sample selection only. They were not converted into the $F_\mathrm{tot}$ values used in the physical calculations. All $F_\mathrm{tot}$ values reported elsewhere in this work are 10--1000~keV time-averaged energy fluxes obtained from the corresponding adopted time-integrated spectral fits over the listed source intervals. For both H2013 and G2015, $F_\mathrm{BB}$ and $F_\mathrm{tot}$ are taken from the same fitted spectral model, with $\phi=f_\mathrm{th}=F_\mathrm{BB}/F_\mathrm{tot}$. The two DIC columns count bins that pass the indicated model comparison; they are not component energy fractions. The last column independently reports the number of accepted bins in which the fitted mBB contribution exceeds half of the mBB+PL energy flux. Redshifts are based on afterglow/host spectroscopy for GRB~130518A \citep{GCN14687}, GRB~210610B \citep{GCN30194}, GRB~210619B \citep{GCN30272}, GRB~240825A \citep{GCN37293}, GRB~241228B \citep{GCN38704}, and GRB~250407A \citep{GCN40175}. For GRB~211211A, $z=0.076$ is the redshift of the associated galaxy rather than a direct afterglow redshift \citep{GCN31221}; for GRB~231129C we adopt the estimated value reported by \citet{2024ApJ...972..132C}. These two cases are flagged because their redshift provenance differs from secure afterglow spectroscopy.

\end{table*}

\begin{deluxetable*}{lccccccccc}
\tablecaption{Excerpt from the Time-resolved mBB+PL Results\label{table:mrt}}
\tablehead{
\colhead{GRB} & \colhead{Time bin} & \colhead{$S$} & \colhead{$m$} & \colhead{$kT_{\min}$} & \colhead{$kT_{\max}$} & \colhead{$F_{\rm mBB}$} & \colhead{$F_{\rm tot}$} & \colhead{$f_{\rm mBB}$} & \colhead{Flag} \\
\colhead{} & \colhead{(s)} & \colhead{} & \colhead{} & \colhead{(keV)} & \colhead{(keV)} & \multicolumn{2}{c}{$(10^{-6}\,\mathrm{erg\,cm^{-2}\,s^{-1}})$} & \colhead{} & \colhead{}
}
\startdata
210619B & 0.07--0.21 & 38 & $-0.370^{+0.100}_{-0.080}$ & $26.11^{+3.63}_{-3.03}$ & $1046.26^{+374.99}_{-153.23}$ & $18.110^{+0.780}_{-0.780}$ & $20.480^{+0.640}_{-0.570}$ & 0.8843 & G \\
210619B & 0.21--0.30 & 40 & $-0.270^{+0.150}_{-0.010}$ & $22.81^{+0.10}_{-6.47}$ & $831.16^{+130.81}_{-175.32}$ & $25.230^{+1.300}_{-1.070}$ & $28.570^{+0.880}_{-0.750}$ & 0.8831 & U \\
210619B & 0.30--0.40 & 54 & $-0.140^{+0.140}_{-0.110}$ & $16.21^{+4.18}_{-3.40}$ & $537.94^{+120.82}_{-110.28}$ & $37.060^{+1.850}_{-1.750}$ & $41.870^{+1.210}_{-1.240}$ & 0.8851 & U \\
210619B & 0.40--0.50 & 70 & $-0.420^{+0.080}_{-0.070}$ & $20.81^{+2.20}_{-2.15}$ & $898.04^{+175.42}_{-166.68}$ & $45.810^{+1.650}_{-1.480}$ & $51.160^{+1.200}_{-1.130}$ &0.8954 & G \\
211211A & 15.26--16.07 & 36 & $0.770^{+0.050}_{-1.890}$ & $1.09^{+5.71}_{-1.11}$ & $25.81^{+494.67}_{-0.11}$ & $0.970^{+1.330}_{-0.310}$ & $2.750^{+0.090}_{-0.100}$ & 0.3527 & G
\enddata
\tablecomments {Table~\ref{table:mrt} is published in its entirety in machine-readable form as \texttt{tab3.txt}. A portion is shown here for guidance regarding its form and content. The complete table contains 215 accepted bins and 45 fixed-width columns, including $kT_{\min}$, all asymmetric errors, the derived PL flux, a prior-configuration identifier, stable added-BB/PL comparison flags, and DIC and PG-stat/dof values for all ten fitted models. G denotes a reported fit; U denotes a bin containing at least one unstable auxiliary compound-model DIC value that is reported but excluded from preference counts.}
\digitalasset
\end{deluxetable*}

\subsection{Distribution of Spectral Parameters}

\begin{figure}[htbp]
\centering
\includegraphics[width=0.46\textwidth]{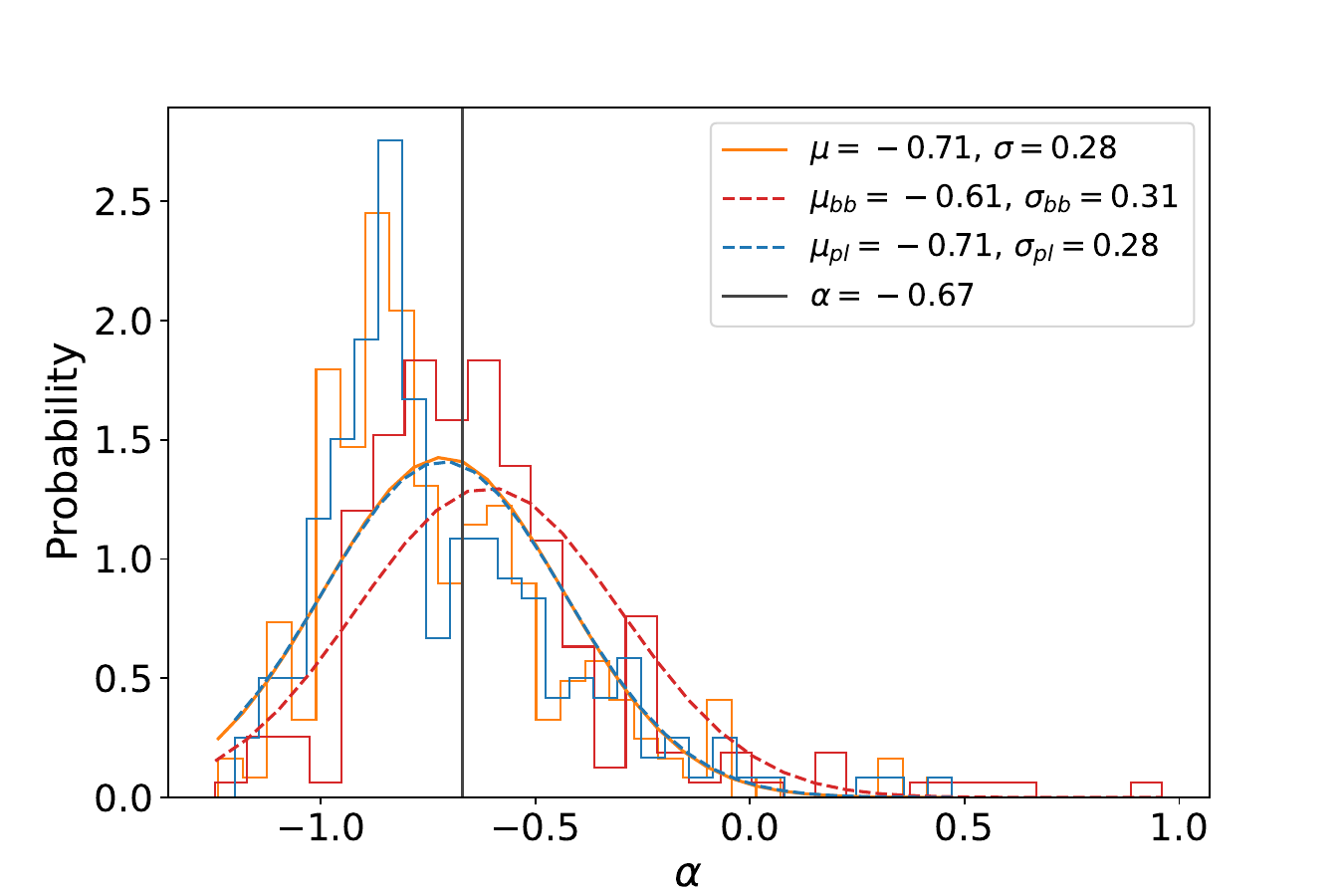}
\includegraphics[width=0.46\textwidth]{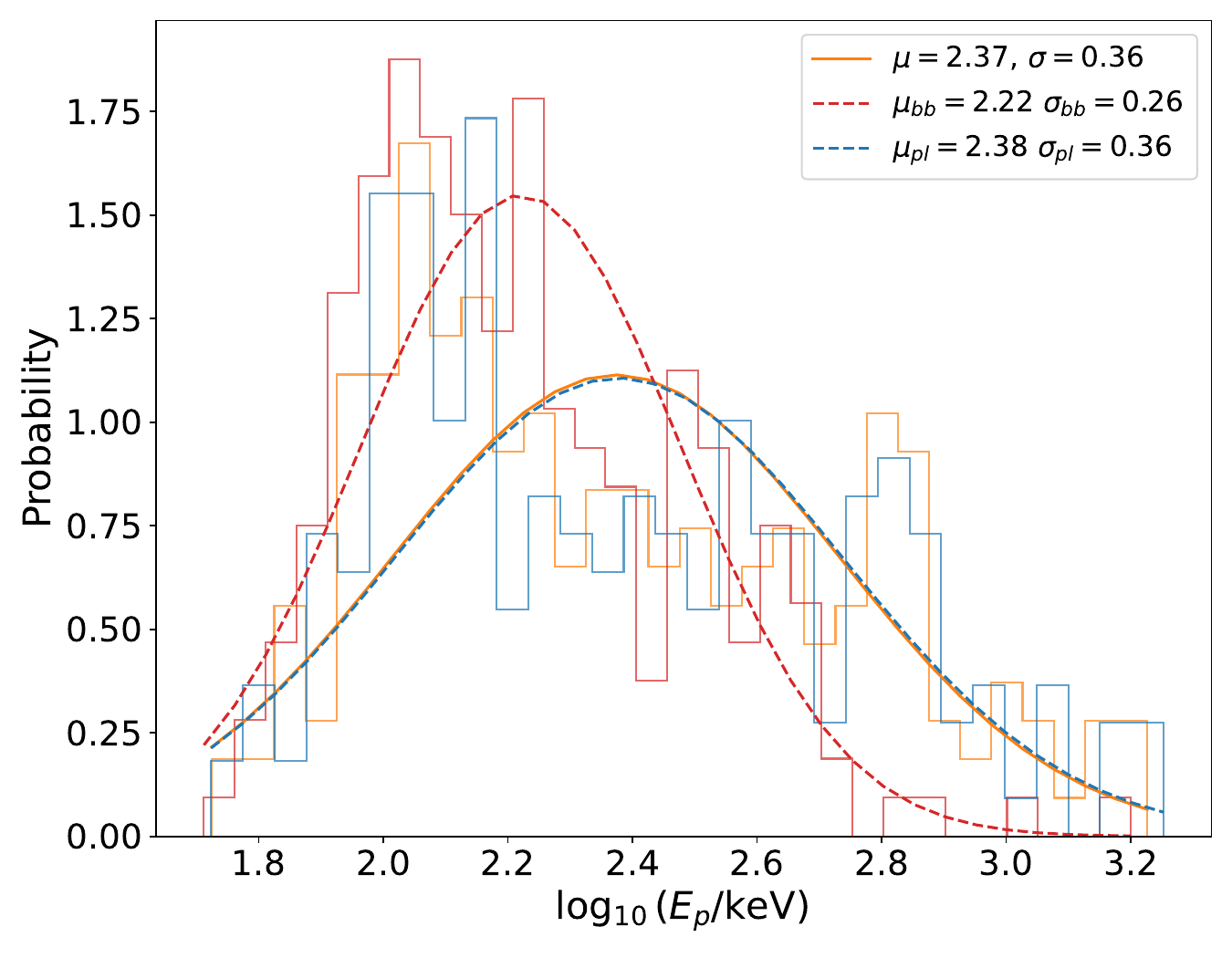}
   \figcaption{Distributions of the low-energy photon index $\alpha$ (left) and peak energy $E_p$ (right) for the full sample. The green solid line marks $\alpha=-2/3$. The black dashed curves represent the time-resolved baseline fits, and the blue and yellow dashed curves represent the fits with additional BB and PL components, respectively.} \label{fig 2}

\end{figure}

The spectral properties of the eight GRBs were further compared through the parameter distributions obtained from the time-resolved fits. Figure~\ref{fig 2} shows the distributions of the low-energy spectral index $\alpha$ and peak energy $E_p$ for the full sample, together with the means and standard deviations derived from Gaussian fits.

\begin{figure}[htbp]
\centering
\includegraphics[width=0.32\textwidth]{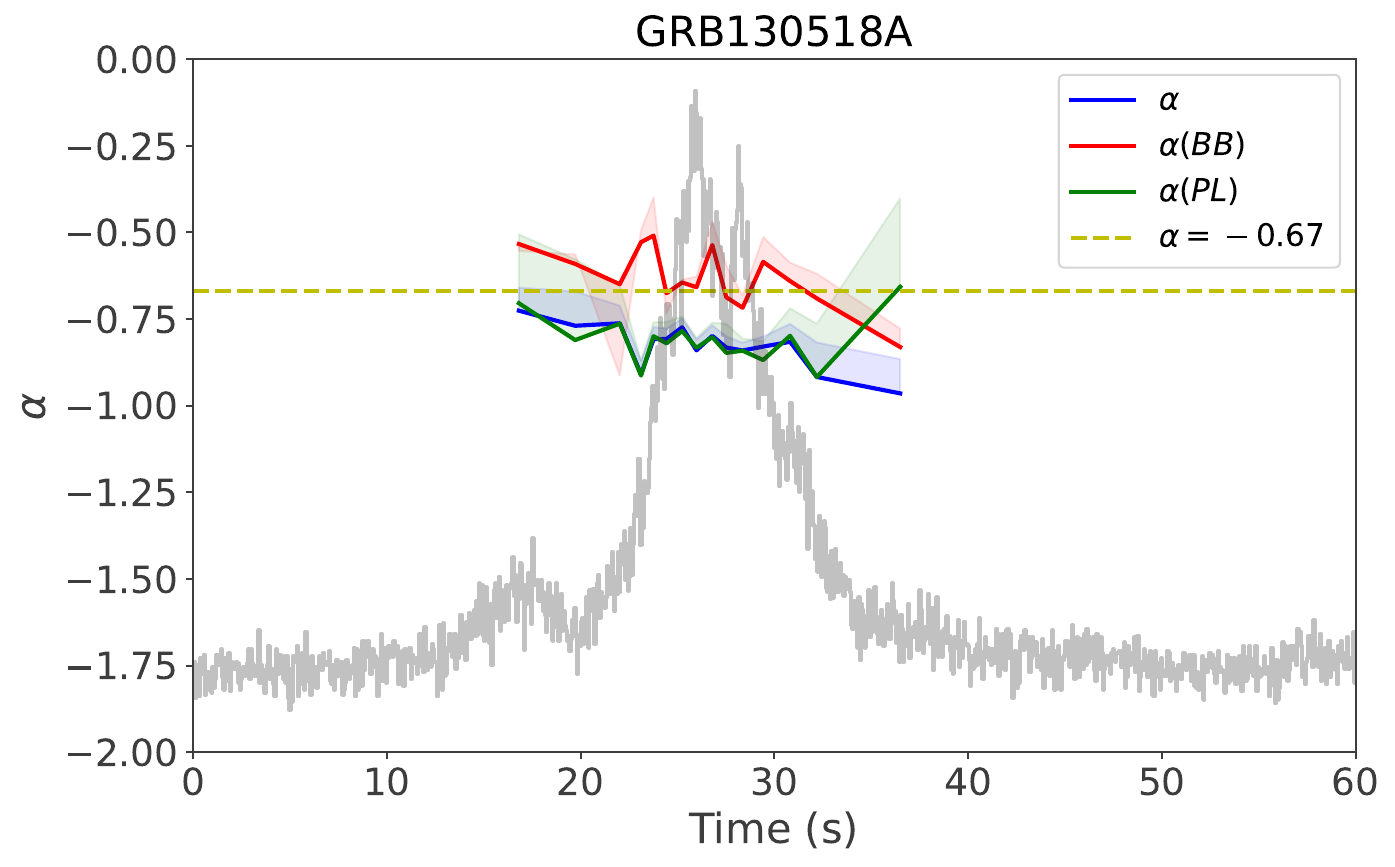}
\includegraphics[width=0.32\textwidth]{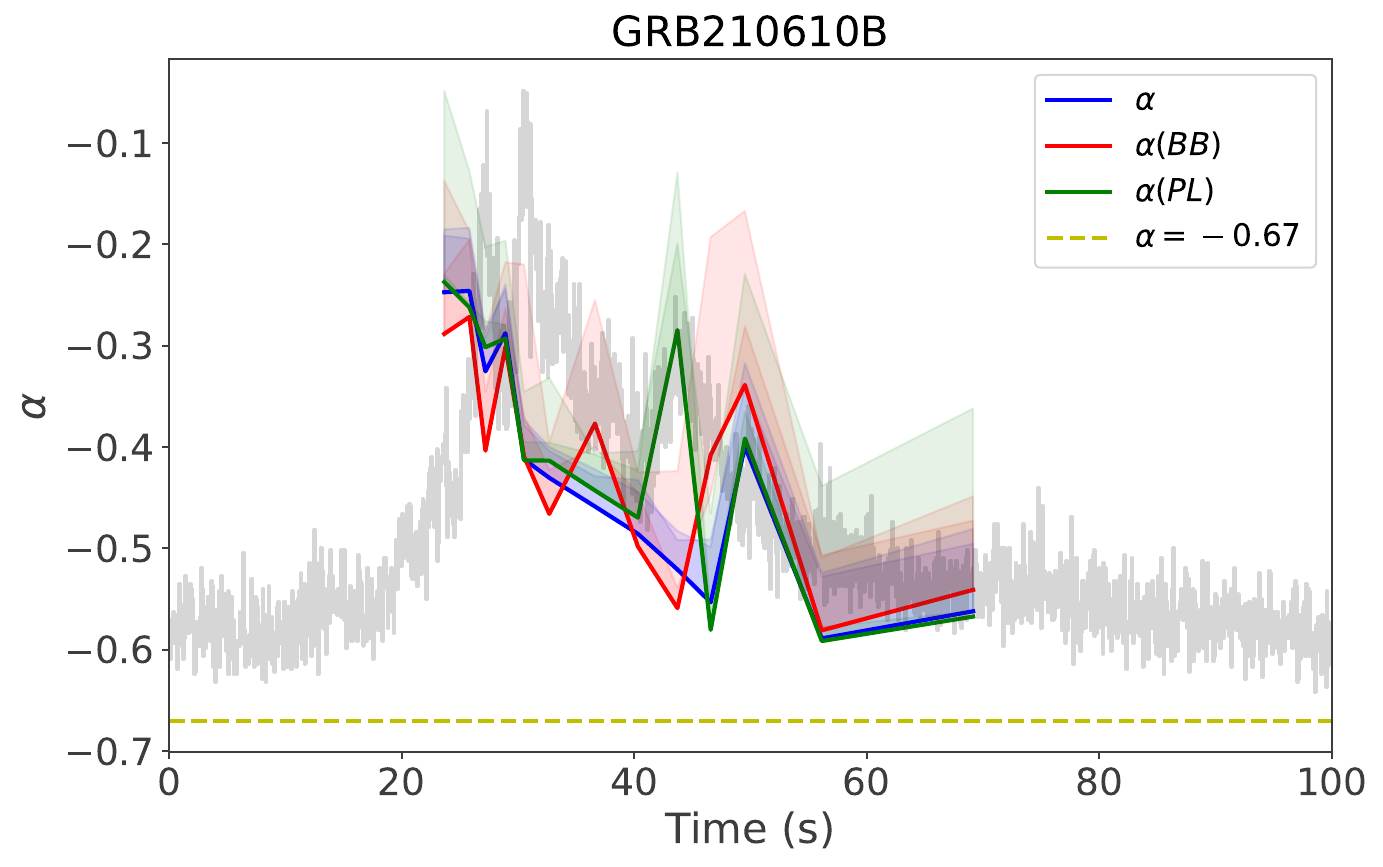}
\includegraphics[width=0.32\textwidth]{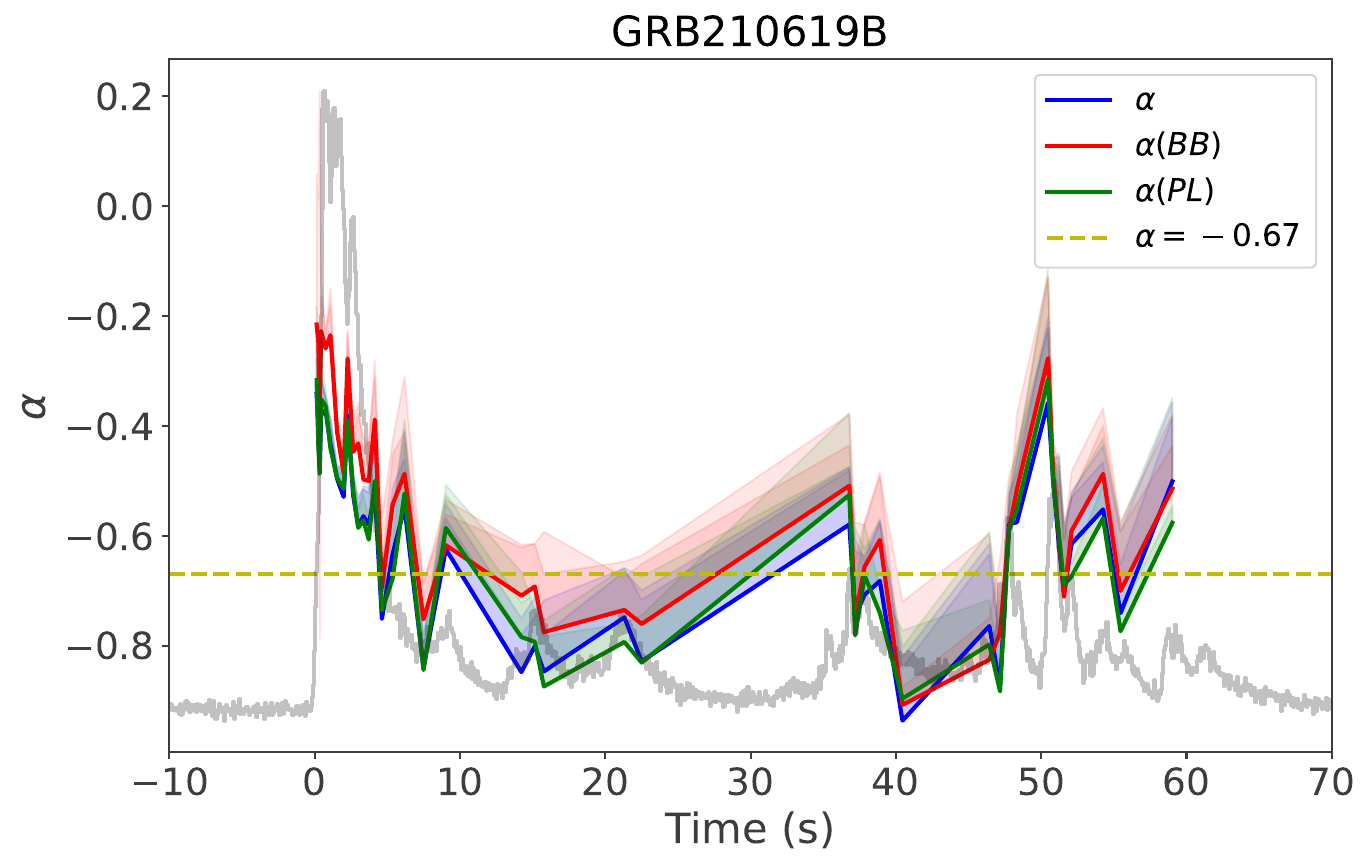}
\includegraphics[width=0.32\textwidth]{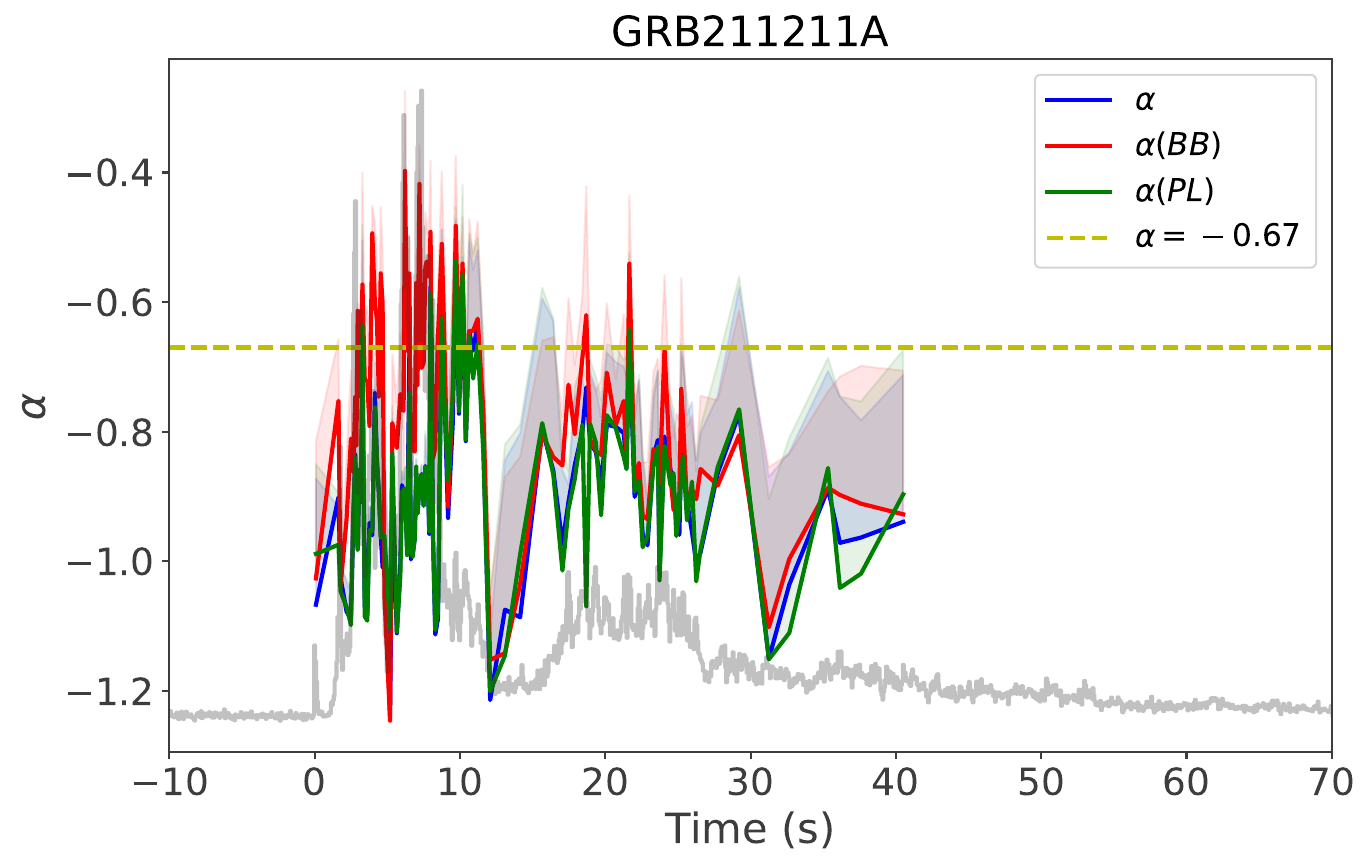}
\includegraphics[width=0.32\textwidth]{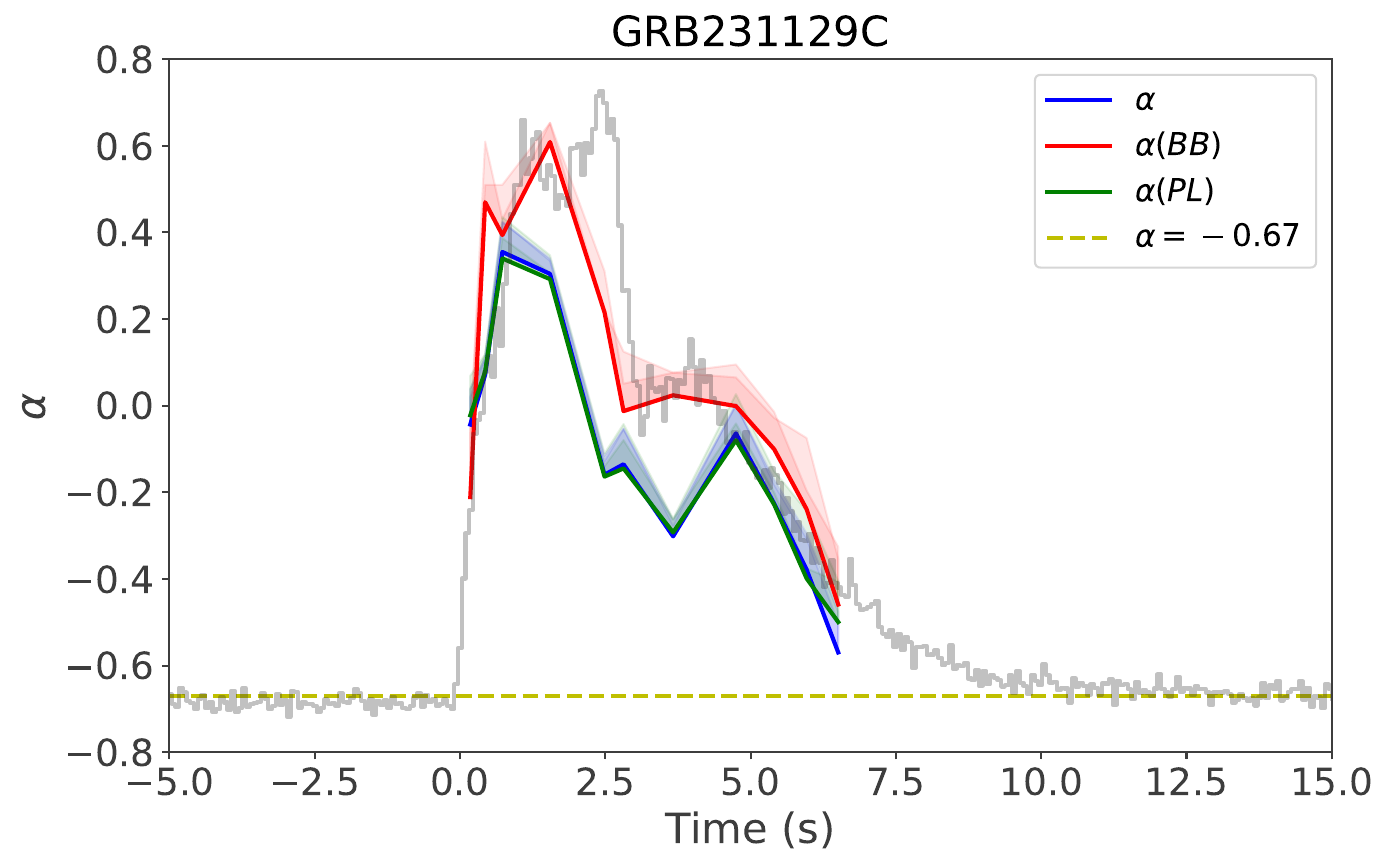}
\includegraphics[width=0.32\textwidth]{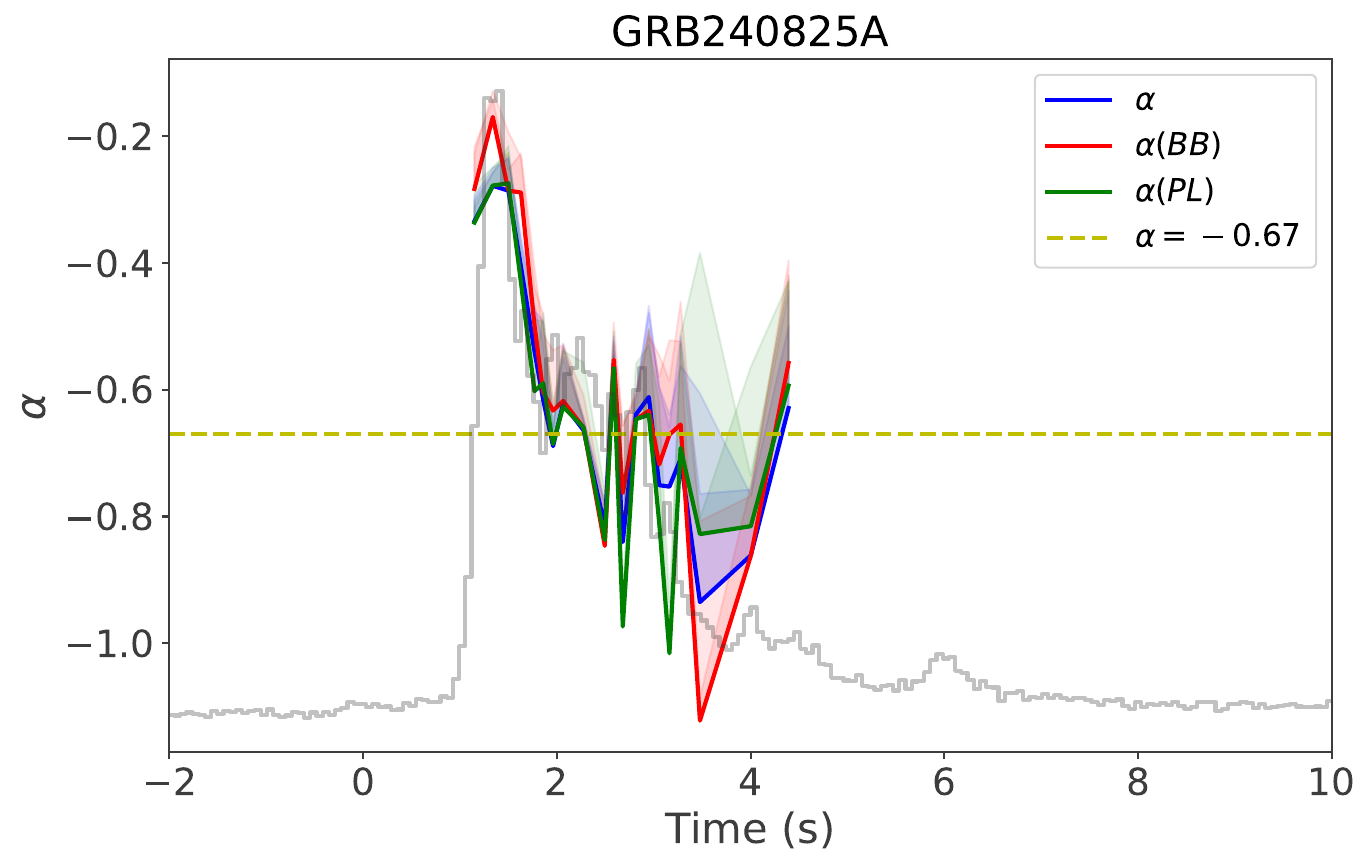}
\includegraphics[width=0.32\textwidth]{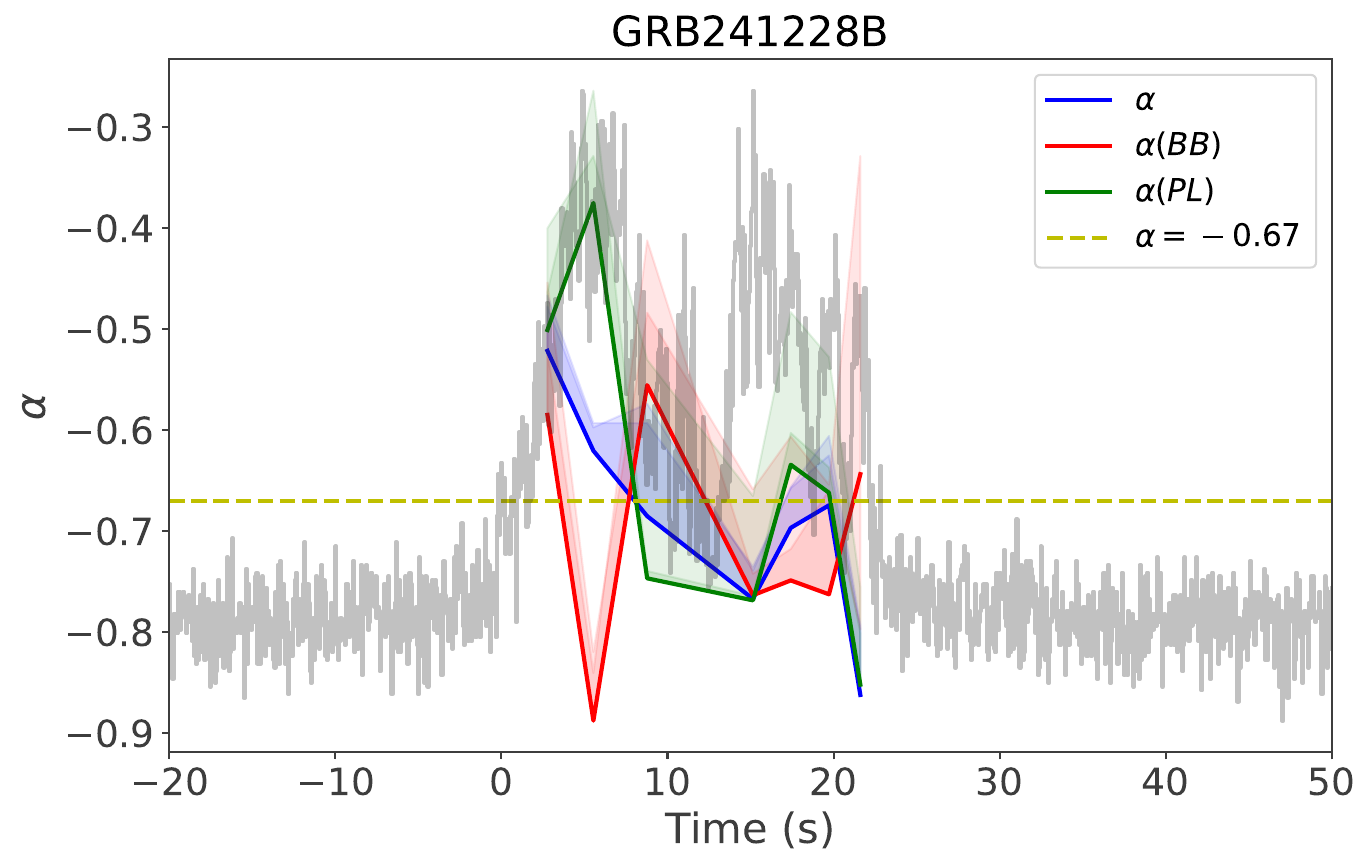}
\includegraphics[width=0.32\textwidth]{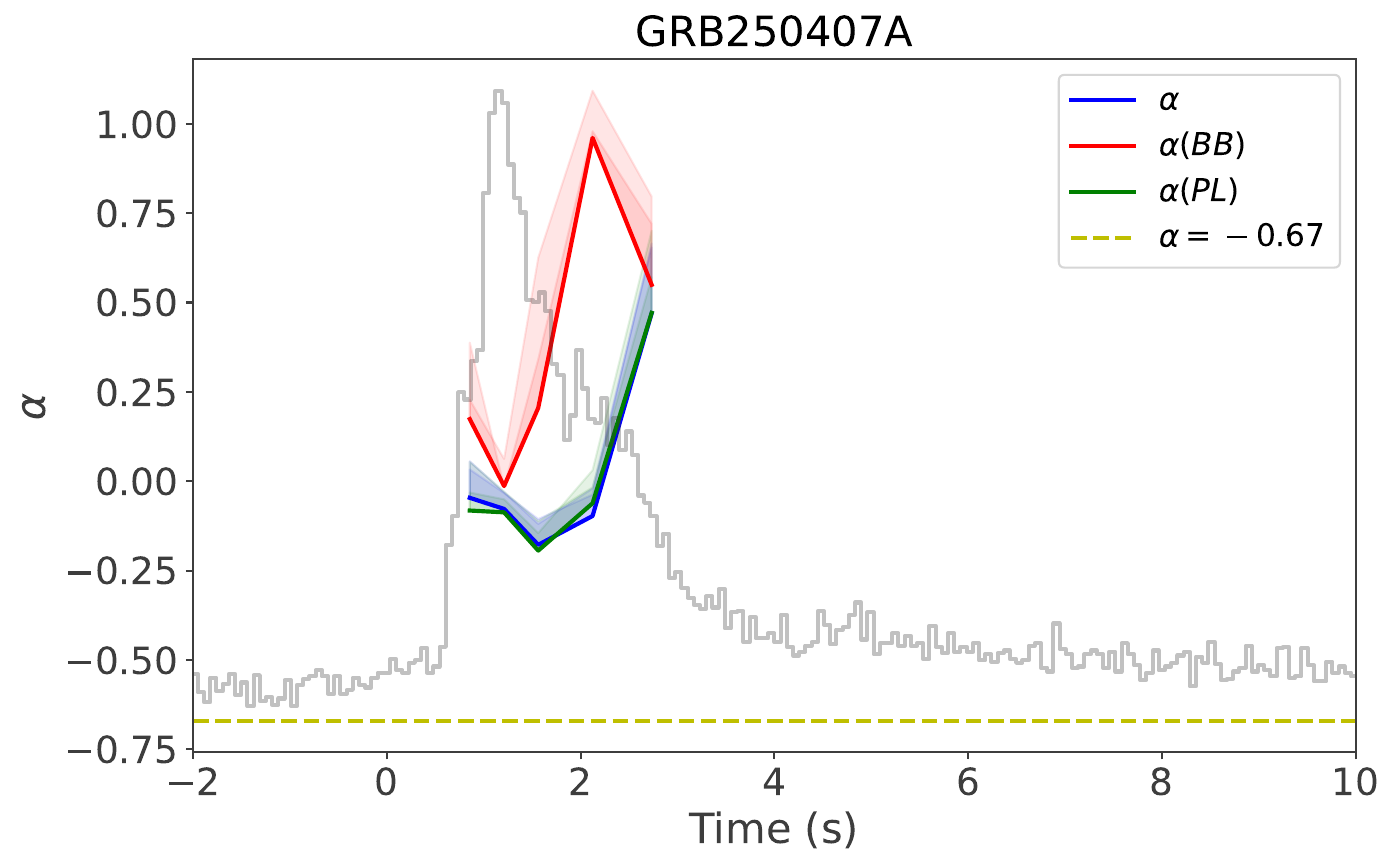}
   \figcaption{Time evolution and distribution of the low-energy photon index $\alpha$ for the eight GRBs. The yellow dashed line marks $\alpha=-2/3$. The blue curves represent the baseline fits, and the red and green curves represent the fits with additional BB and PL components, respectively.} \label{fig 3}
\end{figure}

\begin{figure}[htbp]
\centering
\includegraphics[width=0.32\textwidth]{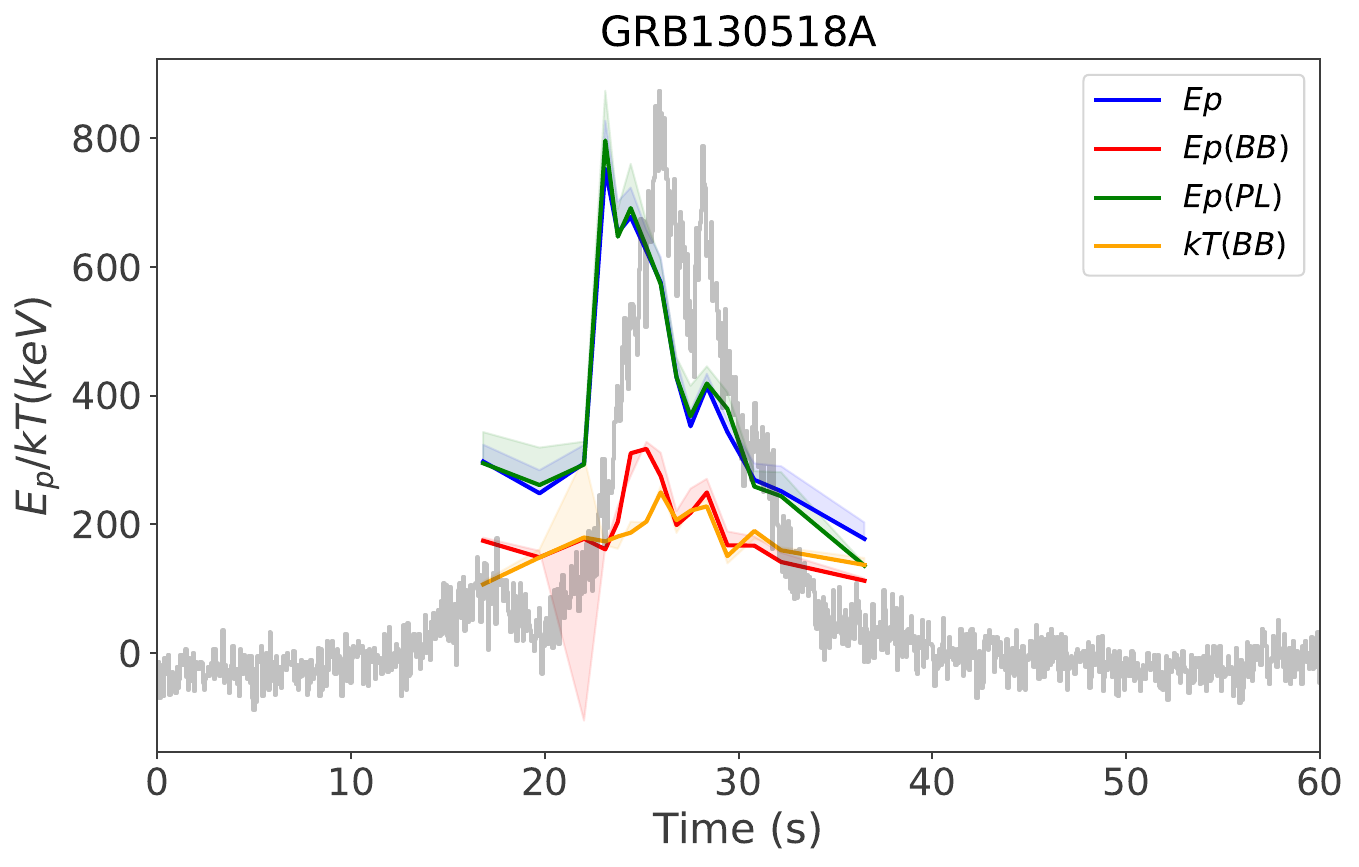}
\includegraphics[width=0.32\textwidth]{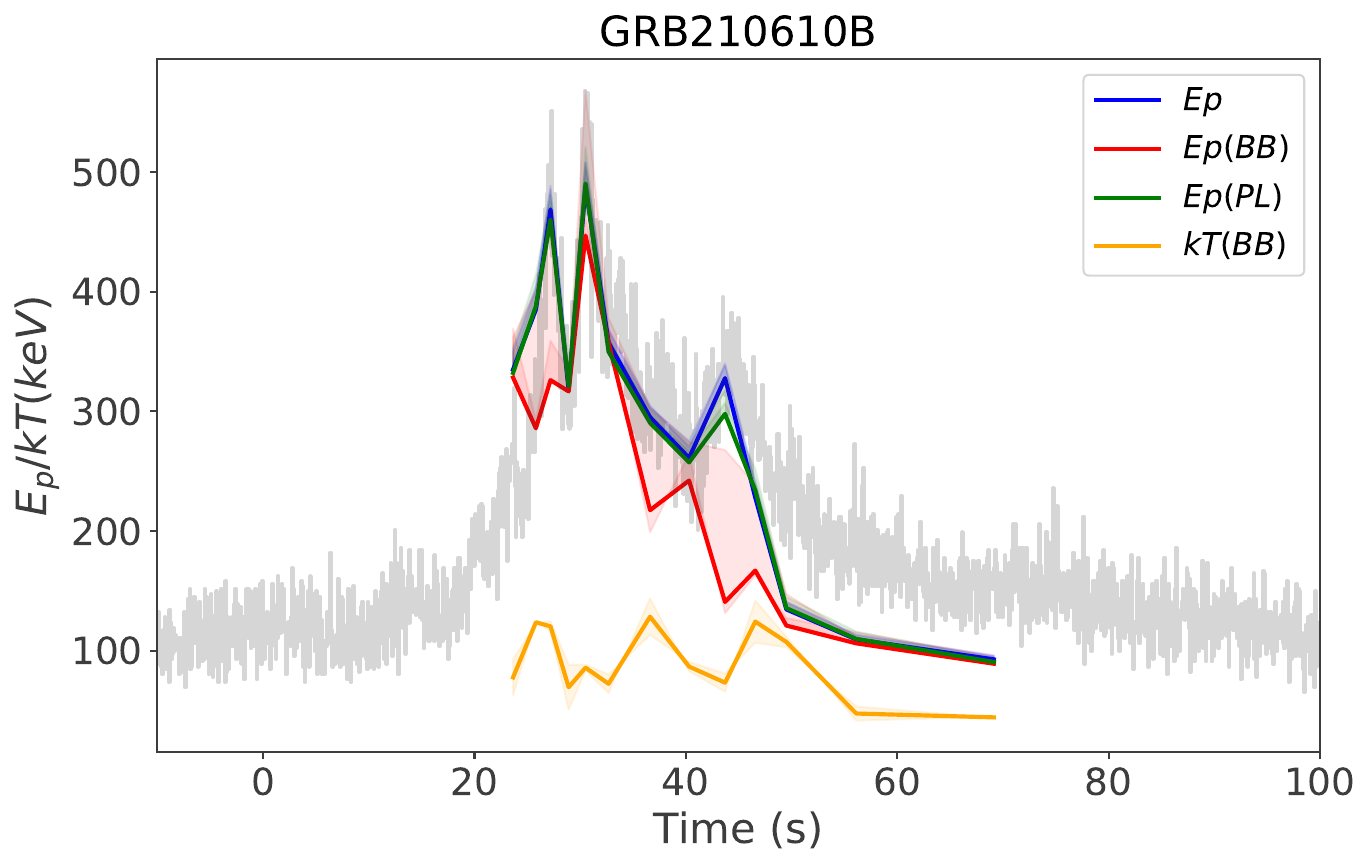}
\includegraphics[width=0.32\textwidth]{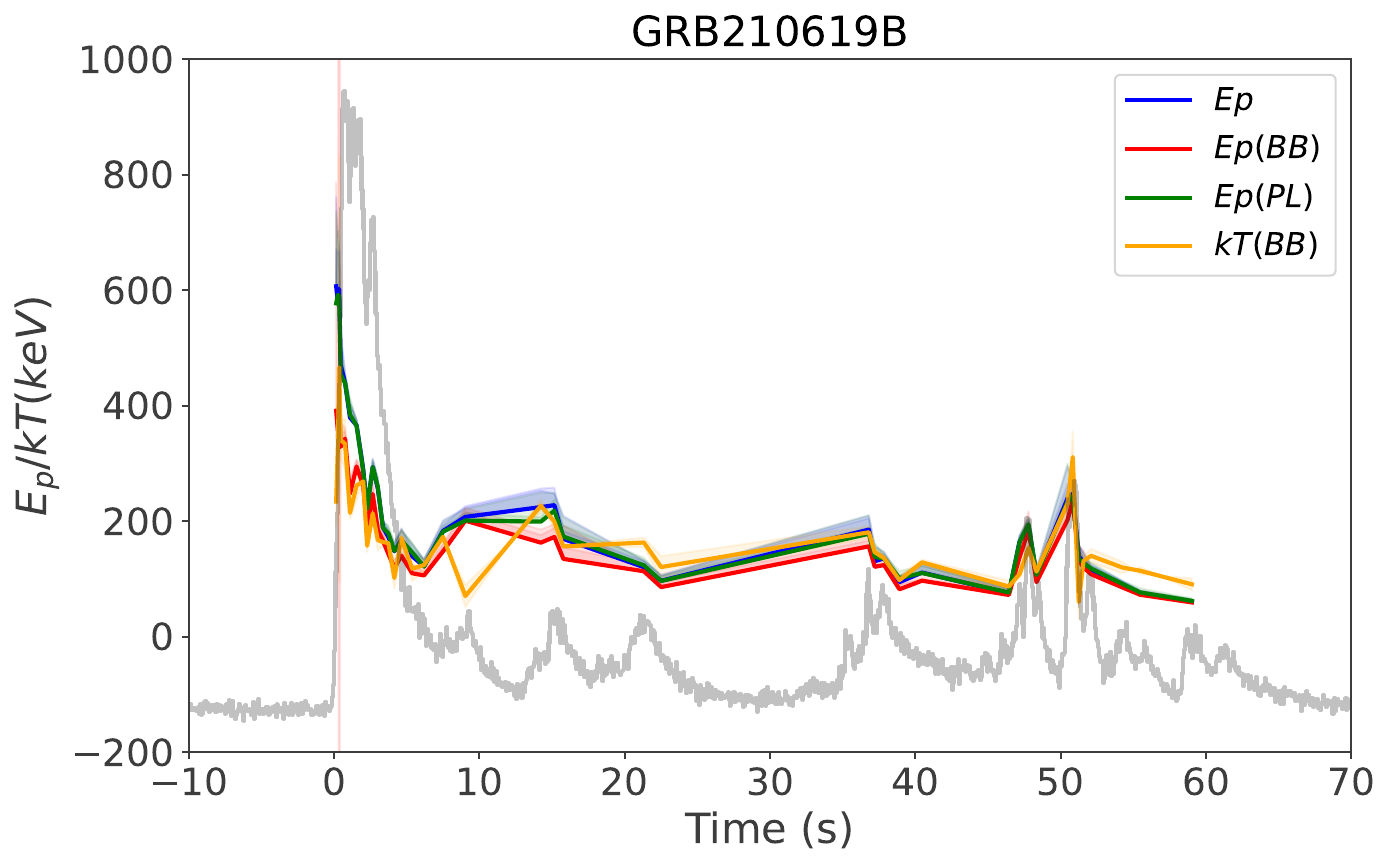}
\includegraphics[width=0.32\textwidth]{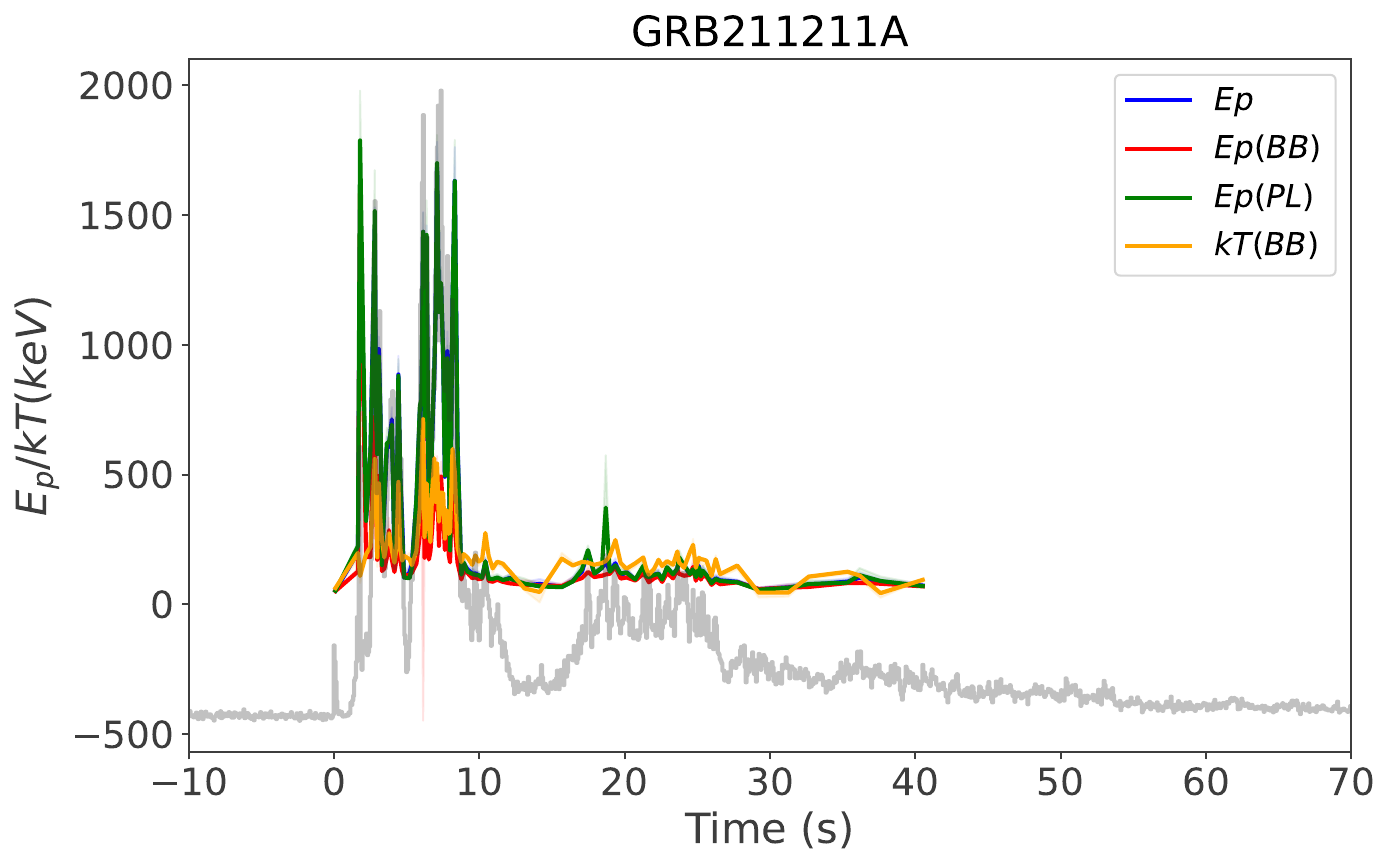}
\includegraphics[width=0.32\textwidth]{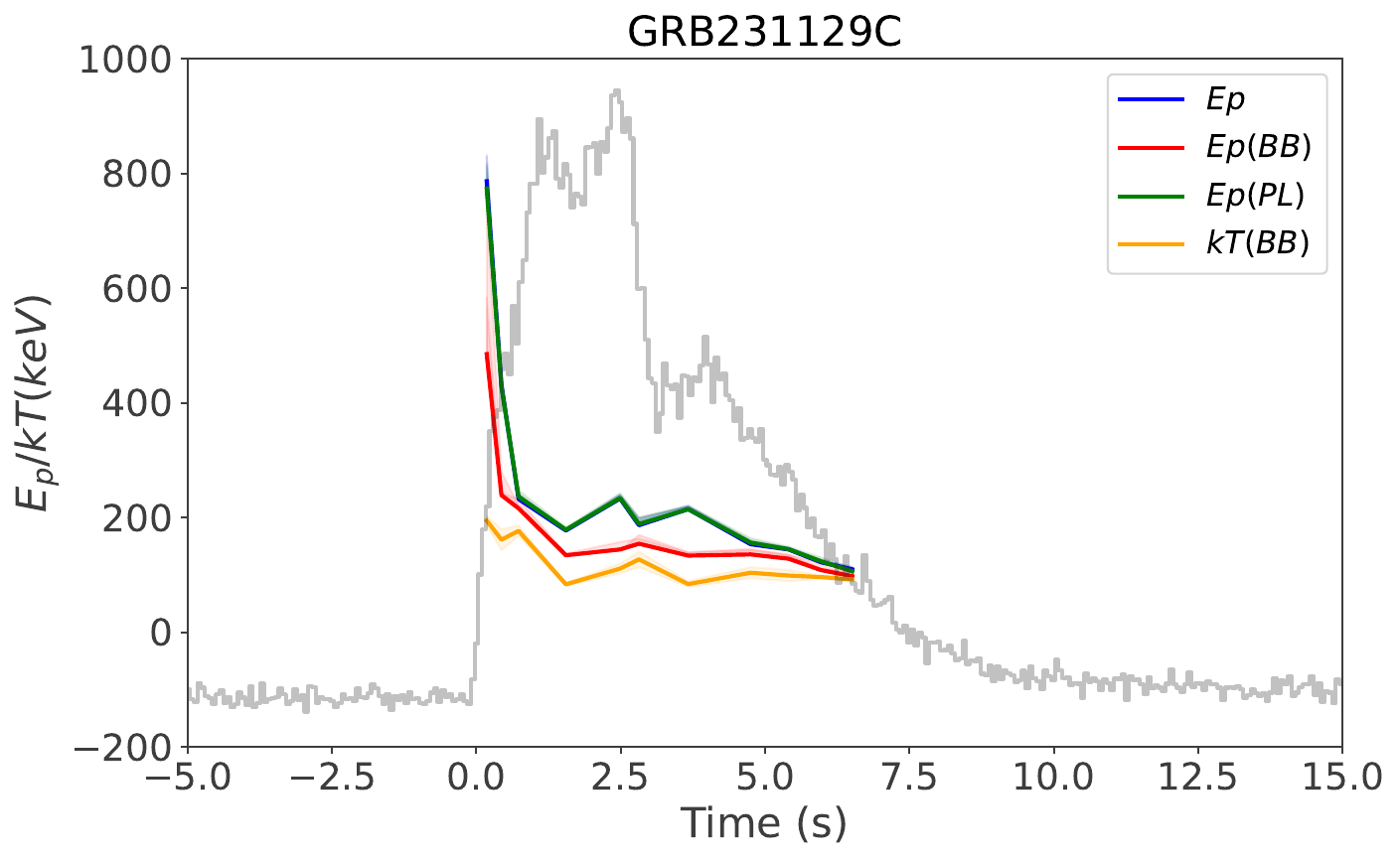}
\includegraphics[width=0.32\textwidth]{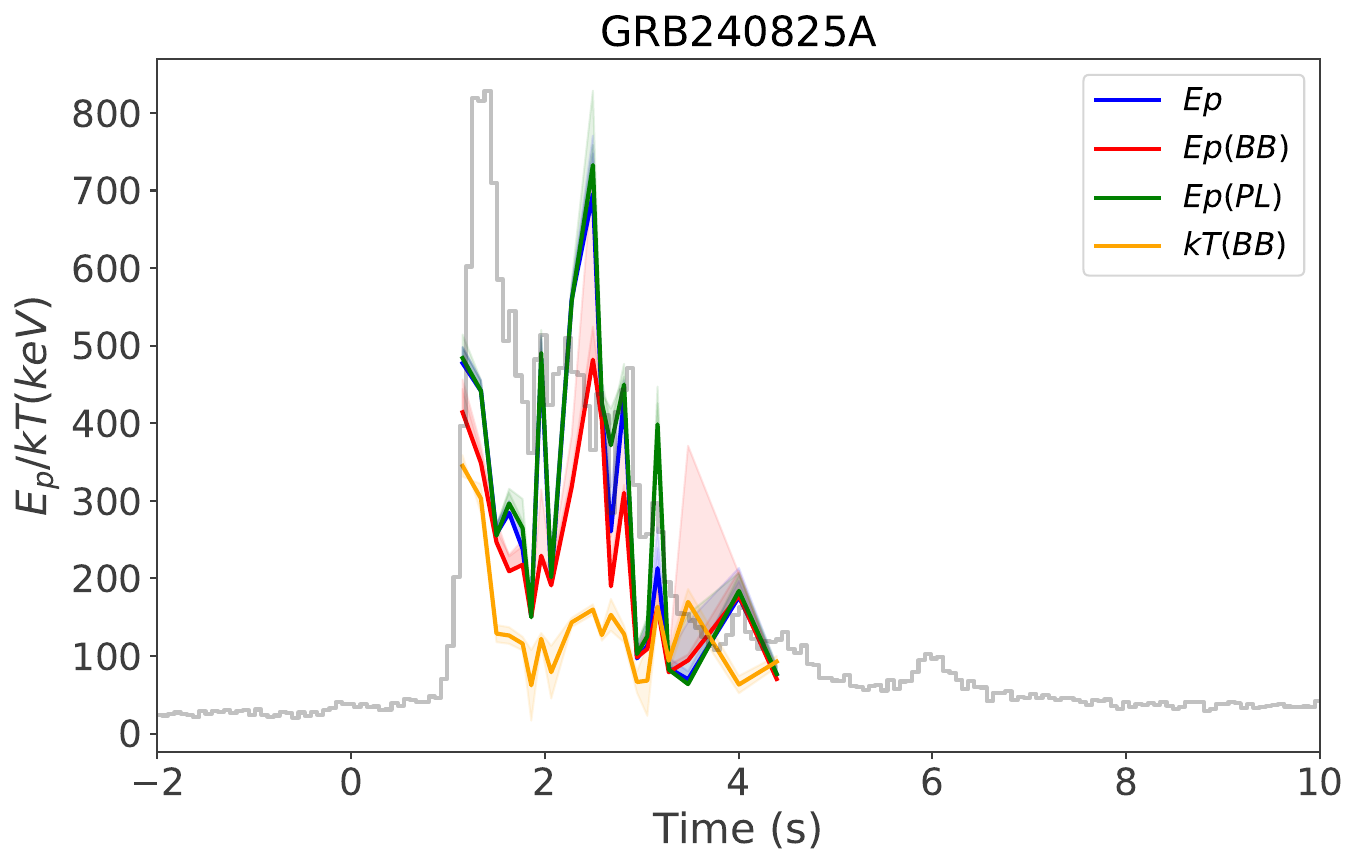}
\includegraphics[width=0.32\textwidth]{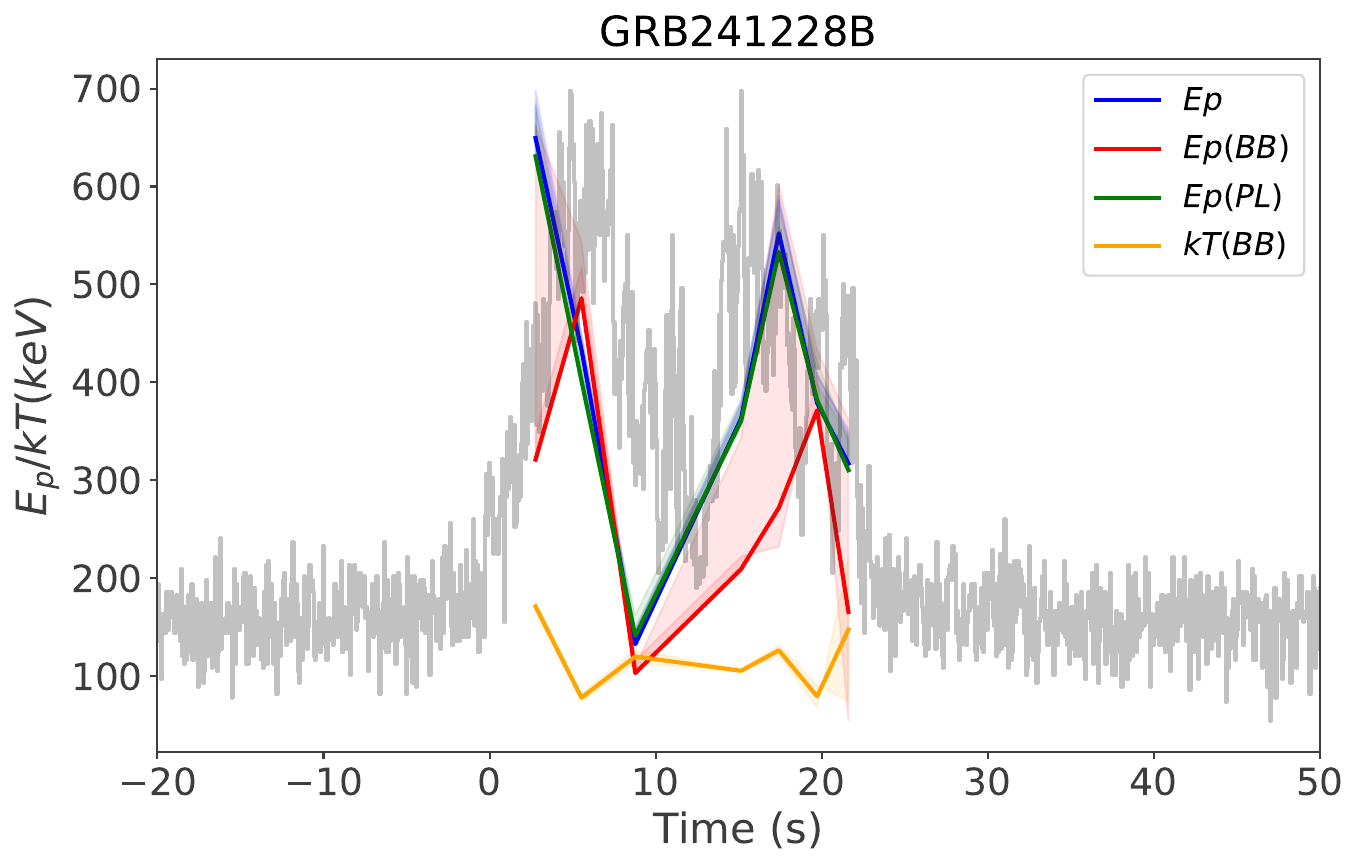}
\includegraphics[width=0.32\textwidth]{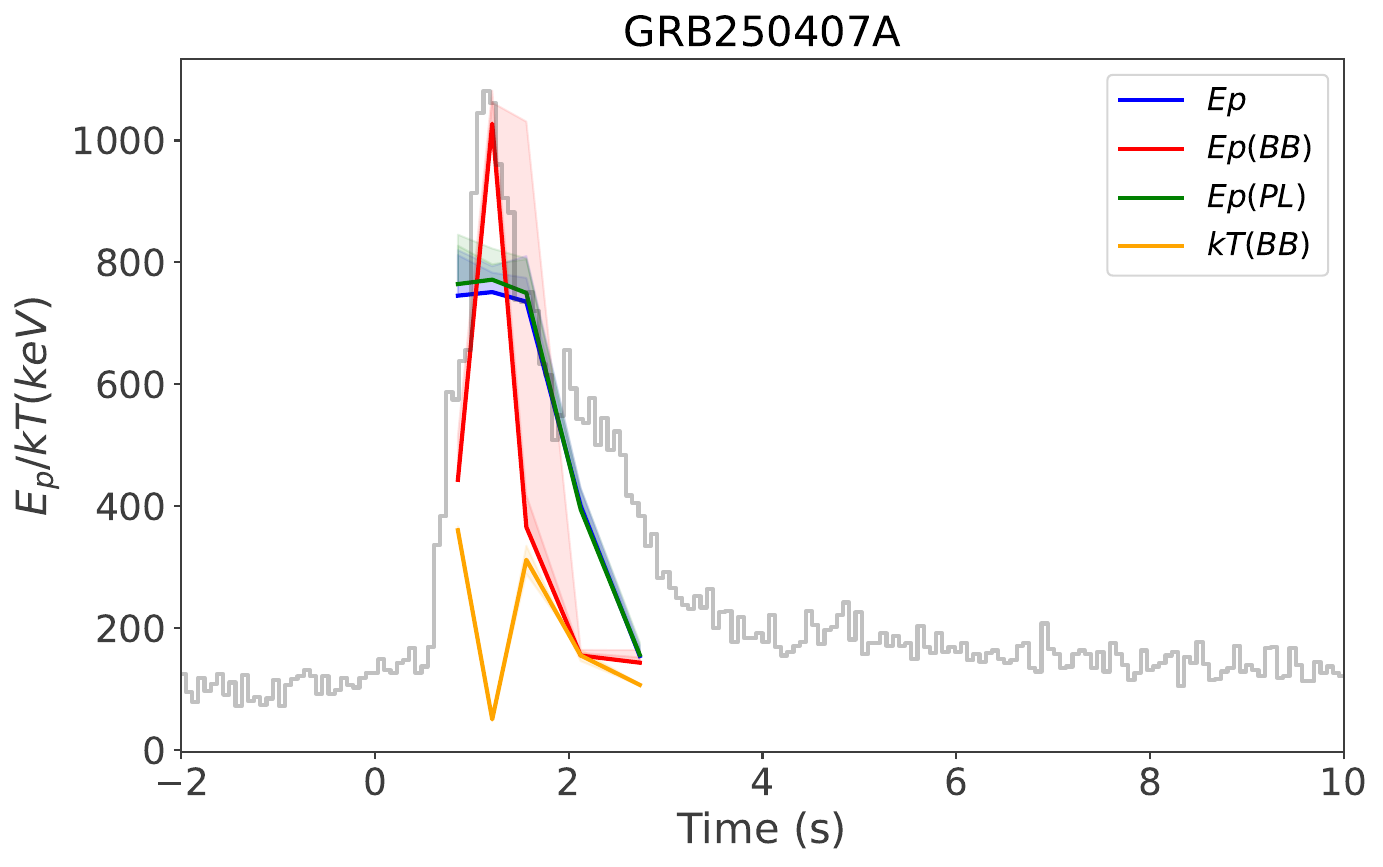}
   \figcaption{Time evolution and distribution of the peak energy $E_p$ for the eight GRBs, with the same model colors as in Figure~\ref{fig 3}. The orange curves show the corresponding blackbody temperature $kT_{\rm BB}$.} \label{fig 4}
\end{figure}

As shown in Figure~\ref{fig 2}, the baseline fits give $\alpha=-0.71\pm0.28$. The value remains $-0.71\pm0.28$ when the PL component is included and changes to $-0.61\pm0.31$ when the BB component is added. For the baseline model, $\log_{10}(E_p/{\rm keV})=2.37\pm0.36$, corresponding to a central value of approximately 234~keV. With the PL component, $\log_{10}(E_p/{\rm keV})=2.38\pm0.36$ (approximately 240~keV); with the BB component, it decreases to $2.22\pm0.26$ (approximately 166~keV). Thus, adding the PL component has little effect on $\alpha$ or $E_p$, whereas adding the BB component hardens $\alpha$ and slightly decreases $E_p$ \citep{2022ApJ...941...27L,2023ApJS..266...31L}. Figures~\ref{fig 3} and \ref{fig 4} show the evolution and distributions of $\alpha$ and $E_p$ for each GRB.

\subsection{Correlation between Model Parameters}

Band and CPL provide adequate empirical descriptions of many accepted bins, while models containing BB, mBB, or an additional PL can be preferred in the nested comparisons described above. The term ``thermal-like'' is used for this spectral curvature. A photospheric identification additionally requires the physical assumptions of Sections~\ref{Sections4} and is therefore not established by DIC alone. Likewise, thermal-flux dominance is assigned only through the mBB+PL flux fraction defined above.

Appendix Figures~\ref{fig 8} and \ref{fig 9} present, for the first time bin (0.07--0.21~s) of GRB~210619B, the photon count spectra and corner plots from the Band, CPL, mBB, and mBB+PL fits.

\begin{figure}[htbp]
\centering
\includegraphics[width=0.48\textwidth]{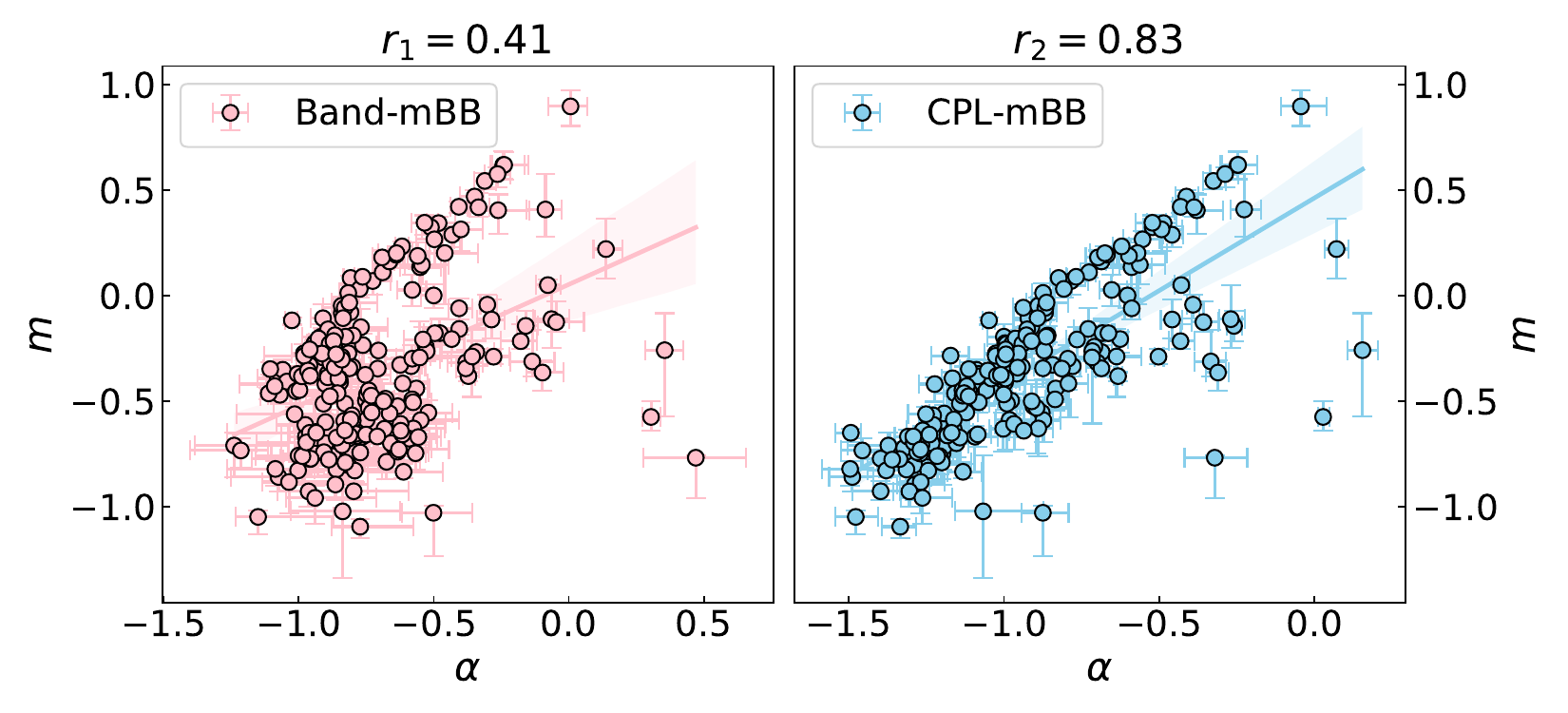}
\includegraphics[width=0.48\textwidth]{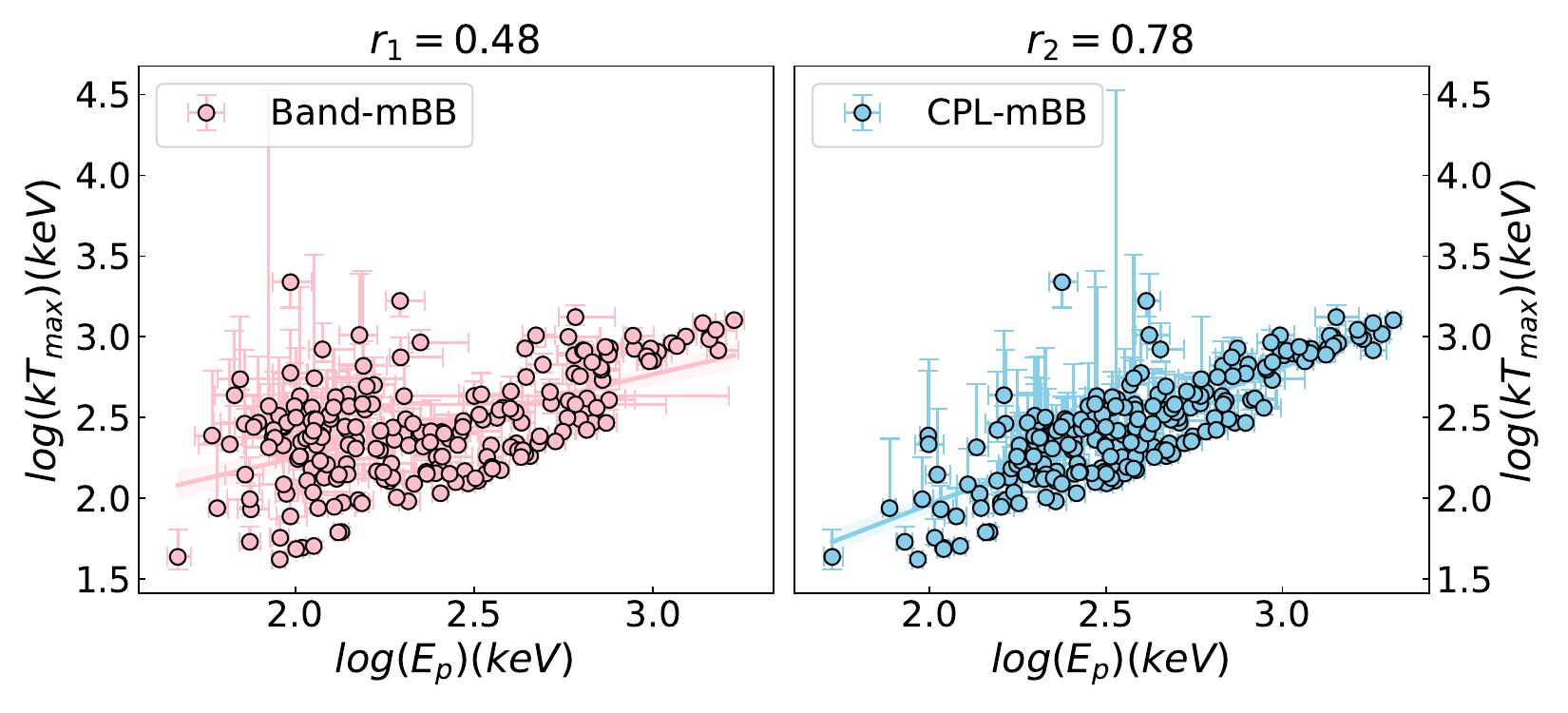}
\includegraphics[width=0.48\textwidth]{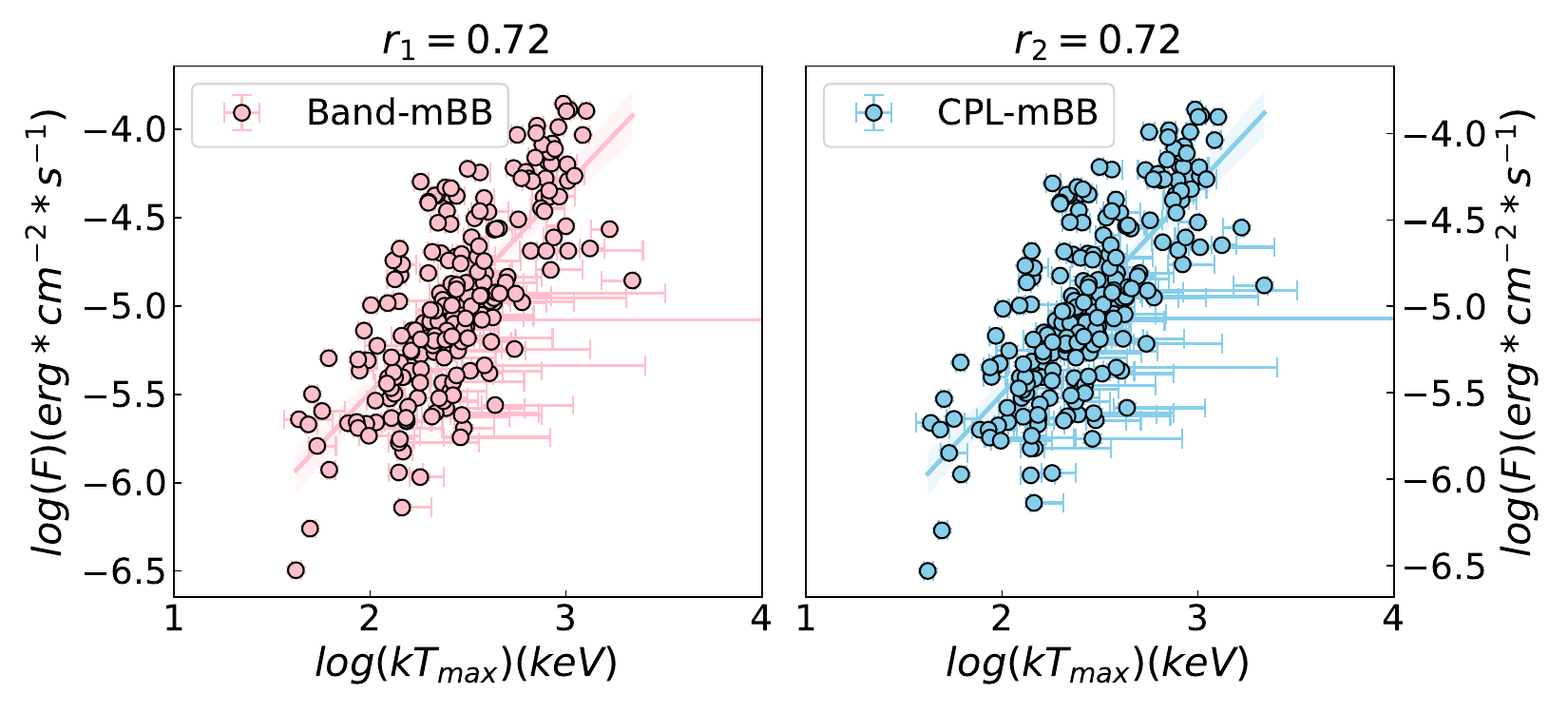}
   \figcaption{Correlations between parameters obtained with the empirical and mBB spectral descriptions.} \label{fig 5}
\end{figure}

Figure~\ref{fig 5} compares parameters obtained with the empirical and mBB descriptions. The quoted $r$ values are Pearson correlation coefficients calculated from the accepted time bins. The CPL--mBB correlations are stronger ($r=0.83$ for $m$--$\alpha$ and $r=0.78$ for $kT_{\max}$--$E_p$) than the corresponding Band--mBB correlations ($r=0.41$ and $0.48$). This difference is expected in part because Band introduces an independently fitted high-energy index, whereas CPL and mBB both impose stronger high-energy curvature; it also shows that mBB cannot be claimed to reproduce Band and CPL equally well. Because multiple bins come from the same burst and the fitted parameters have shared degeneracies, no population-level $p$ values are attached to these coefficients, nor are they used as independent evidence for a photospheric origin. They are reported only as descriptive comparisons and were not used in sample selection.

This baseline dependence is consistent with the broader model-comparison results of \citet{2022ApJ...941...27L}, who showed that using Band or CPL for the same spectrum can shift the inferred $\alpha$ and $E_p$, and with the model-wise Fermi analysis of \citet{2023ApJS..266...31L}. The present result adds a specific diagnostic for thermal-like curvature: the apparent correspondence between empirical and mBB parameters is substantially tighter for CPL than for Band. Consequently, correlations obtained from a single empirical baseline should not be interpreted as model-independent evidence for a unique radiation mechanism.

\section{Time-integrated Spectral Analysis\label{Sections4}}
\subsection{Fitting Results}

GRB~210619B provides an illustrative case for the time-integrated analysis. Its spectrum was fitted with the MultiNest implementation in 3ML, with the results listed in Table~\ref{table3}. Among the three baseline models (Band, CPL, and BB), Band yields the smallest DIC. Its low-energy index $\alpha$ is close to the typical value, whereas the high-energy index $\beta$ is slightly higher than the commonly quoted value ($\alpha=-1$, $\beta=-2.3$; \citealt{2000ApJS..126...19P}). Adding a BB component improves the fit relative to the corresponding Band and CPL baselines, with Band+BB giving the lowest DIC among these models. For the Band+BB fit, the blackbody temperature is $kT_\mathrm{BB}=15.82^{+3.98}_{-2.17}$ keV, the blackbody flux is \( F_{\rm BB}=(6.52\pm2.93)\times10^{-8}\,\mathrm{erg\,cm^{-2}\,s^{-1}} \), the total flux is \( F_{\rm tot}=0.40\times10^{-5}\,\mathrm{erg\,cm^{-2}\,s^{-1}} \), and the peak energy is \( E_p=289.81\,\mathrm{keV} \).

\begin{figure}[htbp]
\centering
\includegraphics[width=0.32\textwidth]{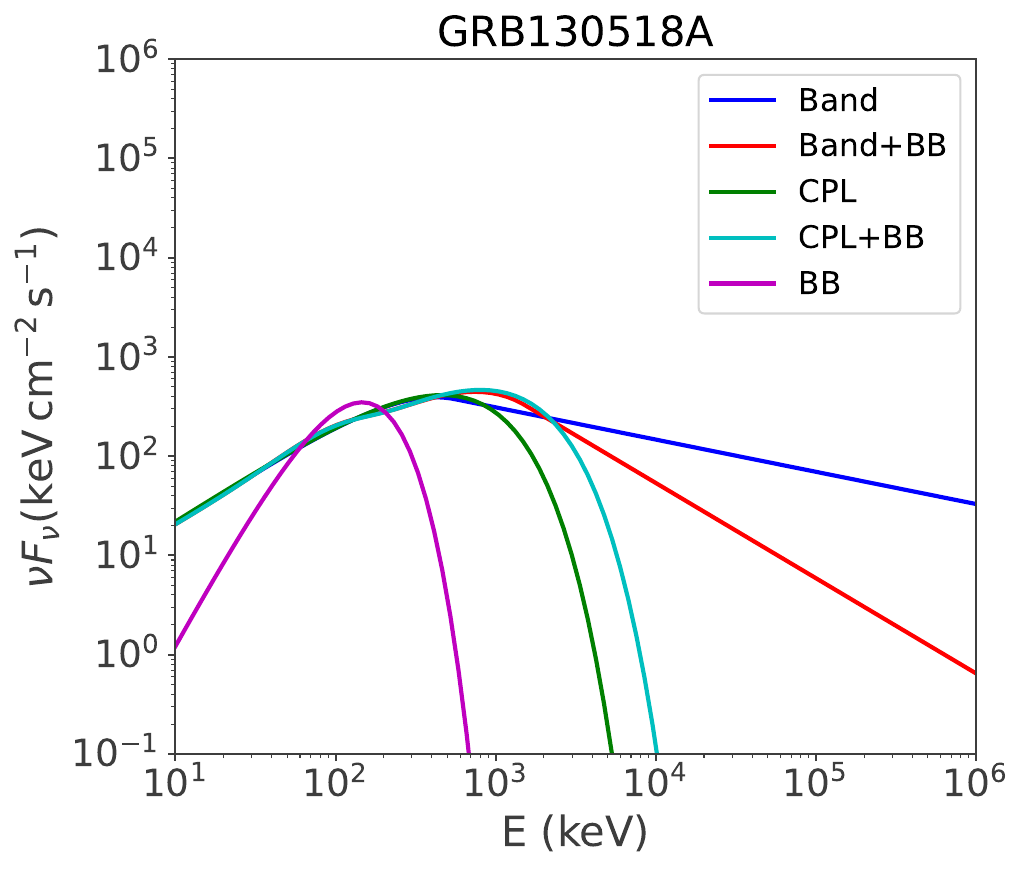}
\includegraphics[width=0.32\textwidth]{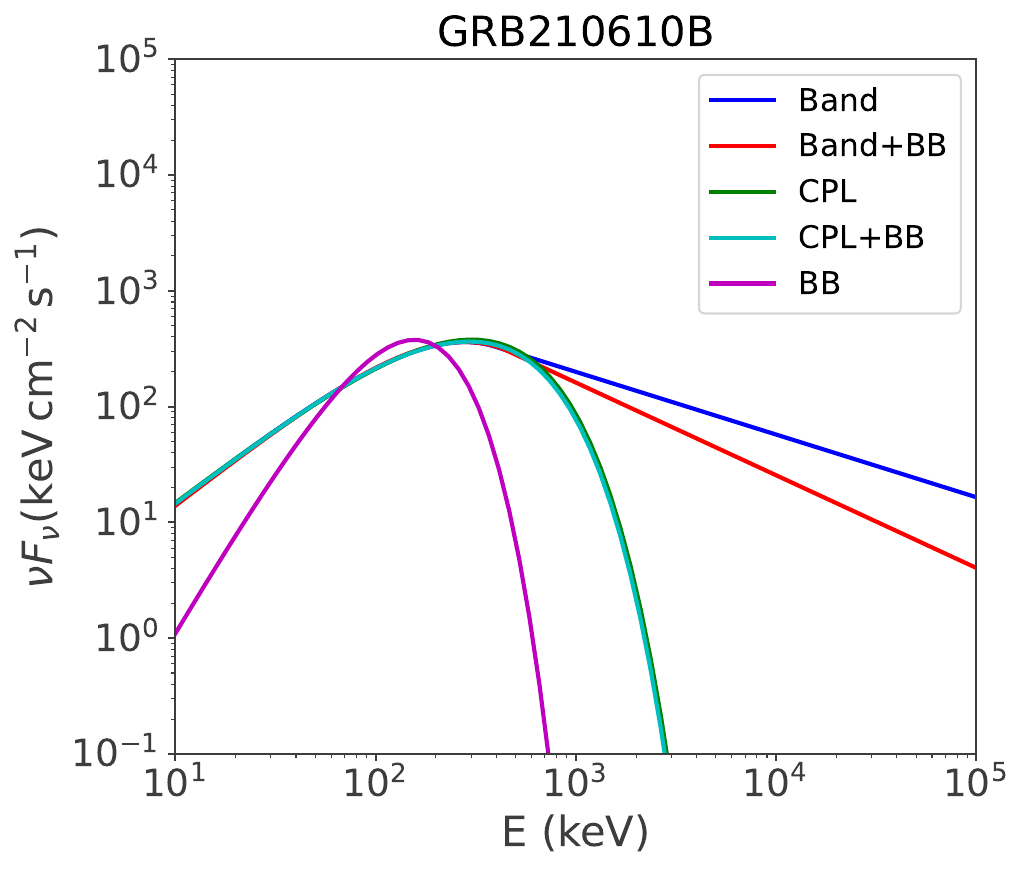}
\includegraphics[width=0.32\textwidth]{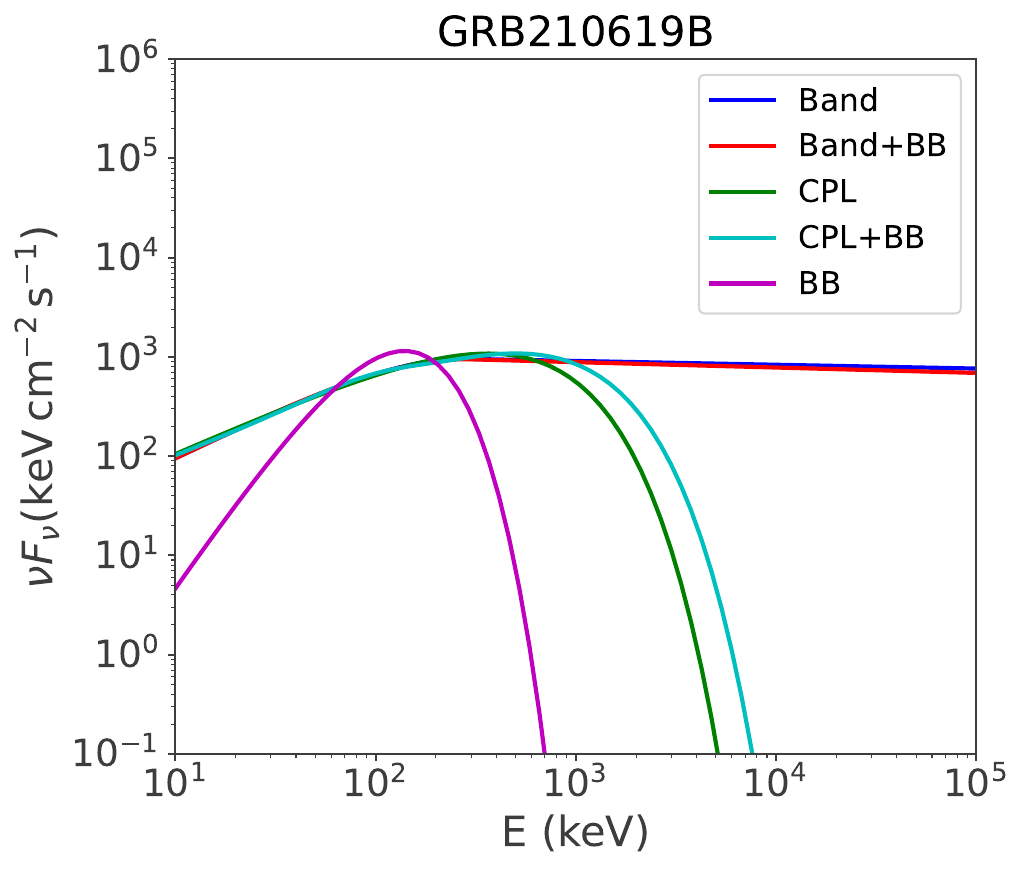}
\includegraphics[width=0.32\textwidth]{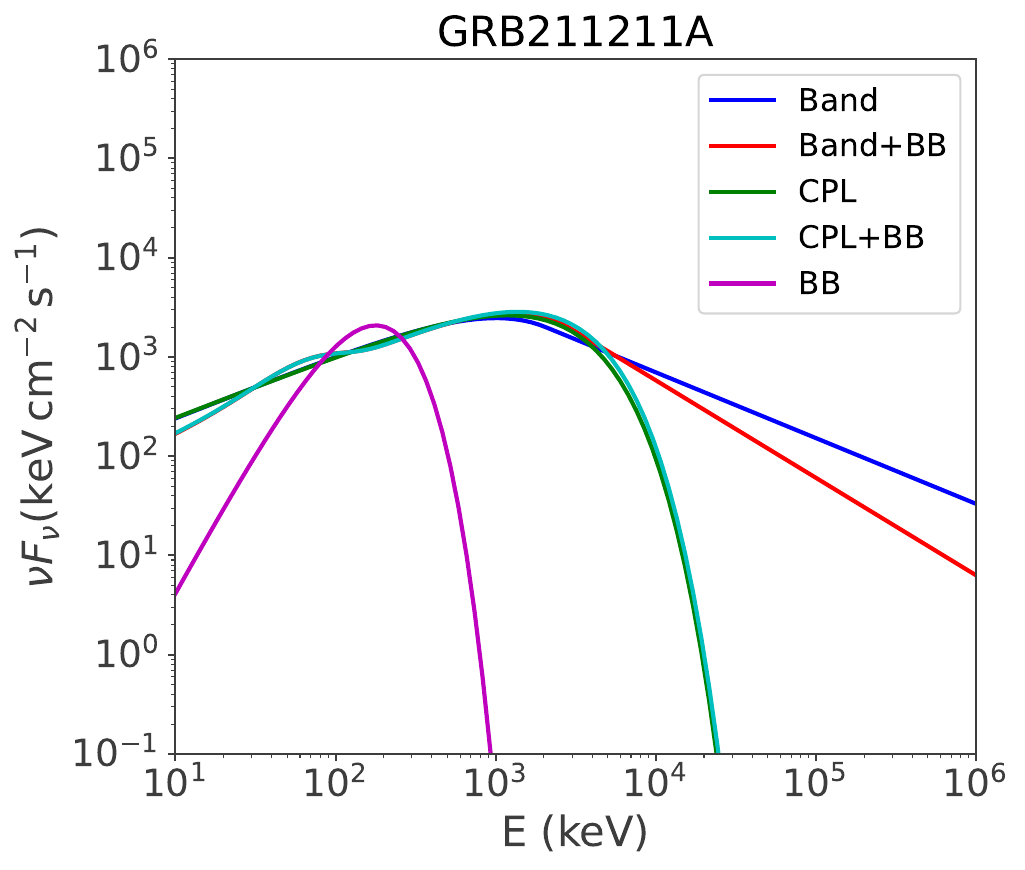}
\includegraphics[width=0.32\textwidth]{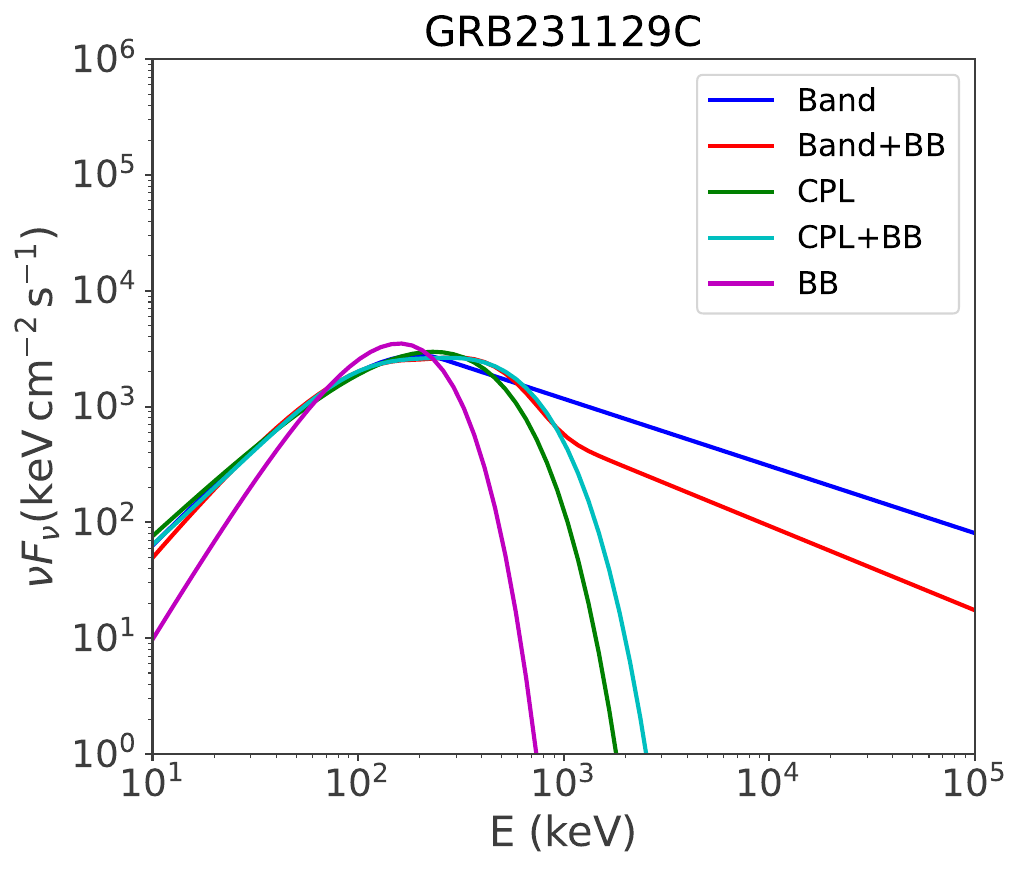}
\includegraphics[width=0.32\textwidth]{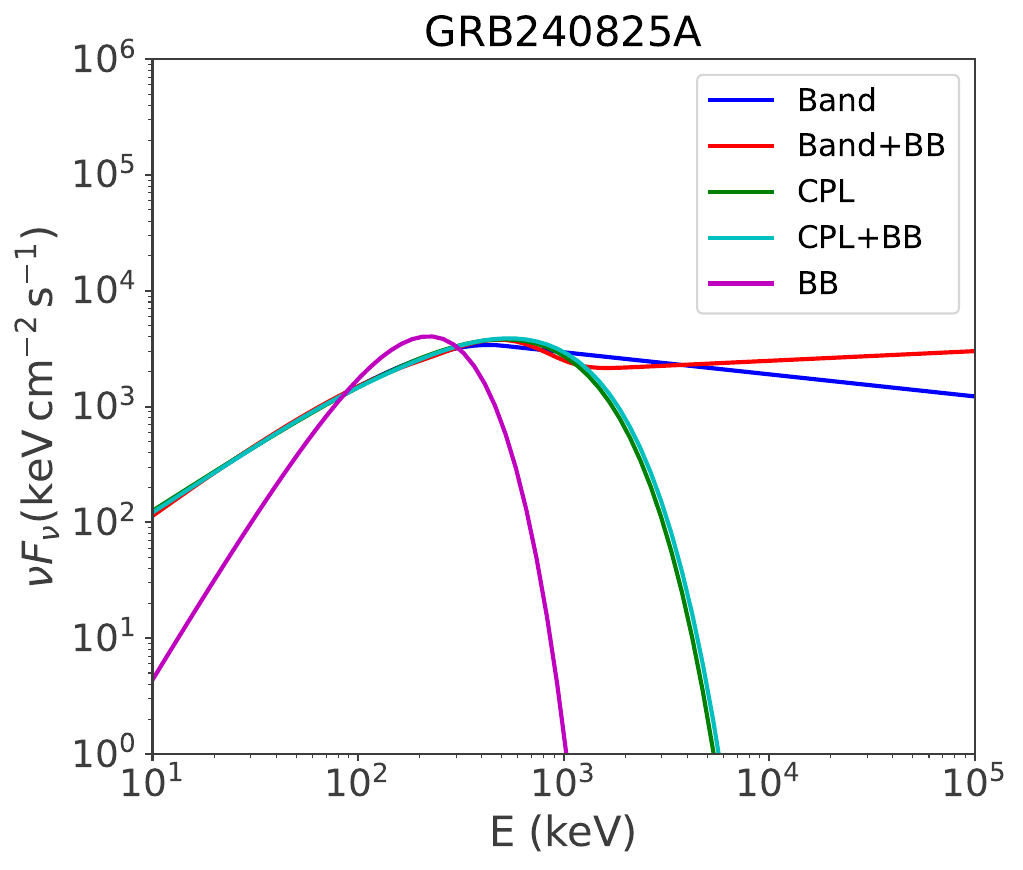}
\includegraphics[width=0.32\textwidth]{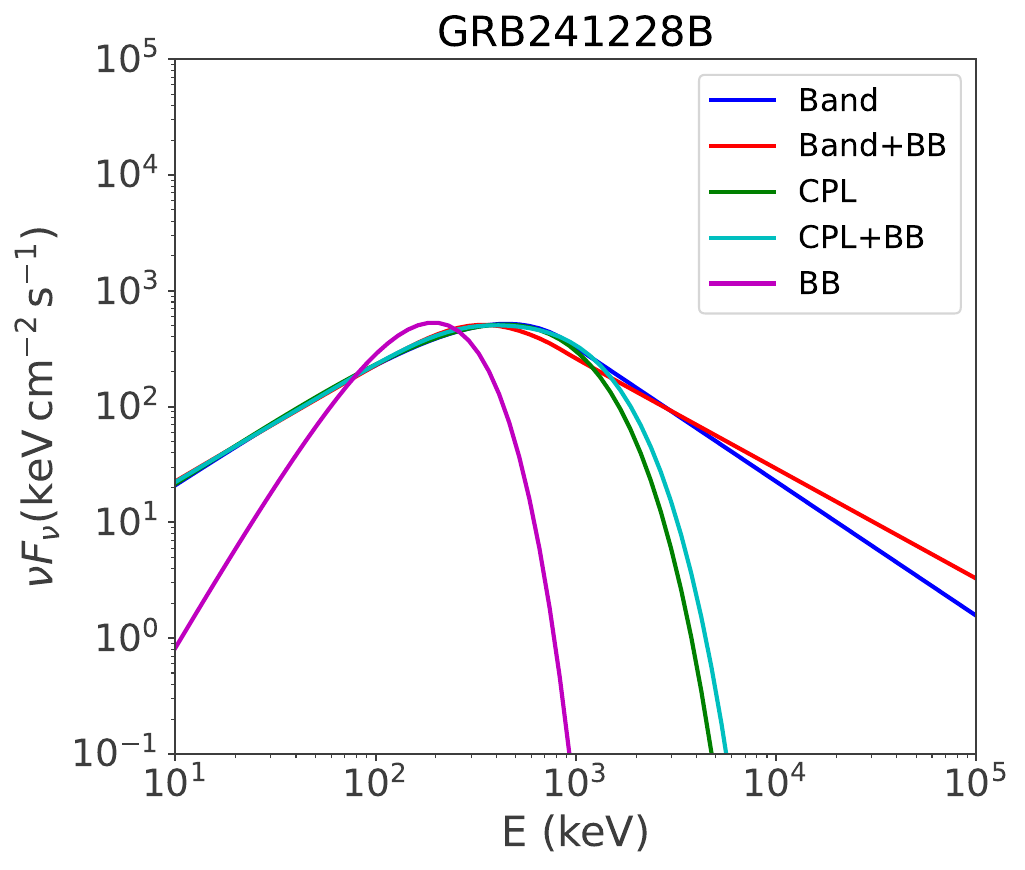}
\includegraphics[width=0.32\textwidth]{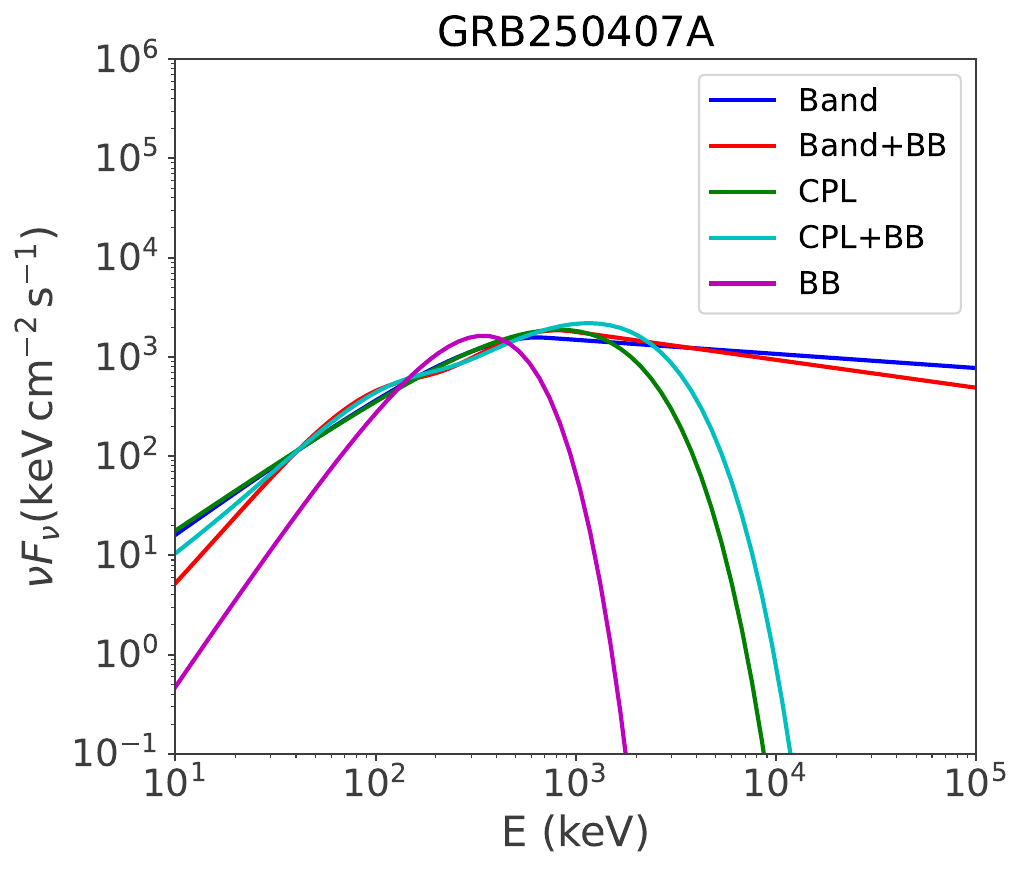}
   \figcaption{Time-integrated $\nu F_{\nu}$ spectra of the eight GRBs.} \label{fig 6}
\end{figure}

The same procedure was applied to the time-integrated spectra of all eight GRBs (Table~\ref{table4} and Figure~\ref{fig 6}). For each burst, $kT_{\rm BB}$, $F_{\rm BB}$, and $F_{\rm tot}$ were taken from the same adopted time-integrated BB-containing spectral fit over the listed source interval. These paired fitted quantities, rather than catalog fluence divided by duration, are the flux inputs used in both the H2013 and G2015 calculations. The resulting narrow-BB fractions $F_{\rm BB}/F_{\rm tot}$ range from 0.005 to 0.311 and therefore do not establish thermal dominance of the time-integrated energy flux. The time-integrated BB parameters are effective quantities: spectral evolution and the superposition of temperatures can broaden or create apparent curvature. They are used below only to obtain conditional, time-averaged estimates within the H2013 and G2015 prescriptions.

Six of the eight time-integrated fractions are below 0.12; only GRBs~231129C and 240825A approach 0.3. This distribution resembles the subdominant thermal components emphasized by \citet{2013A&A...551A.124H} more closely than exceptional, strongly thermal cases such as GRB~090902B. It also demonstrates why the time-resolved mBB criterion and the integrated narrow-BB fraction cannot be used interchangeably: the former describes recurrent broadened curvature within short bins, whereas the latter measures the contribution of a single-temperature component after spectral evolution has been integrated over the burst.

The G2015 calculation was also applied to individual accepted time bins as a qualitative robustness check. Although the bin-by-bin estimates have substantially larger uncertainties, they show the same broad sensitivity to the assumed launch radius as the time-integrated calculation. The time-integrated results are retained for a uniform comparison across all eight events and are interpreted as effective, conditional averages rather than direct measurements of instantaneous central-engine parameters. The narrow-BB and time-resolved mBB descriptions are not treated as interchangeable physical measurements.

Table~\ref{table3} reports posterior medians and 68\% credible intervals. Asymmetric spectral uncertainties, including the $+3.98$~keV upper uncertainty of the GRB~210619B temperature, are used without symmetrization. For a derived quantity available only with a symmetric error, the quoted interval is used when it remains positive; if it crosses zero, a positive log-normal approximation with the same central value and coefficient of variation provides the 16th--84th percentiles. Because joint posterior samples are unavailable, these intervals are treated as positivity-preserving approximations rather than full joint-posterior propagation.

\begin{table}[h!]
\centering
\caption{Time-integrated Spectral Fits for GRB~210619B} \label{tab3}
\label{table3}
\resizebox{\textwidth}{!}{
\begin{tabular}{ccccccccc}
\toprule
$t_{\rm start}-t_{\rm end}$ & \multirow{2}*{Model} & \multirow{2}*{$\alpha$} & \multirow{2}*{$\beta$} & $E_p/E_c$ & $kT$ & $F_{\rm BB}\times10^{-6}$ & \textbf{$F_{\rm tot}\times10^{-6}$} & \multirow{2}*{DIC} \\
 $(s)$& & & &$(keV)$&$(keV)$&$(ergcm^{-2}s^{-1})$&$(ergcm^{-2}s^{-1})$&  \\
\hline
  \midrule

\multirow{5}{*}{\(0 \sim 70\)} & Band & ${-0.99}^{+0.01}_{-0.02}$ & ${-2.05}^{+0.02}_{-0.03}$ & ${254.50}^{+11.70}_{-6.63}$ & ... & ... & ${4.44}^{+0.03}_{-0.02}$ & 9881.95 \\
 & CPL & ${-1.11}^{+0.01}_{-0.01}$ & ... & ${371.65}^{+7.70}_{-8.33}$ & ... & ... & ${4.55}^{+0.03}_{-0.03}$ & 10078.95 \\
 & Band+BB & ${-1.01}^{+0.02}_{-0.04}$ & ${-2.09}^{+0.03}_{-0.04}$ & ${289.81}^{+25.42}_{-18.53}$ & ${15.82}^{+3.98}_{-2.17}$ & ${0.07}^{+0.03}_{-0.03}$ & ${4.46}^{+0.03}_{-0.03}$ & 9860.53 \\
 & CPL+BB & ${-1.19}^{+0.02}_{-0.02}$ & ... & ${525.70}^{+25.34}_{-27.00}$ & ${24.68}^{+1.10}_{-1.42}$ & ${0.27}^{+0.03}_{-0.03}$ & ${4.68}^{+0.03}_{-0.03}$ & 9935.11 \\
 & BB & ... & ... & ... & ${33.84}^{+0.22}_{-0.22}$ & ... & ${2.42}^{+0.01}_{-0.01}$ & 26514.33 \\

\hline
\bottomrule
\end{tabular}
}
\end{table}

\subsection{Constraints on Hybrid Jet Parameters}

The time-integrated fits were interpreted with two methods for outflows containing both thermal and magnetic components and no subphotospheric magnetic dissipation. The first is the H2013 method of \citet{2013A&A...551A.124H}; the second is the G2015 hybrid-jet method of \citet{2015ApJ...801..103G}. Both have been widely applied in studies of GRB jet properties \citep{2013ApJ...770...32G,2015ApJ...814...10G,2016ApJ...833..139A,2020ApJ...894..100L,2022MNRAS.516.4180N,2022ApJ...938..159S,2022ApJ...932...25C,2024ApJ...972..132C}.

\subsubsection{The H2013 Approach}
In the H2013 method, the total energy injected at the initial radius \(R_0\) is partitioned between a thermal fraction \(\epsilon_{\rm Th}\) and a magnetic fraction \(1-\epsilon_{\rm Th}\). This parameterization spans a pure fireball (\(\epsilon_{\rm Th}=1\)), a Poynting-flux-dominated outflow (\(\epsilon_{\rm Th}\to0\)), and hybrid configurations between these limits. Acceleration can be driven by thermal energy, magnetic energy, or both. The flow is assumed to become approximately spherical at \(R_{\rm sph}\) and to complete its acceleration at the saturation radius \(R_{\rm sat}>R_{\rm sph}\). At the photospheric radius \(R_{\rm ph}\), the outflow becomes transparent and releases thermal photons. The model assumes \(R_{\rm sat}<R_{\rm ph}\) and no magnetic dissipation below the photosphere.

At the end of acceleration, $\sigma$ is the ratio of magnetic to kinetic energy flux. The H2013 prescription defines the passive-field value as $\sigma_{\rm passive}=(1-\epsilon_{\rm Th})/\epsilon_{\rm Th}$. Equality corresponds to a field carried passively through complete thermal acceleration. Values of $\sigma<\sigma_{\rm passive}$ indicate efficient conversion of magnetic energy into kinetic energy, whereas $\sigma>\sigma_{\rm passive}$ indicates inefficient magnetic acceleration. These relations describe magnetic-energy conversion and should not be recast as a simple thermal-versus-magnetic dominance test. Internal shocks are expected to become inefficient as $\sigma$ approaches or exceeds unity, at which point magnetic dissipation may become relevant \citep{2013A&A...551A.124H}; the transition is not a sharp observational classifier.

\citet{2013ApJ...770...32G}, following \citet{2002MNRAS.336.1271D} and \citet{2007ApJ...664L...1P}, constrained the initial radius \(R_0\), Lorentz factor \(\Gamma\), and photospheric radius \(R_{\rm ph}\) by interpreting the BB component of GRB~120323A as photospheric emission. \citet{2013A&A...551A.124H} extended the approach of \citet{2007ApJ...664L...1P} to magnetized outflows, assuming \(R_{\rm sat}<R_{\rm ph}\) and no magnetic dissipation below the photosphere. Under these assumptions, \(R_0\), \(\Gamma\), and \(R_{\rm ph}\) are related to the observables by

\begin{equation}
R_0 \simeq \left[ \frac{D_{\rm L} \mathcal{R}}{2(1+z)^2} \left( \frac{\phi}{1-\phi} \right)^{3/2} \right] \times \left[ \frac{f_{\rm NT}}{\epsilon_{\rm Th}} \right]^{3/2},
\end{equation}

\begin{equation}
\Gamma \simeq \left[ \frac{\sigma_{\rm T}}{m_{\rm p} c^3} \frac{(1+z)^2 D_{\rm L} F_{\rm BB}}{\mathcal{R}} \frac{1-\phi}{\phi} \right]^{1/4} \times \left[ (1+\sigma) f_{\rm NT} \right]^{-1/4},
\end{equation}

\begin{equation}
R_{\rm ph} \simeq \left[ \frac{\sigma_{\rm T}}{16 m_{\rm p} c^3} \frac{D_{\rm L}^5 F_{\rm BB} \mathcal{R}^3}{(1+z)^6} \frac{1-\phi}{\phi} \right]^{1/4} \times \left[ (1+\sigma) f_{\rm NT} \right]^{-1/4},
\end{equation}

Following \citet{2007ApJ...664L...1P}, the parameter \(\mathcal{R}\) is

\begin{equation}
\mathcal{R} = \left( \frac{F_{\rm BB}}{\sigma_{\rm SB} T_{\rm BB}^4} \right)^{1/2}.
\end{equation}

In the above equations, $z$ represents the GRB redshift, \(D_{\rm L}\) denotes the luminosity distance, \(T_{\rm BB}\) is the observed temperature of the BB component, and \(F_{\rm BB}\) is its observed flux. The ratio \(\phi=F_{\rm BB}/F_{\rm tot}\) is the blackbody-to-total observed flux ratio. In addition, \(\sigma_{\rm T}\) is the Thomson scattering cross-section, \(m_{\rm p}\) is the proton mass, $c$ is the speed of light, \textbf{\(\sigma_{\rm SB}\)} is the Stefan-Boltzmann constant, and \(f_{\rm NT}\) is the non-thermal emission efficiency. The observed values are listed in Table~\ref{table4}.

The H2013 relations were evaluated at $R_0=2.1\times10^6$, $10^7$, and $10^8$~cm and at $f_{\rm NT}=0.05$, 0.1, and 0.5 (Tables~\ref{table4} and \ref{table5}). These radii and efficiencies are scenario values, not independent measurements. In particular, the formal GRB~210610B estimate extends below a physically plausible compact-object engine scale and is interpreted as tension between the adopted spectral/model assumptions, rather than as support for $R_0=10^4$ or $10^5$~cm launch radii. Time-resolved studies of GRB~160625B and multipulse Fermi GRBs have reported transitions from fireball-like to Poynting-flux-dominated emission \citep{2018NatAs...2...69Z,2019ApJS..242...16L}, but the formal radius obtained here is not sufficient to assign the same transition to GRB~210610B. The efficiency grid brackets a low efficiency typical of many internal-shock calculations and higher efficiencies sometimes associated with magnetic dissipation, but it does not uniquely identify either mechanism \citep{2023ApJ...944L..57L,2024ApJ...972..195L}.

For GRB~210619B, for example, the inferred $\epsilon_{\rm Th}$ at $f_{\rm NT}=0.5$ decreases from 0.89 at $R_0=2.1\times10^6$~cm to 0.31 and 0.07 at $10^7$ and $10^8$~cm, while $\sigma_{\rm passive}$ increases from 0.12 to 2.18 and 13.76. The same data therefore map to a nearly thermal initial energy budget at the smallest assumed radius and a magnetically dominated initial budget at larger radii. Lower $f_{\rm NT}$ values shift the solutions toward smaller $\epsilon_{\rm Th}$ and larger $\sigma_{\rm passive}$. These trends demonstrate sensitivity to the assumptions; they do not select the true $R_0$, $f_{\rm NT}$, or dissipation channel.

\begin{figure}[htbp]
\centering
\includegraphics[width=0.43\textwidth]{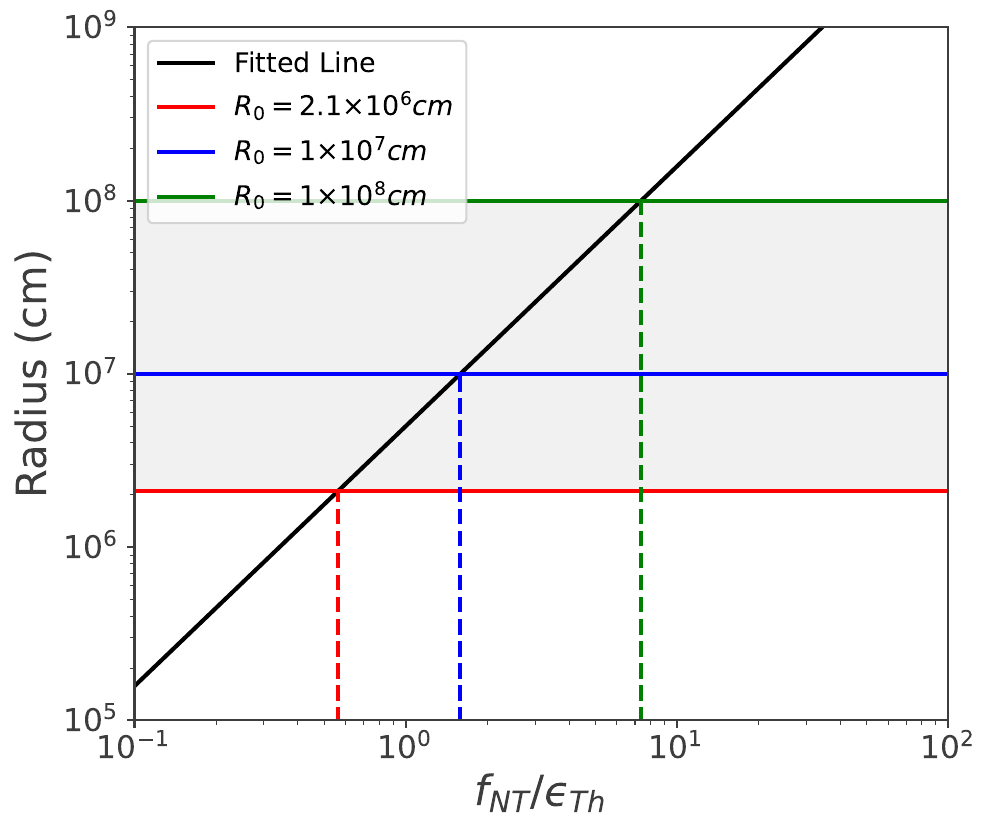}
\vspace{3pt}

\includegraphics[width=0.43\textwidth]{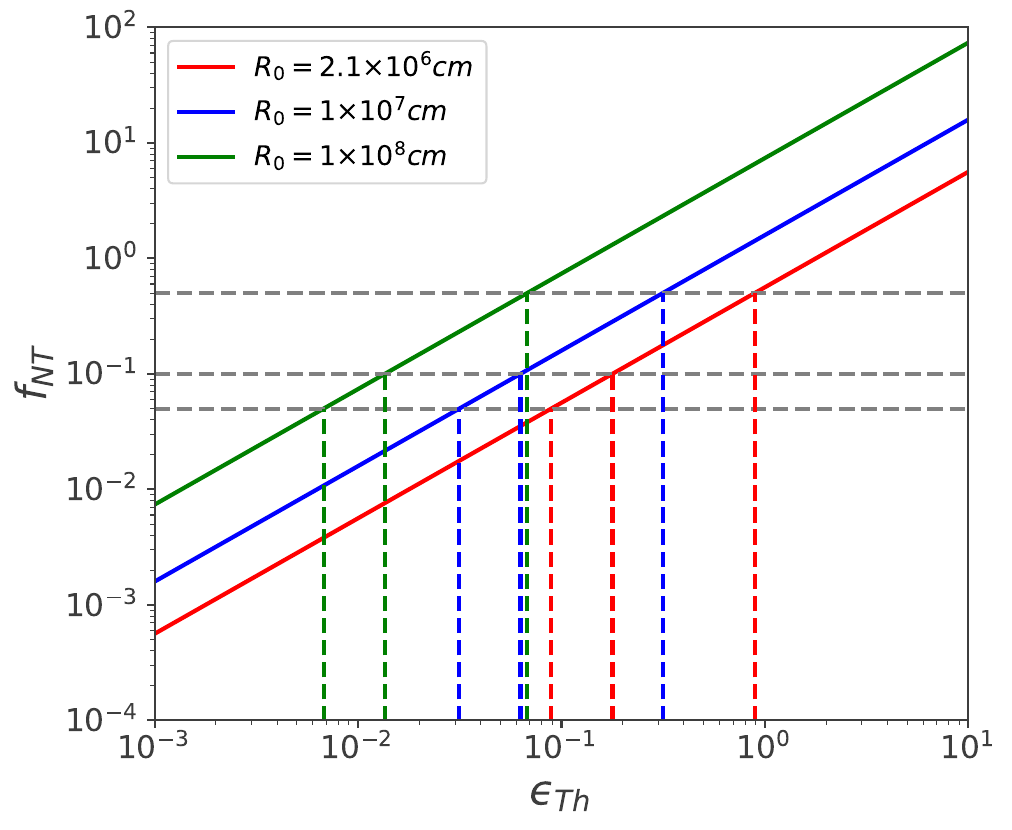}
\vspace{3pt}

\includegraphics[width=0.43\textwidth]{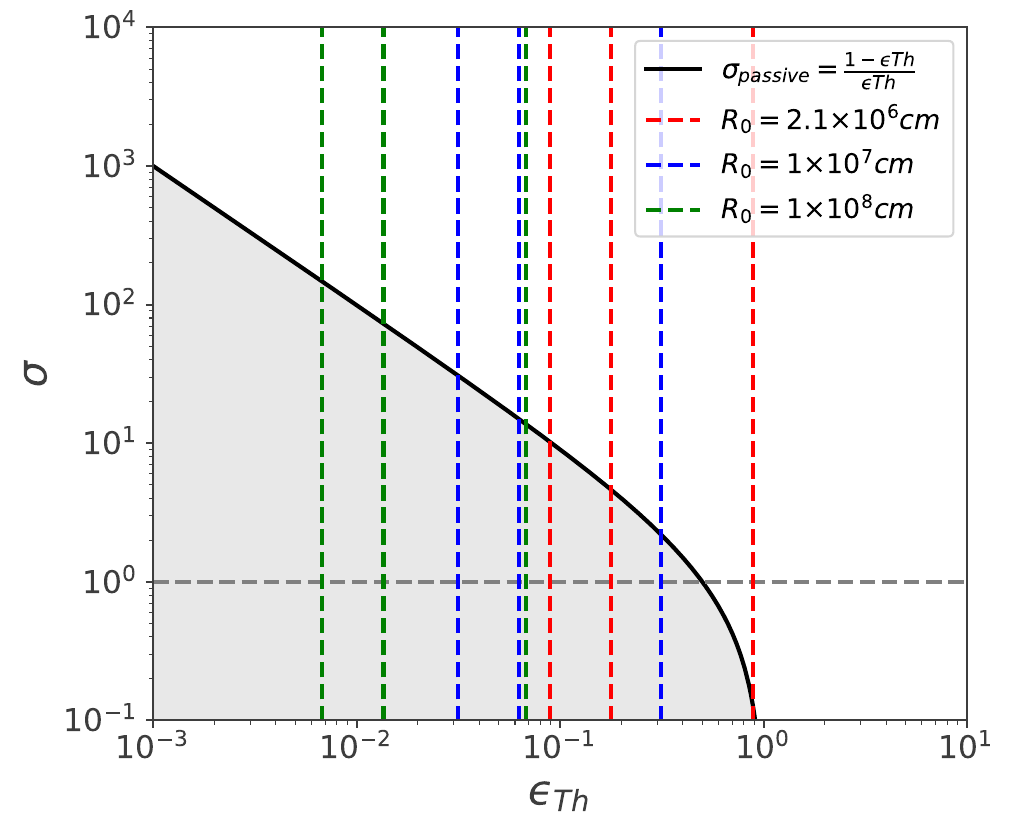}
   \figcaption{Conditional H2013 parameter mapping for GRB~210619B. The upper panel shows the relation between \(R_0\) and \(f_{\rm NT}/\epsilon_{\rm Th}\); the red, blue, and green curves correspond to \(R_0=2.1\times10^6\,\mathrm{cm}\), \(10^7\,\mathrm{cm}\), and \(10^8\,\mathrm{cm}\), respectively. The middle panel shows the non-thermal radiation efficiency \(f_{\rm NT}\) as a function of the initial thermal fraction \(\epsilon_{\rm Th}\) for the three radii. The horizontal lines mark \(f_{\rm NT}=0.5\), \(0.1\), and \(0.05\), and the vertical lines identify the corresponding \(\epsilon_{\rm Th}\) values (Table~\ref{table5}). In the lower panel, the black curve gives the H2013 magnetization \(\sigma\) at the end of acceleration, and the shaded region marks \(\sigma<\sigma_{\rm passive}\), where \(\sigma_{\rm passive}=(1-\epsilon_{\rm Th})/\epsilon_{\rm Th}\). The vertical dashed lines correspond to the \(\epsilon_{\rm Th}\) values for the three adopted efficiencies.} \label{fig 7}
\end{figure}

Figure~\ref{fig 7} visualizes this conditional mapping for GRB~210619B. The upper and middle panels show how the adopted $R_0$ and $f_{\rm NT}$ determine the inferred $\epsilon_{\rm Th}$, while the lower panel compares the magnetization at the end of acceleration with $\sigma_{\rm passive}=(1-\epsilon_{\rm Th})/\epsilon_{\rm Th}$. In particular, the marked $f_{\rm NT}=0.5$ solutions make the change from $\sigma_{\rm passive}<1$ at $R_0=2.1\times10^6$~cm to $\sigma_{\rm passive}>1$ at the two larger radii explicit. The figure is therefore not independent evidence for a particular launch radius; it is a compact representation of the radius--efficiency degeneracy quantified in Table~\ref{table5}.

Table~\ref{table5} reports point estimates over a grid of assumed $R_0$ and $f_{\rm NT}$. The variation across that grid is shown as a model-sensitivity range and is not a posterior credible interval. Accordingly, the table maps conditional trends rather than probabilistic jet classes.

\subsubsection{The G2015 Approach}

\citet{2015ApJ...801..103G} developed a photospheric model for hybrid relativistic outflows comprising a hot fireball component, characterized by the dimensionless entropy $\eta$, and a cold Poynting-flux component, characterized by the central-engine magnetization \(\sigma_0\). For the non-dissipative photosphere, their simplified jet dynamics yields a ``top-down'' inference of central-engine parameters from the observables; this prescription is referred to here as the G2015 method.

\citet{2015ApJ...801..103G} proposed that the initial thermal acceleration is followed by a magnetic acceleration phase. During rapid acceleration, the Lorentz factor \(\Gamma\) increases approximately linearly with radius until the flow reaches \(r_{\rm ra}\). The definition of this radius depends on the relation between \(\eta\) and \((1+\sigma_0)^{1/2}\). If \(\eta>(1+\sigma_0)^{1/2}\), the flow reaches the Alfv\'en Lorentz factor before thermal acceleration ends, and \(r_{\rm ra}\) marks the end of the thermal acceleration phase. If \(\eta<(1+\sigma_0)^{1/2}\), magnetic acceleration extends the rapid phase until \(\Gamma\) reaches the Alfv\'en Lorentz factor at \(r_{\rm ra}\). Beyond \(r_{\rm ra}\), the flow accelerates more slowly, with \(\Gamma\propto r^\delta\), until it reaches the coasting radius \(r_c\), where \(0<\delta\leq1/3\) \citep{2011ApJ...733L..40M,2012ApJ...755...12V}.

Depending on the central-engine parameters, \(\Gamma_{\rm ra}\) is either \(\eta\) or \([\eta(1+\sigma_0)]^{1/3}\). In each branch, the photospheric radius \(r_{\rm ph}\) can lie below \(r_{\rm ra}\), between \(r_{\rm ra}\) and \(r_c\), or above \(r_c\). The G2015 method therefore defines six regimes: (I) \(\eta>(1+\sigma_0)^{1/2}\) and \(r_{\rm ph}<r_{\rm ra}\); (II) \(\eta>(1+\sigma_0)^{1/2}\) and \(r_{\rm ra}<r_{\rm ph}<r_c\); (III) \(\eta>(1+\sigma_0)^{1/2}\) and \(r_{\rm ph}>r_c\); (IV) \(\eta<(1+\sigma_0)^{1/2}\) and \(r_{\rm ph}<r_{\rm ra}\); (V) \(\eta<(1+\sigma_0)^{1/2}\) and \(r_{\rm ra}<r_{\rm ph}<r_c\); and (VI) \(\eta<(1+\sigma_0)^{1/2}\) and \(r_{\rm ph}>r_c\).

The non-dissipative G2015 top-down relations were applied at the same three assumed launch radii, with $f_\gamma=0.5$. The observational inputs are $kT_{\rm ob}$, $F_{\rm BB}$, $F_{\rm tot}$, and $z$, with $f_{\rm th}=F_{\rm BB}/F_{\rm tot}$. In this notation, $\sigma_0=L_{\rm P}/L_{\rm h}$ is the initial ratio of Poynting to hot-component luminosity and $\mu_0=\eta(1+\sigma_0)$ is the total energy per baryon. The inequality $\eta>(1+\sigma_0)^{1/2}$ selects a dynamical acceleration branch; it is not a thermal-dominance criterion. The initial energy budget is described as hot-component dominated when $\sigma_0<1$, Poynting-flux dominated when $\sigma_0>1$, and hybrid when both components are non-negligible. Table~\ref{table5} reports $\sigma$ rather than $1+\sigma$; any formal inversion below $\sigma=0$ is a boundary or physically inadmissible solution and must not be interpreted as a measured negative magnetization.

For GRB~210619B, the $R_0=2.1\times10^6$ and $10^7$~cm cases fall in G2015 Region~III, while the $10^8$~cm case falls in Region~II. The corresponding point estimates are $(\sigma_0,\eta)\simeq(0,413)$, $(1.79,146)$, and $(11.94,48)$. Under the energy-flux-ratio definition above, these cases progress from a hot-component-dominated boundary solution to Poynting-flux-dominated initial conditions as $R_0$ increases. The large value of $\eta$ indicates high dimensionless entropy but does not reverse the classification set by $\sigma_0$. At $R_0=10^8$~cm, the reported $\sigma_{\rm ph}\simeq0.77$ and boundary value $\sigma_{15}\simeq0$ would permit internal shocks at $10^{15}$~cm; they do not establish internal shocks as the unique emission mechanism.

Values printed as $\sigma_0=0$ and $\epsilon_{\rm Th}=1$ in Table~\ref{table5} are boundary flags, not measurements; the latter include formal inversions with $\epsilon_{\rm Th}>1$. All such rows are excluded from threshold counts and categorical interpretation.

Taken together, the H2013 and G2015 calculations reveal a common directional trend despite their different parameter definitions. Increasing the assumed launch radius shifts the inferred initial energy budget toward a larger magnetic contribution, whereas the smallest adopted radii favor hotter and less magnetized initial conditions. Because the two methods share the observational inputs but encode different stages of the flow, agreement in this trend is more informative than a numerical comparison of their individual magnetization parameters. The cross-framework result identifies $R_0$ as a dominant control variable and shows explicitly why a jet composition inferred at one assumed radius cannot be treated as an intrinsic, model-independent property of the burst.

Table~\ref{table5} separates valid from boundary/inadmissible inversions. At $f_{\rm NT}=0.5$, H2013 gives $N(\epsilon_{\rm Th}>0.5)/N_{\rm valid}=1/2$, 1/4, and 2/8 at 21, 100, and 1000~km, with 6, 4, and 0 invalid solutions, respectively. At $f_\gamma=0.5$, G2015 gives $N(\sigma_0<1)/N_{\rm valid}=0/1$, 1/4, and 1/8, with 7, 4, and 0 boundary solutions. The changing valid denominator precludes interpreting these as a monotonic population fraction. The robust result is instead parameter-directional: for bursts with valid solutions at multiple radii, larger $R_0$ generally implies lower $\epsilon_{\rm Th}$ and higher $\sigma_0$.

The dependence of these counts on the two redshift-provenance cases was evaluated by leave-one-out tests. Excluding GRB~231129C, whose $z=0.5$ value is estimated, leaves the H2013 counts unchanged at 1/2 and 1/4 for 21 and 100~km and changes the 1000~km count from 2/8 to 1/7. The corresponding G2015 counts remain 0/1 and 1/4 and change from 1/8 to 1/7. Excluding the merger-associated GRB~211211A likewise leaves the direction of the radius dependence unchanged. These tests show that neither redshift-provenance case generates the cross-framework trend with $R_0$; given the small and heterogeneous remaining sample, the leave-one-out ratios are used only as robustness diagnostics and not as population fractions.

\begin{table}[h!]
\centering
\caption{Observed Spectral Inputs and H2013 Outflow Parameters}
\label{table4}
\resizebox{\textwidth}{!}{
\begin{tabular}{cccccc}
\toprule
\multirow{2}*{GRB} & Time interval & $kT_{\rm BB}$ & $F_{\rm BB}$ & $\phi$ & $\mathcal{R}$  \\
 &$(\mathrm{s})$&$(\mathrm{keV})$&$(\times10^{-7}\,\mathrm{erg\,cm^{-2}\,s^{-1}})$&\textbf{$(F_{\rm BB}/F_{\rm tot})$}&$(\times10^{-19})$ \\

\hline
  \midrule

GRB130518A & $0-60$ & $27.13 \pm 0.83$ & $1.10 \pm 0.10$ & $0.069 \pm 0.006$ & $4.44 \pm 0.33$ \\
GRB210610B & $10-95$ & $22.77 \pm 5.11$ & $0.06 \pm 0.05$ & $0.005 \pm 0.004$ & $1.49 \pm 0.91$ \\
GRB210619B & $0-70$ & $15.82^{+3.98}_{-2.17}$ & $0.65 \pm 0.29$ & $0.015 \pm 0.007$ & $10.06 \pm 3.57$ \\
GRB211211A & $0-55$ & $17.84 \pm 0.35$ & $7.28 \pm 0.24$ & $0.084 \pm 0.003$ & $26.45 \pm 1.11$ \\
GRB231129C & $0.1-8$ & $87.85 \pm 2.48$ & $30.00 \pm 1.62$ & $0.311 \pm 0.016$ & $2.26 \pm 0.14$ \\
GRB240825A & $0.1-8$ & $127.01 \pm 5.50$ & $40.00 \pm 2.43$ & $0.289 \pm 0.019$ & $1.17 \pm 0.11$ \\
GRB241228B & $0-25$ & $72.08 \pm 18.64$ & $1.91 \pm 1.02$ & $0.104 \pm 0.055$ & $0.83 \pm 0.48$ \\
GRB250407A & $0.1-10$ & $27.17 \pm 0.61$ & $5.14 \pm 0.47$ & $0.111 \pm 0.010$ & $9.58 \pm 0.61$ \\

\hline
\multicolumn{2}{l}{H2013 approach} \\

\multirow{2}*{GRB} & $\mathcal{R}$ & $R_0$ & $\Gamma$ & $R_{\rm ph}$ \\

 &$(\times 10^{-19})$ &$(\times 10^7 [f_{\rm NT}/\epsilon_{\rm Th}]^{3/2}\,\mathrm{cm})$ &$(\times [(1+\sigma)f_{\rm NT}]^{-1/4})$ &$(\times 10^{11}[(1+\sigma)f_{\rm NT}]^{-1/4}\,\mathrm{cm})$ \\

    \midrule
    GRB130518A & $4.44 \pm 0.33$ & $2.38 \pm 0.38$ & $441.87 \pm 16.46$ & $5.17 \pm 0.34$ \\
    GRB210610B & $1.49 \pm 0.91$ & $0.012^{+0.022}_{-0.008}$ & $349.01 \pm 115.39$ & $1.40 \pm 0.76$ \\
    GRB210619B & $10.06 \pm 3.57$ & $0.50 \pm 0.38$ & $402.58 \pm 73.67$ & $11.11 \pm 3.45$ \\
    GRB211211A & $26.45 \pm 1.11$ & $3.49 \pm 0.24$ & $86.71 \pm 1.40$ & $1.09 \pm 0.04$ \\
    GRB231129C & $2.26 \pm 0.14$ & $13.77 \pm 1.77$ & $309.65 \pm 8.55$ & $1.41 \pm 0.07$ \\
    GRB240825A & $1.17 \pm 0.11$ & $6.95 \pm 1.17$ & $445.53 \pm 16.42$ & $1.20 \pm 0.09$ \\
    GRB241228B & $0.83 \pm 0.48$ & $0.84^{+1.17}_{-0.49}$ & $725.46 \pm 179.66$ & $1.56 \pm 0.75$ \\
    GRB250407A & $9.58 \pm 0.61$ & $11.72 \pm 1.97$ & $322.18 \pm 12.27$ & $8.50 \pm 0.50$ \\

\hline
\multicolumn{2}{l}{G2015 approach} \\

GRB & &21km & 100km & 1000km  \\
    
    \midrule
    \multirow{2}{*}{GRB130518A} & $\Gamma_{\rm ph}$ & $451.96 \pm 13.07$ & $451.96 \pm 13.07$ & $451.00 \pm 10.73$ \\
    & $R_{\rm ph}\ (10^{11}\ \rm cm)$ & $10.30 \pm 0.77$ & $10.30 \pm 0.77$ & $10.27 \pm 0.95$ \\
    \midrule
    \multirow{2}{*}{GRB210610B} & $\Gamma_{\rm ph}$ & $351.02 \pm 122.89$ & $253.94 \pm 57.99$ & $129.74 \pm 29.63$ \\
    & $R_{\rm ph}\ (10^{11}\ \rm cm)$ & $2.74 \pm 2.65$ & $1.98^{+3.29}_{-0.74}$ & $1.01^{+1.68}_{-0.38}$ \\
    \midrule
    \multirow{2}{*}{GRB210619B} & $\Gamma_{\rm ph}$ & $405.93 \pm 58.09$ & $405.93 \pm 58.09$ & $352.43 \pm 43.10$ \\
    & $R_{\rm ph}\ (10^{11}\ \rm cm)$ & $21.81 \pm 7.98$ & $21.81 \pm 7.98$ & $18.92 \pm 16.63$ \\
    \midrule
    \multirow{2}{*}{GRB211211A} & $\Gamma_{\rm ph}$ & $89.04 \pm 1.19$ & $89.04 \pm 1.19$ & $89.04 \pm 1.19$ \\
    & $R_{\rm ph}\ (10^{11}\ \rm cm)$ & $2.18 \pm 0.08$ & $2.18 \pm 0.08$ & $2.18 \pm 0.08$ \\
    \midrule
    \multirow{2}{*}{GRB231129C} & $\Gamma_{\rm ph}$ & $341.43 \pm 6.90$ & $341.43 \pm 6.90$ & $341.43 \pm 6.90$ \\
    & $R_{\rm ph}\ (10^{11}\ \rm cm)$ & $3.02 \pm 0.17$ & $3.02 \pm 0.17$ & $3.02 \pm 0.17$ \\
    \midrule
    \multirow{2}{*}{GRB240825A} & $\Gamma_{\rm ph}$ & $487.43 \pm 13.91$ & $487.43 \pm 13.91$ & $452.12 \pm 8.26$ \\
    & $R_{\rm ph}\ (10^{11}\ \rm cm)$ & $2.54 \pm 0.20$ & $2.54 \pm 0.20$ & $2.36 \pm 0.24$ \\
    \midrule
    \multirow{2}{*}{GRB241228B} & $\Gamma_{\rm ph}$ & $749.01 \pm 148.06$ & $749.01 \pm 148.06$ & $503.88 \pm 73.79$ \\
    & $R_{\rm ph}\ (10^{11}\ \rm cm)$ & $3.13 \pm 1.66$ & $3.13 \pm 1.66$ & $2.11 \pm 1.43$ \\
    \midrule
    \multirow{2}{*}{GRB250407A} & $\Gamma_{\rm ph}$ & $333.37 \pm 9.34$ & $333.37 \pm 9.34$ & $333.37 \pm 9.34$ \\
    & $R_{\rm ph}\ (10^{11}\ \rm cm)$ & $17.12 \pm 1.20$ & $17.12 \pm 1.20$ & $17.12 \pm 1.20$ \\

\hline
\bottomrule
\end{tabular}
}

\end{table}

\section{Discussion and Conclusions\label{Sections5}}

The analysis establishes a traceable hierarchy from observed spectral structure to conditional jet physics. After the stability exclusions, an added BB is favored by $\Delta\mathrm{DIC}>10$ in 203 of 215 accepted bins (60--100\% per burst), while the independent mBB+PL decomposition gives $f_{\rm mBB}>0.5$ in 214 of 215 bins. The lower 3/5 added-BB count for GRB~250407A shows why a statistical component claim must not be inferred from unstable DIC values. The sole exception to mBB-flux dominance, the 15.26--16.07~s interval of GRB~211211A, has $f_{\rm mBB}=0.353$ and a weakly constrained mBB+PL fit. Thus, recurrent thermal-like curvature and mBB-flux dominance are directly demonstrated for this selected data set, but neither statistic alone proves a photospheric origin. The distinction is essential because the high 214/215 fraction is partly imposed by the event-level selection and cannot be interpreted as the prevalence of thermal dominance in the GBM population.

Related time-resolved studies provide useful context. For GRB~210610B, \citet{2022ApJ...932...25C} reported that adding a thermal component improved 76.47\% of their time slices and interpreted the event with a hybrid-jet model. All 13 bins retained here pass the added-BB comparison and the mBB-flux criterion, but the percentages are not directly comparable because the binning, significance threshold, and baseline comparisons differ. \citet{2019ApJS..242...16L} instead identified pulse-to-pulse transitions from thermal-like to non-thermal emission in 9 of 43 bright multipulse bursts. The present sample addresses a different question: it selects bursts with recurrent time-resolved mBB-flux dominance and tests how a common spectral data product maps into two jet prescriptions. It therefore does not establish a temporal fireball-to-Poynting transition for any individual event.

The empirical-model comparison identifies an additional source of uncertainty. The CPL--mBB correlations ($r=0.83$ for $m$--$\alpha$ and $r=0.78$ for $kT_{\max}$--$E_p$) are appreciably stronger than the Band--mBB correlations ($r=0.41$ and $0.48$). This pattern is consistent with CPL and mBB sharing stronger imposed high-energy curvature, whereas Band has an independently fitted high-energy index. It also extends the model-dependence demonstrated by \citet{2022ApJ...941...27L} and \citet{2023ApJS..266...31L}: choosing Band or CPL can alter $\alpha$, $E_p$, their correlations, and consequently the observables propagated into a physical model. The correlations here are therefore descriptive diagnostics of spectral shape, not independent proof that the mBB component is photospheric.

The time-integrated results sharpen this distinction. The narrow-BB fractions span 0.005--0.311, and six of eight are below 0.12. This is consistent with the warning of \citet{2013A&A...551A.124H} that an identifiable photospheric component can remain subdominant and need not resemble a single Planck spectrum. The result does not contradict the time-resolved mBB-flux counts: mBB represents a broadened temperature distribution within short bins, whereas the integrated BB is a single-temperature component fitted after spectral evolution has been combined. The contrast between these quantities is itself informative, because it shows that statistical component evidence, time-resolved energetic dominance, and the effective integrated input to a photospheric calculation are three different measurements.

Within the non-dissipative photosphere assumption, H2013 and G2015 translate the effective integrated observables into complementary parameter-space constraints. H2013 relates them to the initial thermal fraction and passive-field magnetization at the end of acceleration, whereas G2015 relates them to the dimensionless entropy and magnetization at the initial, photospheric, and larger radii. After boundary solutions are removed, H2013 yields 1/2, 1/4, and 2/8 valid solutions above $\epsilon_{\rm Th}=0.5$ at 21, 100, and 1000~km; G2015 yields 0/1, 1/4, and 1/8 valid solutions below $\sigma_0=1$. Because the number of valid inversions changes with radius, these ratios are diagnostics rather than occurrence fractions. The scientifically robust cross-framework result is that increasing $R_0$ lowers $\epsilon_{\rm Th}$ and generally raises $\sigma_0$ where both radii yield valid solutions.

This result both agrees with and qualifies earlier hybrid-jet interpretations. Gao \& Zhang showed that a wide variety of thermal contributions can be reproduced by varying $\eta$ and $\sigma_0$, often requiring appreciable magnetization \citep{2015ApJ...801..103G}. The strong magnetization inferred here for GRB~210610B over the full radius grid is broadly compatible with the hybrid interpretation of \citet{2022ApJ...932...25C}. For most of the other bursts, however, the inferred initial dominance changes across the adopted radius grid. The present comparison therefore advances beyond a burst-by-burst categorical label by identifying which assumption drives the change and by applying the same test to both H2013 and G2015.

The adopted efficiencies introduce a second controlled degeneracy. Detailed energy-budget studies of GRB~190114C obtained radiative efficiencies only after combining prompt thermal emission with independent early-afterglow constraints \citep{2023ApJ...944L..57L,2024ApJ...972..195L}. Comparable external constraints are unavailable for the eight events analyzed here. Consequently, $f_{\rm NT}$ and $f_\gamma$ are varied or fixed as scenario parameters rather than measured quantities. A low $\sigma_{15}$ permits efficient internal shocks at large radius, while a higher magnetization leaves magnetic dissipation viable; neither condition uniquely selects the observed non-thermal mechanism.

The principal contribution is therefore not a universal jet-composition fraction, and the abundance of invalid small-radius inversions supplies an additional constraint on the joint spectral--dynamical assumptions. It is an evidence-linked demonstration that (1) recurrent time-resolved mBB-flux dominance can coexist with a subdominant time-integrated narrow BB, (2) the inferred thermal-like parameters depend measurably on the empirical baseline, and (3) two physically distinct photospheric mappings show the same sample-wide shift toward stronger initial magnetic contributions as the assumed launch radius increases. The limitations are correspondingly clear: the sample is small and selected, the time bins within a burst are not independent, the integrated BB parameters are effective quantities, and the jet uncertainties inherit the adopted posterior approximation.

Future work can test the result by propagating the complete time-resolved posterior through both frameworks and by combining prompt spectra with afterglow, polarization, or independent engine-size constraints. Such measurements would determine whether the radius dependence survives within individual ejecta shells and could break the present degeneracy among launch radius, efficiency, and spectral decomposition. Until then, the selected GRBs support diverse thermal, magnetic, and hybrid initial conditions organized by a reproducible conditional dependence, rather than a single model-independent jet classification.

\section*{Acknowledgments}

We acknowledge the use of public data from the Fermi archive. This work was supported by the National Natural Science Foundation of China (grants 12163007 and 11763009), the Key Laboratory of Colleges and Universities in Yunnan Province for High-energy Astrophysics, and the National Astronomical Observatory--Yunnan Normal University Astronomy Science Education Base.


\section*{Appendix}
\begin{figure}[htbp]
\centering
\includegraphics[width=0.47\textwidth]{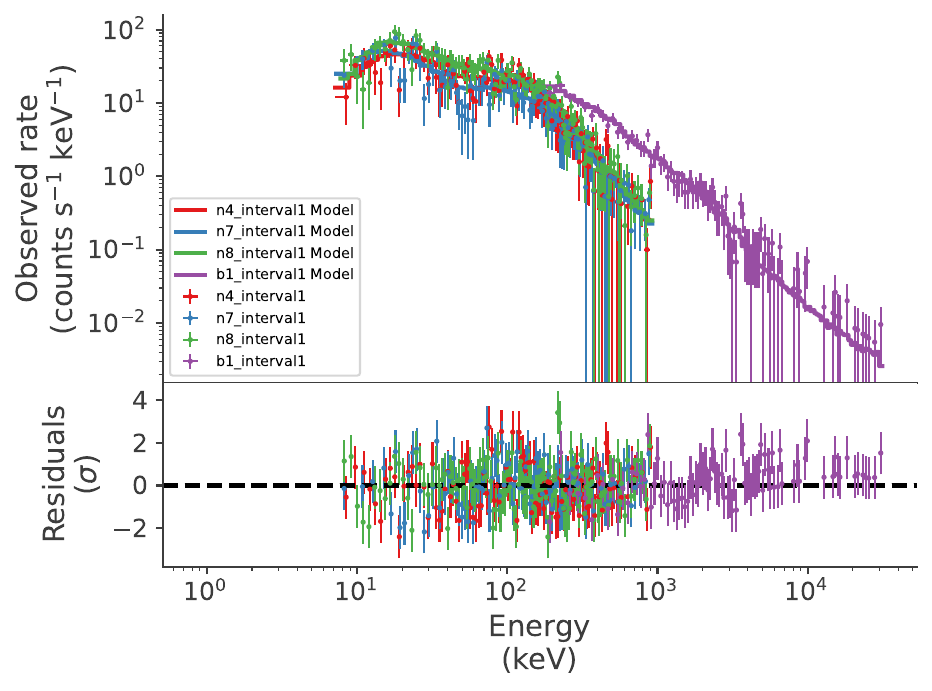}
\includegraphics[width=0.47\textwidth]{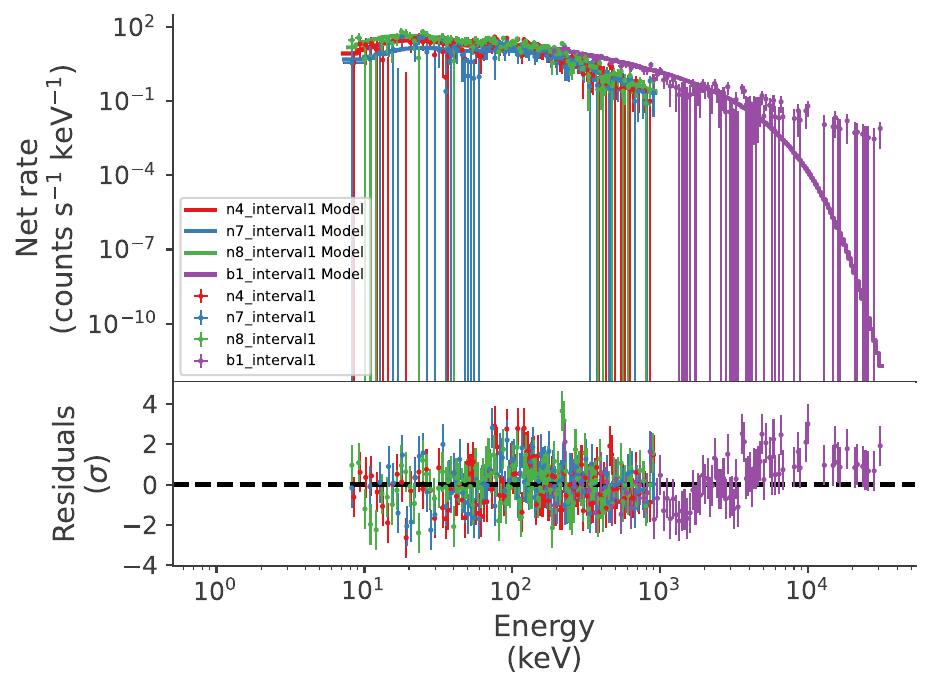}
\includegraphics[width=0.47\textwidth]{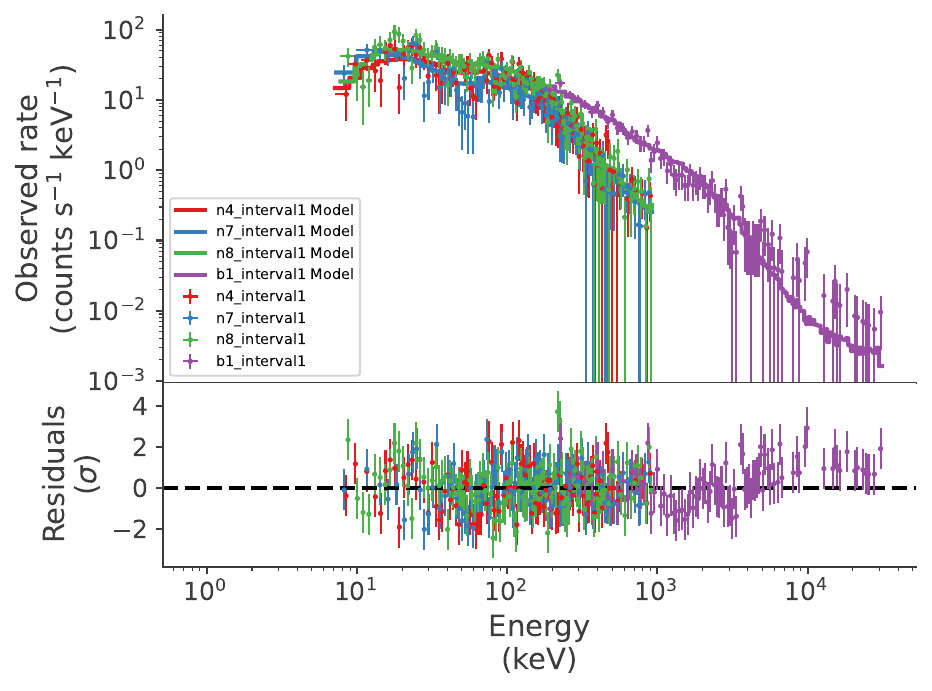}
\includegraphics[width=0.47\textwidth]{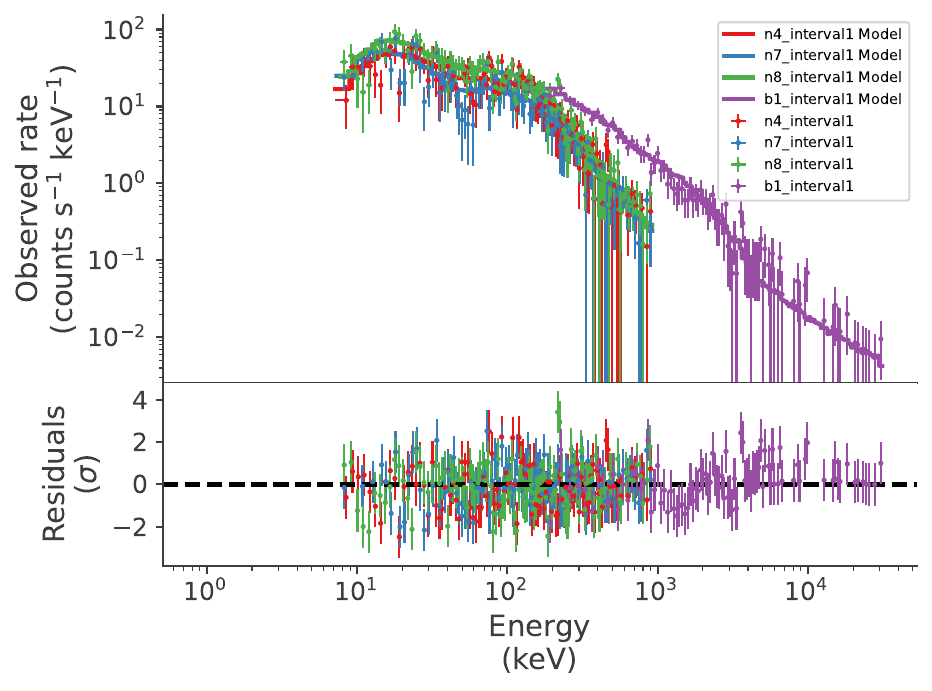}
\figcaption{Photon count spectra and residuals for the first time bin (0.07--0.21~s) of GRB~210619B. From left to right and top to bottom, the panels correspond to the Band, CPL, mBB, and mBB+PL models. \label{fig 8}}
\end{figure}

\begin{figure}[htbp]
\centering
\includegraphics[width=0.47\textwidth]{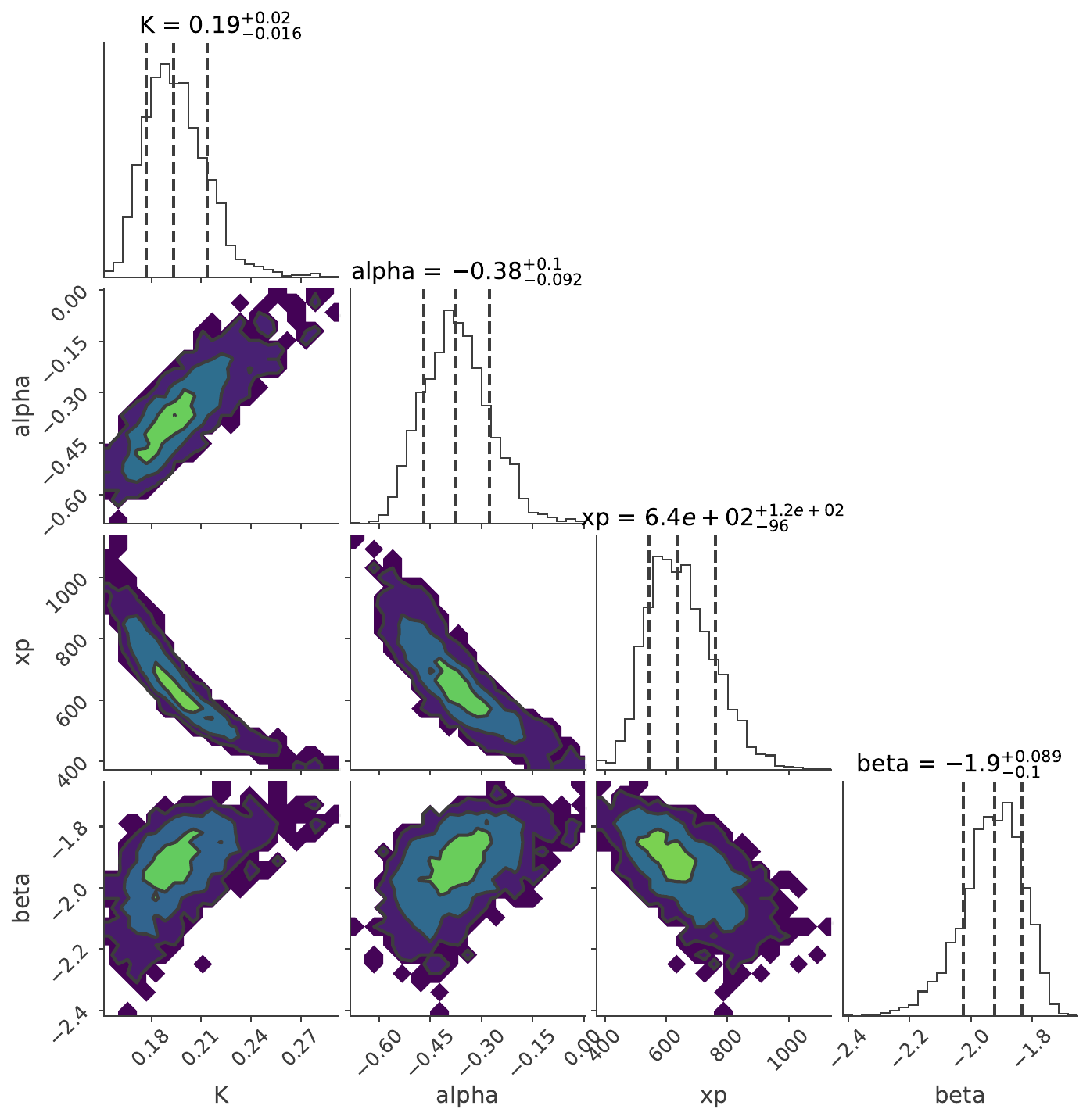}
\includegraphics[width=0.47\textwidth]{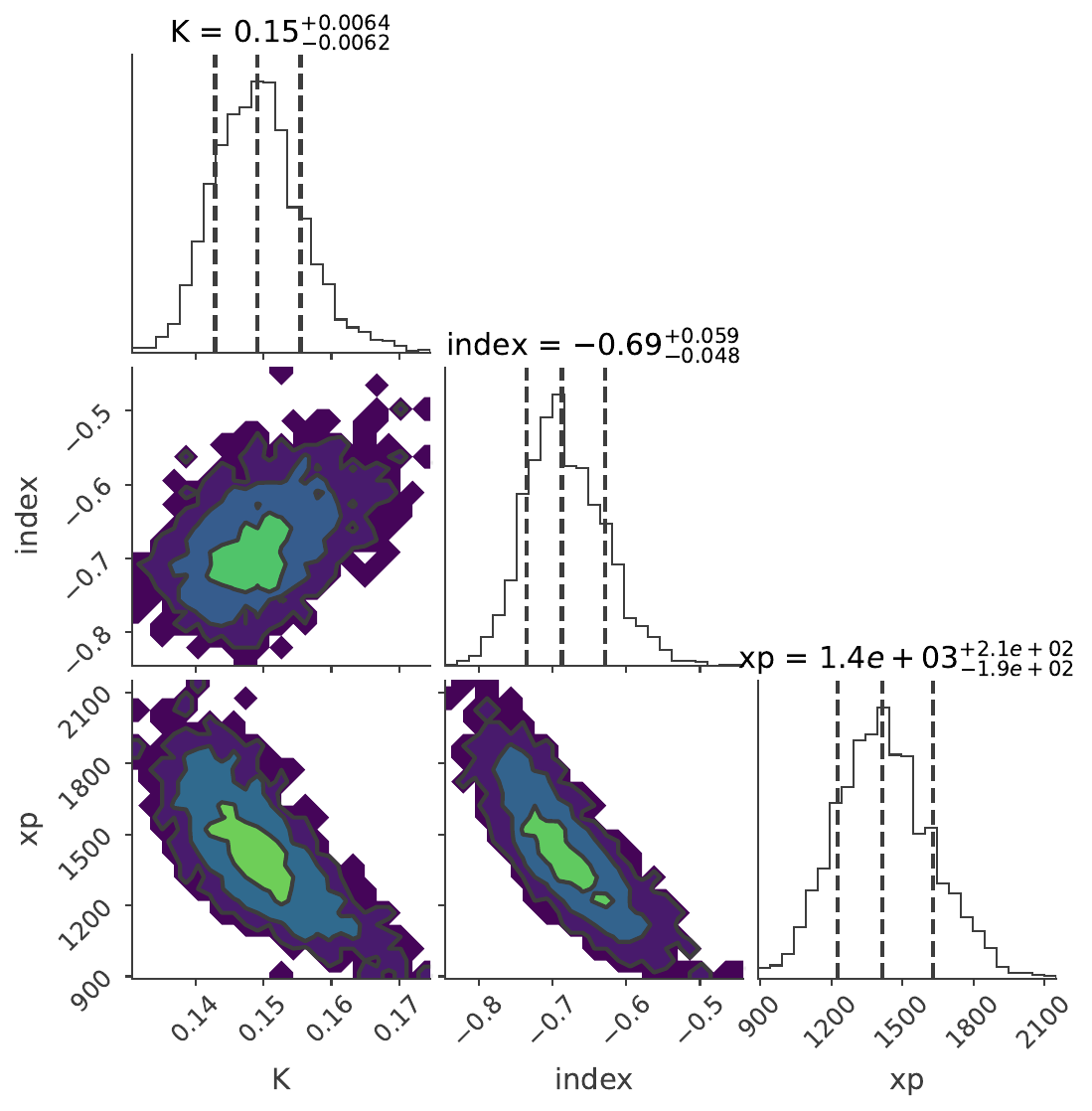}
\includegraphics[width=0.47\textwidth]{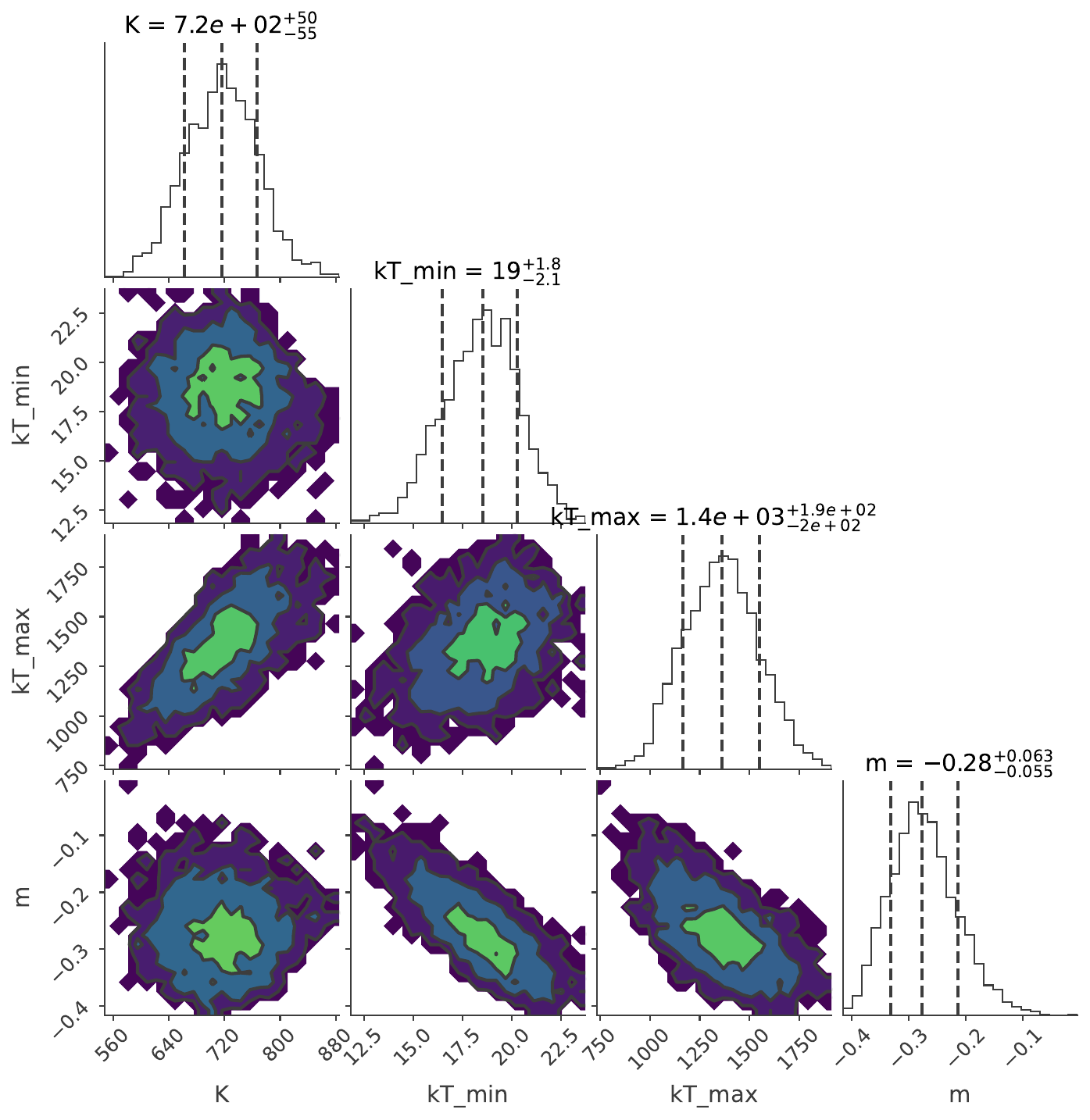}
\includegraphics[width=0.47\textwidth]{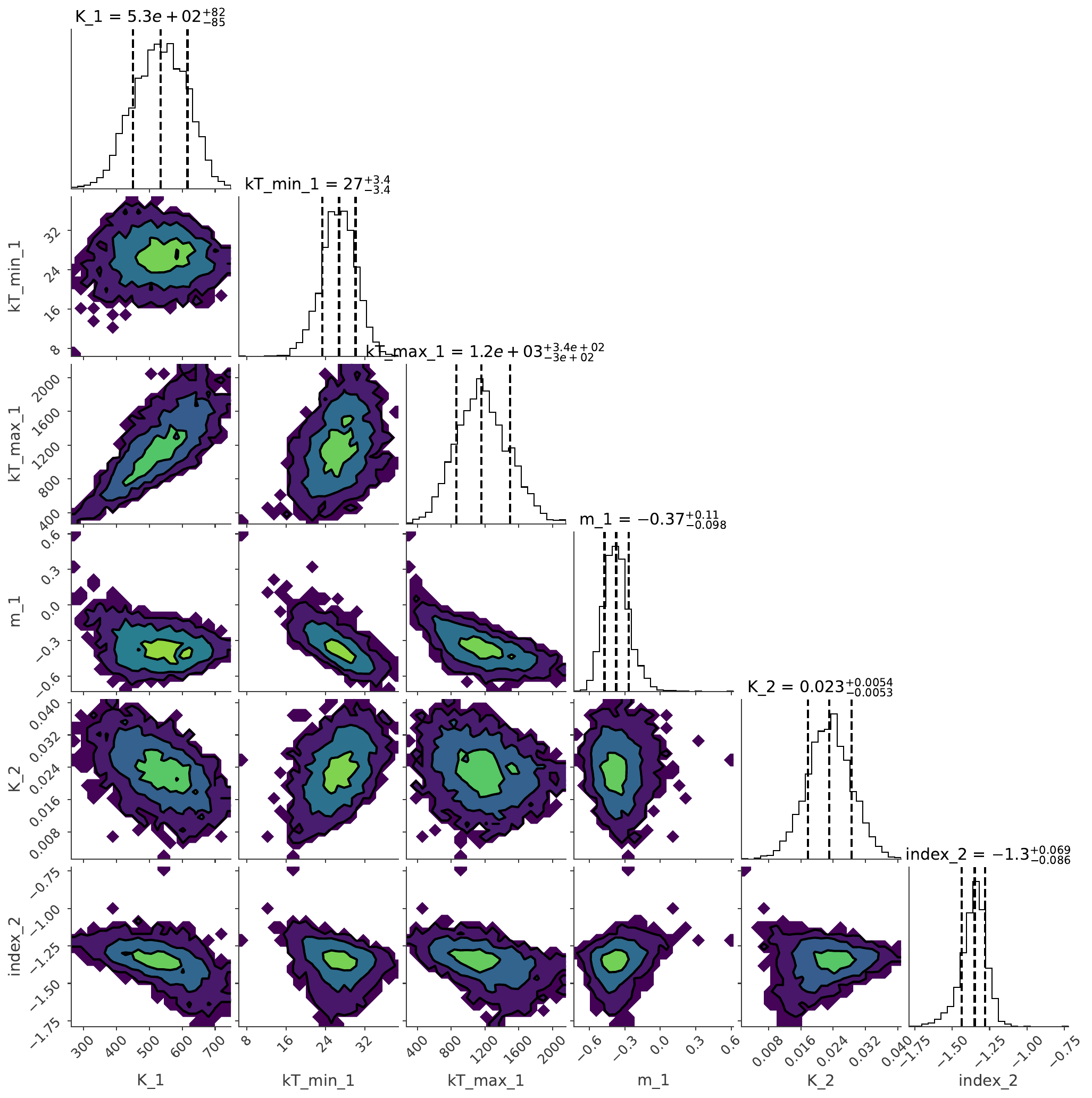}
\figcaption{Posterior corner plots for the same GRB~210619B time bin. From left to right and top to bottom, the panels correspond to the Band, CPL, mBB, and mBB+PL models. \label{fig 9}}
\end{figure}

This appendix presents (1) the photon count spectra and corner plots of the first time bin (0.07--0.21 s) of GRB 210619B, fitted with the Band, CPL, mBB, and mBB+PL models (Figures \ref{fig 8} and \ref{fig 9}), and (2) the jet characteristic parameters of the eight GRBs derived with the H2013 and G2015 methods from the time-integrated spectral fitting results (Table \ref{table5}).

\begin{table*}[htbp]
\centering
\caption{Conditional H2013 and G2015 Parameter-Space Results\label{tab5}}
\label{table5}
\resizebox{\textwidth}{!}{
\begin{tabular}{cccccccccccccccc}
\toprule
\multirow{3}*{GRB} &\multirow{3}*{Radius}& &\multicolumn{7}{c}{H2013 approach} & & \multicolumn{5}{c}{G2015 approach}  \\
\cline{4-10}
\cline{12-16}
 & &\multirow{2}*{$\frac{f_{\rm NT}}{\epsilon_{\rm Th}}$} &\multicolumn{3}{c}{$\epsilon_{\rm Th}$}& &\multicolumn{3}{c}{$\sigma_{\rm passive}$}& &\multirow{2}*{Regime}&$\sigma_0$&$\eta$&$\sigma_{\rm ph}$&$\sigma_{15}$ \\
\cline{4-6}
\cline{8-10}
 &$(\mathrm{km})$& &$f_{\rm NT}=0.5$&$f_{\rm NT}=0.1$&$f_{\rm NT}=0.05$& &$f_{\rm NT}=0.5$&$f_{\rm NT}=0.1$&$f_{\rm NT}=0.05$& & &$f_\gamma=0.5$  \\

\hline
  \midrule

\multirow{3}{*}{GRB130518A} & 21   & 0.20 & 1 & 0.51 & 0.25 & & 0 & 0.98 & 2.96 & & III & 0 & 1232.37 & \dots & \dots \\
& 100  & 0.56 & 0.89 & 0.18 & 0.09 & & 0.12 & 4.60 & 10.21 & & III & 0.04 & 435.40 & \dots & \dots \\
& 1000 & 2.60 & 0.19 & 0.04 & 0.02 & & 4.20 & 25.01 & 51.02 & & II & 3.83 & 94.36 & 0.45 & 0 \\
\hline
\multirow{3}{*}{GRB210610B} & 21   & 6.59 & 0.08 & 0.02 & 0    & & 12.18 & 64.89 & 130.78 & & III & 10.44 & 30.76 & \dots & \dots \\
& 100  & 18.65 & 0.03 & 0.01 & 0    & & 36.30 & 185.49 & 371.99 & & II & 31.38 & 28.69 & 2.67 & 0 \\
& 1000 & 86.56 & 0.01 & 0    & 0    & & 172.12 & 864.62 & 1730.25 & & II & 149.28 & 46.35 & 52.81 & 1.52 \\
\hline
\multirow{3}{*}{GRB210619B} & 21   & 0.56 & 0.89 & 0.18 & 0.09 & & 0.12 & 4.62 & 10.23 & & III & 0 & 413.06 & \dots & \dots \\
& 100  & 1.59 & 0.31 & 0.06 & 0.03 & & 2.18 & 14.9 & 30.8 & & III & 1.79 & 145.93 & \dots & \dots \\
& 1000 & 7.38 & 0.07 & 0.01 & 0.007 & & 13.76 & 72.79 & 146.59 & & II & 11.94 & 48.02 & 0.77 & 0 \\
\hline
\multirow{3}{*}{GRB211211A} & 21   & 0.15 & 1 & 0.65 & 0.33 & & 0 & 0.54 & 2.07 & & III & 0 & 308.22 & \dots & \dots \\
& 100  & 0.43 & 1 & 0.23 & 0.12 & & 0 & 3.35 & 7.69 & & III & 0 & 108.90 & \dots & \dots \\
& 1000 & 2.02 & 0.25 & 0.05 & 0.02 & & 3.03 & 19.17 & 39.34 & & III & 2.80 & 23.46 & \dots & \dots \\
\hline
\multirow{3}{*}{GRB231129C} & 21   & 0.06 & 1 & 1 & 0.81 & & 0 & 0 & 0.23 & & III & 0 & 2219.02 & \dots & \dots \\
& 100  & 0.17 & 1 & 0.57 & 0.29 & & 0 & 0.74 & 2.48 & & III & 0 & 783.98 & \dots & \dots \\
& 1000 & 0.81 & 0.62 & 0.12 & 0.06 & & 0.62 & 7.08 & 15.16 & & III & 1.03 & 168.90 & \dots & \dots \\
\hline
\multirow{3}{*}{GRB240825A} & 21   & 0.10 & 1 & 1 & 0.52 & & 0 & 0 & 0.94 & & III & 0 & 2072.93 & \dots & \dots \\
& 100  & 0.27 & 1 & 0.36 & 0.18 & & 0 & 1.74 & 4.49 & & III & 0 & 732.37 & \dots & \dots \\
& 1000 & 1.27 & 0.39 & 0.08 & 0.04 & & 1.55 & 11.74 & 24.48 & & II & 2.10 & 197.62 & 1.36 & 0.08 \\
\hline
\multirow{3}{*}{GRB241228B} & 21   & 0.40 & 1 & 0.25 & 0.13 & & 0 & 2.95 & 6.91 & & III & 0 & 984.81 & \dots & \dots \\
& 100  & 1.12 & 0.45 & 0.09 & 0.04 & & 1.24 & 10.19 & 21.39 & & III & 1.16 & 347.93 & \dots & \dots \\
& 1000 & 5.20 & 0.10 & 0.02 & 0.01 & & 9.39 & 50.95 & 102.91 & & II & 9.02 & 246.10 & 3.90 & 0 \\
\hline
\multirow{3}{*}{GRB250407A} & 21   & 0.07 & 1 & 1 & 0.73 & & 0 & 0 & 0.37 & & III & 0 & 2509.39 & \dots & \dots \\
& 100  & 0.19 & 1 & 0.52 & 0.26 & & 0 & 0.94 & 2.88 & & III & 0 & 886.57 & \dots & \dots \\
& 1000 & 0.90 & 0.56 & 0.11 & 0.06 & & 0.80 & 8.00 & 17.00 & & III & 0.75 & 191.01 & \dots & \dots \\

\hline
\bottomrule
\end{tabular}
}

\tablecomments{All entries are conditional on the listed launch radius and efficiencies. The G2015 columns report $\sigma_0$, not $1+\sigma_0$. Values shown as $\sigma_0=0$ or $\epsilon_{\rm Th}=1$ are boundary flags; the latter category includes formal $\epsilon_{\rm Th}>1$ inversions. They are excluded from all threshold counts and physical classifications. At $R_0=21,100,1000$~km the numbers of invalid H2013 solutions are 6, 4, and 0, and the numbers of boundary G2015 solutions are 7, 4, and 0, respectively.}
\end{table*}
\FloatBarrier

\providecommand{\dodoi}[1]{doi:~\href{https://doi.org/#1}{\nolinkurl{#1}}}
\bibliographystyle{aasjournal}
\bibliography{reference}

@ARTICLE{2015PhR...561....1K,
       author = {{Kumar}, Pawan and {Zhang}, Bing},
        title = "{The physics of gamma-ray bursts \& relativistic jets}",
      journal = {\physrep},
         year = 2015,
        month = feb,
       volume = {561},
        pages = {1-109},
          doi = {10.1016/j.physrep.2014.09.008},
archivePrefix = {arXiv},
       eprint = {1410.0679},
 primaryClass = {astro-ph.HE},
       adsurl = {https://ui.adsabs.harvard.edu/abs/2015PhR...561....1K}
}

@BOOK{2018pgrb.book.....Z,
       author = {{Zhang}, Bing},
        title = "{The Physics of Gamma-Ray Bursts}",
         year = 2018,
          doi = {10.1017/9781139226530},
       adsurl = {https://ui.adsabs.harvard.edu/abs/2018pgrb.book.....Z}
}

@ARTICLE{2019MmSAI..90...57M,
       author = {{M{\'e}sz{\'a}ros}, P.},
        title = "{Gamma-Ray Bursts: theoretical issues and developments}",
      journal = {\memsai},
         year = 2019,
        month = jan,
       volume = {90},
        pages = {57},
          doi = {10.48550/arXiv.1904.10488},
archivePrefix = {arXiv},
       eprint = {1904.10488},
 primaryClass = {astro-ph.HE},
       adsurl = {https://ui.adsabs.harvard.edu/abs/2019MmSAI..90...57M}
}

@ARTICLE{1978MNRAS.183..359C,
       author = {{Cavallo}, G. and {Rees}, M.~J.},
        title = "{A qualitative study of cosmic fireballs and gamma -ray bursts.}",
      journal = {\mnras},
         year = 1978,
        month = may,
       volume = {183},
        pages = {359-365},
          doi = {10.1093/mnras/183.3.359},
       adsurl = {https://ui.adsabs.harvard.edu/abs/1978MNRAS.183..359C}
}

@ARTICLE{1986ApJ...308L..47G,
       author = {{Goodman}, J.},
        title = "{Are gamma-ray bursts optically thick?}",
      journal = {\apjl},
         year = 1986,
        month = sep,
       volume = {308},
        pages = {L47},
          doi = {10.1086/184741},
       adsurl = {https://ui.adsabs.harvard.edu/abs/1986ApJ...308L..47G}
}

@ARTICLE{1986ApJ...308L..43P,
       author = {{Paczynski}, B.},
        title = "{Gamma-ray bursters at cosmological distances}",
      journal = {\apjl},
         year = 1986,
        month = sep,
       volume = {308},
        pages = {L43-L46},
          doi = {10.1086/184740},
       adsurl = {https://ui.adsabs.harvard.edu/abs/1986ApJ...308L..43P}
}

@ARTICLE{1993ApJ...413..281B,
       author = {{Band}, D. and {Matteson}, J. and {Ford}, L. and {Schaefer}, B. and {Palmer}, D. and {Teegarden}, B. and {Cline}, T. and {Briggs}, M. and {Paciesas}, W. and {Pendleton}, G. and {Fishman}, G. and {Kouveliotou}, C. and {Meegan}, C. and {Wilson}, R. and {Lestrade}, P.},
        title = "{BATSE Observations of Gamma-Ray Burst Spectra. I. Spectral Diversity}",
      journal = {\apj},
         year = 1993,
        month = aug,
       volume = {413},
        pages = {281},
          doi = {10.1086/172995},
       adsurl = {https://ui.adsabs.harvard.edu/abs/1993ApJ...413..281B}
}

@ARTICLE{1990ApJ...365L..55S,
       author = {{Shemi}, Amotz and {Piran}, Tsvi},
        title = "{The Appearance of Cosmic Fireballs}",
      journal = {\apjl},
         year = 1990,
        month = dec,
       volume = {365},
        pages = {L55},
          doi = {10.1086/185887},
       adsurl = {https://ui.adsabs.harvard.edu/abs/1990ApJ...365L..55S}
}

@ARTICLE{1993ApJ...405..278M,
       author = {{Meszaros}, P. and {Rees}, M.~J.},
        title = "{Relativistic Fireballs and Their Impact on External Matter: Models for Cosmological Gamma-Ray Bursts}",
      journal = {\apj},
         year = 1993,
        month = mar,
       volume = {405},
        pages = {278},
          doi = {10.1086/172360},
       adsurl = {https://ui.adsabs.harvard.edu/abs/1993ApJ...405..278M}
}

@ARTICLE{1993MNRAS.263..861P,
       author = {{Piran}, T. and {Shemi}, A. and {Narayan}, R.},
        title = "{Hydrodynamics of Relativistic Fireballs}",
      journal = {\mnras},
         year = 1993,
        month = aug,
       volume = {263},
        pages = {861},
          doi = {10.1093/mnras/263.4.861},
archivePrefix = {arXiv},
       eprint = {astro-ph/9301004},
 primaryClass = {astro-ph},
       adsurl = {https://ui.adsabs.harvard.edu/abs/1993MNRAS.263..861P}
}

@ARTICLE{1990ApJ...363..218P,
       author = {{Paczynski}, Bohdan},
        title = "{Super-Eddington Winds from Neutron Stars}",
      journal = {\apj},
         year = 1990,
        month = nov,
       volume = {363},
        pages = {218},
          doi = {10.1086/169332},
       adsurl = {https://ui.adsabs.harvard.edu/abs/1990ApJ...363..218P}
}

@ARTICLE{2000ApJ...530..292M,
       author = {{M{\'e}sz{\'a}ros}, P. and {Rees}, M.~J.},
        title = "{Steep Slopes and Preferred Breaks in Gamma-Ray Burst Spectra: The Role of Photospheres and Comptonization}",
      journal = {\apj},
         year = 2000,
        month = feb,
       volume = {530},
       number = {1},
        pages = {292-298},
          doi = {10.1086/308371},
archivePrefix = {arXiv},
       eprint = {astro-ph/9908126},
 primaryClass = {astro-ph},
       adsurl = {https://ui.adsabs.harvard.edu/abs/2000ApJ...530..292M}
}

@ARTICLE{2006ApJ...642..995P,
       author = {{Pe'er}, Asaf and {M{\'e}sz{\'a}ros}, Peter and {Rees}, Martin J.},
        title = "{The Observable Effects of a Photospheric Component on GRB and XRF Prompt Emission Spectrum}",
      journal = {\apj},
         year = 2006,
        month = may,
       volume = {642},
       number = {2},
        pages = {995-1003},
          doi = {10.1086/501424},
archivePrefix = {arXiv},
       eprint = {astro-ph/0510114},
 primaryClass = {astro-ph},
       adsurl = {https://ui.adsabs.harvard.edu/abs/2006ApJ...642..995P}
}

@ARTICLE{2004ApJ...614..827R,
       author = {{Ryde}, Felix},
        title = "{The Cooling Behavior of Thermal Pulses in Gamma-Ray Bursts}",
      journal = {\apj},
         year = 2004,
        month = oct,
       volume = {614},
       number = {2},
        pages = {827-846},
          doi = {10.1086/423782},
archivePrefix = {arXiv},
       eprint = {astro-ph/0406674},
 primaryClass = {astro-ph},
       adsurl = {https://ui.adsabs.harvard.edu/abs/2004ApJ...614..827R}
}

@ARTICLE{1994ApJ...430L..93R,
       author = {{Rees}, M.~J. and {Meszaros}, P.},
        title = "{Unsteady Outflow Models for Cosmological Gamma-Ray Bursts}",
      journal = {\apjl},
         year = 1994,
        month = aug,
       volume = {430},
        pages = {L93},
          doi = {10.1086/187446},
archivePrefix = {arXiv},
       eprint = {astro-ph/9404038},
 primaryClass = {astro-ph},
       adsurl = {https://ui.adsabs.harvard.edu/abs/1994ApJ...430L..93R}
}

@ARTICLE{1997ApJ...490...92K,
       author = {{Kobayashi}, Shiho and {Piran}, Tsvi and {Sari}, Re'em},
        title = "{Can Internal Shocks Produce the Variability in Gamma-Ray Bursts?}",
      journal = {\apj},
         year = 1997,
        month = nov,
       volume = {490},
        pages = {92},
          doi = {10.1086/512791},
archivePrefix = {arXiv},
       eprint = {astro-ph/9705013},
 primaryClass = {astro-ph},
       adsurl = {https://ui.adsabs.harvard.edu/abs/1997ApJ...490...92K}
}

@ARTICLE{1998MNRAS.296..275D,
       author = {{Daigne}, F. and {Mochkovitch}, R.},
        title = "{Gamma-ray bursts from internal shocks in a relativistic wind: temporal and spectral properties}",
      journal = {\mnras},
         year = 1998,
        month = may,
       volume = {296},
       number = {2},
        pages = {275-286},
          doi = {10.1046/j.1365-8711.1998.01305.x},
archivePrefix = {arXiv},
       eprint = {astro-ph/9801245},
 primaryClass = {astro-ph},
       adsurl = {https://ui.adsabs.harvard.edu/abs/1998MNRAS.296..275D}
}

@ARTICLE{1996ApJ...466..768T,
       author = {{Tavani}, M.},
        title = "{A Shock Emission Model for Gamma-Ray Bursts. II. Spectral Properties}",
      journal = {\apj},
         year = 1996,
        month = aug,
       volume = {466},
        pages = {768},
          doi = {10.1086/177551},
       adsurl = {https://ui.adsabs.harvard.edu/abs/1996ApJ...466..768T}
}

@ARTICLE{2000ApJ...543..722L,
       author = {{Lloyd}, Nicole M. and {Petrosian}, Vah{\'e}},
        title = "{Synchrotron Radiation as the Source of Gamma-Ray Burst Spectra}",
      journal = {\apj},
         year = 2000,
        month = nov,
       volume = {543},
       number = {2},
        pages = {722-732},
          doi = {10.1086/317125},
archivePrefix = {arXiv},
       eprint = {astro-ph/0007061},
 primaryClass = {astro-ph},
       adsurl = {https://ui.adsabs.harvard.edu/abs/2000ApJ...543..722L}
}

@ARTICLE{2020NatAs...4..210Z,
       author = {{Zhang}, Bing},
        title = "{Synchrotron radiation in {\ensuremath{\gamma}}-ray bursts prompt emission}",
      journal = {Nature Astronomy},
         year = 2020,
        month = feb,
       volume = {4},
        pages = {210-211},
          doi = {10.1038/s41550-020-1041-3},
archivePrefix = {arXiv},
       eprint = {2002.09638},
 primaryClass = {astro-ph.HE},
       adsurl = {https://ui.adsabs.harvard.edu/abs/2020NatAs...4..210Z}
}

@ARTICLE{2001A&A...369..694S,
       author = {{Spruit}, H.~C. and {Daigne}, F. and {Drenkhahn}, G.},
        title = "{Large scale magnetic fields and their dissipation in GRB fireballs}",
      journal = {\aap},
         year = 2001,
        month = apr,
       volume = {369},
        pages = {694-705},
          doi = {10.1051/0004-6361:20010131},
archivePrefix = {arXiv},
       eprint = {astro-ph/0004274},
 primaryClass = {astro-ph},
       adsurl = {https://ui.adsabs.harvard.edu/abs/2001A&A...369..694S}
}

@ARTICLE{2006A&A...457..763G,
       author = {{Giannios}, D.},
        title = "{Prompt emission spectra from the photosphere of a GRB}",
      journal = {\aap},
         year = 2006,
        month = oct,
       volume = {457},
       number = {3},
        pages = {763-770},
          doi = {10.1051/0004-6361:20065000},
archivePrefix = {arXiv},
       eprint = {astro-ph/0602397},
 primaryClass = {astro-ph},
       adsurl = {https://ui.adsabs.harvard.edu/abs/2006A&A...457..763G}
}

@ARTICLE{2002ApJ...581.1236Z,
       author = {{Zhang}, Bing and {M{\'e}sz{\'a}ros}, Peter},
        title = "{An Analysis of Gamma-Ray Burst Spectral Break Models}",
      journal = {\apj},
         year = 2002,
        month = dec,
       volume = {581},
       number = {2},
        pages = {1236-1247},
          doi = {10.1086/344338},
archivePrefix = {arXiv},
       eprint = {astro-ph/0206158},
 primaryClass = {astro-ph},
       adsurl = {https://ui.adsabs.harvard.edu/abs/2002ApJ...581.1236Z}
}

@ARTICLE{2006A&A...450..887G,
       author = {{Giannios}, D. and {Spruit}, H.~C.},
        title = "{The role of kink instability in Poynting-flux dominated jets}",
      journal = {\aap},
         year = 2006,
        month = may,
       volume = {450},
       number = {3},
        pages = {887-898},
          doi = {10.1051/0004-6361:20054107},
archivePrefix = {arXiv},
       eprint = {astro-ph/0601172},
 primaryClass = {astro-ph},
       adsurl = {https://ui.adsabs.harvard.edu/abs/2006A&A...450..887G}
}

@ARTICLE{2011ApJ...726...90Z,
       author = {{Zhang}, Bing and {Yan}, Huirong},
        title = "{The Internal-collision-induced Magnetic Reconnection and Turbulence (ICMART) Model of Gamma-ray Bursts}",
      journal = {\apj},
         year = 2011,
        month = jan,
       volume = {726},
       number = {2},
          eid = {90},
        pages = {90},
          doi = {10.1088/0004-637X/726/2/90},
archivePrefix = {arXiv},
       eprint = {1011.1197},
 primaryClass = {astro-ph.HE},
       adsurl = {https://ui.adsabs.harvard.edu/abs/2011ApJ...726...90Z}
}

@ARTICLE{2014MNRAS.445.3892B,
       author = {{Beniamini}, Paz and {Piran}, Tsvi},
        title = "{The emission mechanism in magnetically dominated gamma-ray burst outflows}",
      journal = {\mnras},
         year = 2014,
        month = dec,
       volume = {445},
       number = {4},
        pages = {3892-3907},
          doi = {10.1093/mnras/stu2032},
archivePrefix = {arXiv},
       eprint = {1402.4113},
 primaryClass = {astro-ph.HE},
       adsurl = {https://ui.adsabs.harvard.edu/abs/2014MNRAS.445.3892B}
}

@ARTICLE{2014ApJ...782...92Z,
       author = {{Zhang}, Bo and {Zhang}, Bing},
        title = "{Gamma-Ray Burst Prompt Emission Light Curves and Power Density Spectra in the ICMART Model}",
      journal = {\apj},
         year = 2014,
        month = feb,
       volume = {782},
       number = {2},
          eid = {92},
        pages = {92},
          doi = {10.1088/0004-637X/782/2/92},
archivePrefix = {arXiv},
       eprint = {1312.7701},
 primaryClass = {astro-ph.HE},
       adsurl = {https://ui.adsabs.harvard.edu/abs/2014ApJ...782...92Z}
}

@ARTICLE{2015ApJ...805..163D,
       author = {{Deng}, Wei and {Li}, Hui and {Zhang}, Bing and {Li}, Shengtai},
        title = "{Relativistic MHD Simulations of Collision-induced Magnetic Dissipation in Poynting-flux-dominated Jets/outflows}",
      journal = {\apj},
         year = 2015,
        month = jun,
       volume = {805},
       number = {2},
          eid = {163},
        pages = {163},
          doi = {10.1088/0004-637X/805/2/163},
archivePrefix = {arXiv},
       eprint = {1501.07595},
 primaryClass = {astro-ph.HE},
       adsurl = {https://ui.adsabs.harvard.edu/abs/2015ApJ...805..163D}
}

@ARTICLE{2022ApJ...927..173S,
       author = {{Shao}, Xueying and {Gao}, He},
        title = "{Gamma-Ray Burst Prompt Emission Spectrum and E $_{p}$ Evolution Patterns in the ICMART Model}",
      journal = {\apj},
         year = 2022,
        month = mar,
       volume = {927},
       number = {2},
          eid = {173},
        pages = {173},
          doi = {10.3847/1538-4357/ac46a8},
archivePrefix = {arXiv},
       eprint = {2201.03750},
 primaryClass = {astro-ph.HE},
       adsurl = {https://ui.adsabs.harvard.edu/abs/2022ApJ...927..173S}
}

@ARTICLE{2000ApJS..126...19P,
       author = {{Preece}, R.~D. and {Briggs}, M.~S. and {Mallozzi}, R.~S. and {Pendleton}, G.~N. and {Paciesas}, W.~S. and {Band}, D.~L.},
        title = "{The BATSE Gamma-Ray Burst Spectral Catalog. I. High Time Resolution Spectroscopy of Bright Bursts Using High Energy Resolution Data}",
      journal = {\apjs},
         year = 2000,
        month = jan,
       volume = {126},
       number = {1},
        pages = {19-36},
          doi = {10.1086/313289},
archivePrefix = {arXiv},
       eprint = {astro-ph/9908119},
 primaryClass = {astro-ph},
       adsurl = {https://ui.adsabs.harvard.edu/abs/2000ApJS..126...19P}
}

@ARTICLE{2016ApJ...816...72Z,
       author = {{Zhang}, Bin-Bin and {Uhm}, Z. Lucas and {Connaughton}, Valerie and {Briggs}, Michael S. and {Zhang}, Bing},
        title = "{Synchrotron Origin of the Typical GRB Band Function{\textemdash}A Case Study of GRB 130606B}",
      journal = {\apj},
         year = 2016,
        month = jan,
       volume = {816},
       number = {2},
          eid = {72},
        pages = {72},
          doi = {10.3847/0004-637X/816/2/72},
archivePrefix = {arXiv},
       eprint = {1505.05858},
 primaryClass = {astro-ph.HE},
       adsurl = {https://ui.adsabs.harvard.edu/abs/2016ApJ...816...72Z}
}

@ARTICLE{2003A&A...406..879G,
       author = {{Ghirlanda}, G. and {Celotti}, A. and {Ghisellini}, G.},
        title = "{Extremely hard GRB spectra prune down the forest of emission models}",
      journal = {\aap},
         year = 2003,
        month = aug,
       volume = {406},
        pages = {879-892},
          doi = {10.1051/0004-6361:20030803},
archivePrefix = {arXiv},
       eprint = {astro-ph/0210693},
 primaryClass = {astro-ph},
       adsurl = {https://ui.adsabs.harvard.edu/abs/2003A&A...406..879G}
}

@ARTICLE{2002A&A...393..409G,
       author = {{Ghirlanda}, G. and {Celotti}, A. and {Ghisellini}, G.},
        title = "{Time resolved spectral analysis of bright gamma ray bursts}",
      journal = {\aap},
         year = 2002,
        month = oct,
       volume = {393},
        pages = {409-423},
          doi = {10.1051/0004-6361:20021038},
archivePrefix = {arXiv},
       eprint = {astro-ph/0206377},
 primaryClass = {astro-ph},
       adsurl = {https://ui.adsabs.harvard.edu/abs/2002A&A...393..409G}
}

@ARTICLE{2005ApJ...625L..95R,
       author = {{Ryde}, Felix},
        title = "{Is Thermal Emission in Gamma-Ray Bursts Ubiquitous?}",
      journal = {\apjl},
         year = 2005,
        month = jun,
       volume = {625},
       number = {2},
        pages = {L95-L98},
          doi = {10.1086/431239},
archivePrefix = {arXiv},
       eprint = {astro-ph/0504450},
 primaryClass = {astro-ph},
       adsurl = {https://ui.adsabs.harvard.edu/abs/2005ApJ...625L..95R}
}

@ARTICLE{2009ApJ...702.1211R,
       author = {{Ryde}, Felix and {Pe'er}, Asaf},
        title = "{Quasi-blackbody Component and Radiative Efficiency of the Prompt Emission of Gamma-ray Bursts}",
      journal = {\apj},
         year = 2009,
        month = sep,
       volume = {702},
       number = {2},
        pages = {1211-1229},
          doi = {10.1088/0004-637X/702/2/1211},
archivePrefix = {arXiv},
       eprint = {0811.4135},
 primaryClass = {astro-ph},
       adsurl = {https://ui.adsabs.harvard.edu/abs/2009ApJ...702.1211R}
}

@ARTICLE{2010ApJ...709L.172R,
       author = {{Ryde}, F. and {Axelsson}, M. and {Zhang}, B.~B. and {McGlynn}, S. and {Pe'er}, A. and {Lundman}, C. and {Larsson}, S. and {Battelino}, M. and {Zhang}, B. and {Bissaldi}, E. and {Bregeon}, J. and {Briggs}, M.~S. and {Chiang}, J. and {de Palma}, F. and {Guiriec}, S. and {Larsson}, J. and {Longo}, F. and {McBreen}, S. and {Omodei}, N. and {Petrosian}, V. and {Preece}, R. and {van der Horst}, A.~J.},
        title = "{Identification and Properties of the Photospheric Emission in GRB090902B}",
      journal = {\apjl},
         year = 2010,
        month = feb,
       volume = {709},
       number = {2},
        pages = {L172-L177},
          doi = {10.1088/2041-8205/709/2/L172},
archivePrefix = {arXiv},
       eprint = {0911.2025},
 primaryClass = {astro-ph.HE},
       adsurl = {https://ui.adsabs.harvard.edu/abs/2010ApJ...709L.172R}
}

@ARTICLE{2014ApJ...785..112D,
       author = {{Deng}, Wei and {Zhang}, Bing},
        title = "{Low Energy Spectral Index and E$_{p}$ Evolution of Quasi-thermal Photosphere Emission of Gamma-Ray Bursts}",
      journal = {\apj},
         year = 2014,
        month = apr,
       volume = {785},
       number = {2},
          eid = {112},
        pages = {112},
          doi = {10.1088/0004-637X/785/2/112},
archivePrefix = {arXiv},
       eprint = {1402.5364},
 primaryClass = {astro-ph.HE},
       adsurl = {https://ui.adsabs.harvard.edu/abs/2014ApJ...785..112D}
}

@ARTICLE{2017IJMPD..2630018P,
       author = {{Pe'Er}, Asaf and {Ryde}, Felix},
        title = "{Photospheric emission in gamma-ray bursts}",
      journal = {International Journal of Modern Physics D},
         year = 2017,
        month = jan,
       volume = {26},
       number = {10},
          eid = {1730018-296},
        pages = {1730018-296},
          doi = {10.1142/S021827181730018X},
archivePrefix = {arXiv},
       eprint = {1603.05058},
 primaryClass = {astro-ph.HE},
       adsurl = {https://ui.adsabs.harvard.edu/abs/2017IJMPD..2630018P}
}

@ARTICLE{2013MNRAS.432.3237G,
       author = {{Ghirlanda}, G. and {Pescalli}, A. and {Ghisellini}, G.},
        title = "{Photospheric emission throughout GRB 100507 detected by Fermi}",
      journal = {\mnras},
         year = 2013,
        month = jul,
       volume = {432},
       number = {4},
        pages = {3237-3244},
          doi = {10.1093/mnras/stt681},
archivePrefix = {arXiv},
       eprint = {1305.3287},
 primaryClass = {astro-ph.HE},
       adsurl = {https://ui.adsabs.harvard.edu/abs/2013MNRAS.432.3237G}
}

@ARTICLE{2013ApJ...770...32G,
       author = {{Guiriec}, S. and {Daigne}, F. and {Hasco{\"e}t}, R. and {Vianello}, G. and {Ryde}, F. and {Mochkovitch}, R. and {Kouveliotou}, C. and {Xiong}, S. and {Bhat}, P.~N. and {Foley}, S. and {Gruber}, D. and {Burgess}, J.~M. and {McGlynn}, S. and {McEnery}, J. and {Gehrels}, N.},
        title = "{Evidence for a Photospheric Component in the Prompt Emission of the Short GRB 120323A and Its Effects on the GRB Hardness-Luminosity Relation}",
      journal = {\apj},
         year = 2013,
        month = jun,
       volume = {770},
       number = {1},
          eid = {32},
        pages = {32},
          doi = {10.1088/0004-637X/770/1/32},
archivePrefix = {arXiv},
       eprint = {1210.7252},
 primaryClass = {astro-ph.HE},
       adsurl = {https://ui.adsabs.harvard.edu/abs/2013ApJ...770...32G}
}

@ARTICLE{2017ApJ...846..138G,
       author = {{Guiriec}, S. and {Gehrels}, N. and {McEnery}, J. and {Kouveliotou}, C. and {Hartmann}, D.~H.},
        title = "{Photospheric Emission in the Joint GBM and Konus Prompt Spectra of GRB 120323A}",
      journal = {\apj},
         year = 2017,
        month = sep,
       volume = {846},
       number = {2},
          eid = {138},
        pages = {138},
          doi = {10.3847/1538-4357/aa81c2},
archivePrefix = {arXiv},
       eprint = {1703.07846},
 primaryClass = {astro-ph.HE},
       adsurl = {https://ui.adsabs.harvard.edu/abs/2017ApJ...846..138G}
}

@ARTICLE{2016MNRAS.456.2157I,
       author = {{Iyyani}, S. and {Ryde}, F. and {Burgess}, J.~M. and {Pe'er}, A. and {B{\'e}gu{\'e}}, D.},
        title = "{Synchrotron emission in GRBs observed by Fermi: its limitations and the role of the photosphere}",
      journal = {\mnras},
         year = 2016,
        month = feb,
       volume = {456},
       number = {2},
        pages = {2157-2171},
          doi = {10.1093/mnras/stv2751},
archivePrefix = {arXiv},
       eprint = {1511.07390},
 primaryClass = {astro-ph.HE},
       adsurl = {https://ui.adsabs.harvard.edu/abs/2016MNRAS.456.2157I}
}

@ARTICLE{2022ApJ...932...25C,
       author = {{Chen}, Jia-Ming and {Peng}, Zhao-Yang and {Du}, Tan-Tan and {Yin}, Yue},
        title = "{Spectral Properties and Hybrid Jet Model Constraints of Fermi GRB 210610B}",
      journal = {\apj},
         year = 2022,
        month = jun,
       volume = {932},
       number = {1},
          eid = {25},
        pages = {25},
          doi = {10.3847/1538-4357/ac6c2a},
       adsurl = {https://ui.adsabs.harvard.edu/abs/2022ApJ...932...25C}
}

@ARTICLE{2011ApJ...727L..33G,
       author = {{Guiriec}, Sylvain and {Connaughton}, Valerie and {Briggs}, Michael S. and {Burgess}, Michael and {Ryde}, Felix and {Daigne}, Fr{\'e}d{\'e}ric and {M{\'e}sz{\'a}ros}, Peter and {Goldstein}, Adam and {McEnery}, Julie and {Omodei}, Nicola and {Bhat}, P.~N. and {Bissaldi}, Elisabetta and {Camero-Arranz}, Ascensi{\'o}n and {Chaplin}, Vandiver and {Diehl}, Roland and {Fishman}, Gerald and {Foley}, Suzanne and {Gibby}, Melissa and {Giles}, Misty M. and {Greiner}, Jochen and {Gruber}, David and {von Kienlin}, Andreas and {Kippen}, Marc and {Kouveliotou}, Chryssa and {McBreen}, Sheila and {Meegan}, Charles A. and {Paciesas}, William and {Preece}, Robert and {Rau}, Arne and {Tierney}, Dave and {van der Horst}, Alexander J. and {Wilson-Hodge}, Colleen},
        title = "{Detection of a Thermal Spectral Component in the Prompt Emission of GRB 100724B}",
      journal = {\apjl},
         year = 2011,
        month = feb,
       volume = {727},
       number = {2},
          eid = {L33},
        pages = {L33},
          doi = {10.1088/2041-8205/727/2/L33},
archivePrefix = {arXiv},
       eprint = {1010.4601},
 primaryClass = {astro-ph.HE},
       adsurl = {https://ui.adsabs.harvard.edu/abs/2011ApJ...727L..33G}
}

@ARTICLE{2012ApJ...757L..31A,
       author = {{Axelsson}, M. and {Baldini}, L. and {Barbiellini}, G. and {Baring}, M.~G. and {Bellazzini}, R. and {Bregeon}, J. and {Brigida}, M. and {Bruel}, P. and {Buehler}, R. and {Caliandro}, G.~A. and {Cameron}, R.~A. and {Caraveo}, P.~A. and {Cecchi}, C. and {Chaves}, R.~C.~G. and {Chekhtman}, A. and {Chiang}, J. and {Claus}, R. and {Conrad}, J. and {Cutini}, S. and {D'Ammando}, F. and {de Palma}, F. and {Dermer}, C.~D. and {Silva}, E. do Couto e. and {Drell}, P.~S. and {Favuzzi}, C. and {Fegan}, S.~J. and {Ferrara}, E.~C. and {Focke}, W.~B. and {Fukazawa}, Y. and {Fusco}, P. and {Gargano}, F. and {Gasparrini}, D. and {Gehrels}, N. and {Germani}, S. and {Giglietto}, N. and {Giroletti}, M. and {Godfrey}, G. and {Guiriec}, S. and {Hadasch}, D. and {Hanabata}, Y. and {Hayashida}, M. and {Hou}, X. and {Iyyani}, S. and {Jackson}, M.~S. and {Kocevski}, D. and {Kuss}, M. and {Larsson}, J. and {Larsson}, S. and {Longo}, F. and {Loparco}, F. and {Lundman}, C. and {Mazziotta}, M.~N. and {McEnery}, J.~E. and {Mizuno}, T. and {Monzani}, M.~E. and {Moretti}, E. and {Morselli}, A. and {Murgia}, S. and {Nuss}, E. and {Nymark}, T. and {Ohno}, M. and {Omodei}, N. and {Pesce-Rollins}, M. and {Piron}, F. and {Pivato}, G. and {Racusin}, J.~L. and {Rain{\`o}}, S. and {Razzano}, M. and {Razzaque}, S. and {Reimer}, A. and {Roth}, M. and {Ryde}, F. and {Sanchez}, D.~A. and {Sgr{\`o}}, C. and {Siskind}, E.~J. and {Spandre}, G. and {Spinelli}, P. and {Stamatikos}, M. and {Tibaldo}, L. and {Tinivella}, M. and {Usher}, T.~L. and {Vandenbroucke}, J. and {Vasileiou}, V. and {Vianello}, G. and {Vitale}, V. and {Waite}, A.~P. and {Winer}, B.~L. and {Wood}, K.~S. and {Burgess}, J.~M. and {Bhat}, P.~N. and {Bissaldi}, E. and {Briggs}, M.~S. and {Connaughton}, V. and {Fishman}, G. and {Fitzpatrick}, G. and {Foley}, S. and {Gruber}, D. and {Kippen}, R.~M. and {Kouveliotou}, C. and {Jenke}, P. and {McBreen}, S. and {McGlynn}, S. and {Meegan}, C. and {Paciesas}, W.~S. and {Pelassa}, V. and {Preece}, R. and {Tierney}, D. and {von Kienlin}, A. and {Wilson-Hodge}, C. and {Xiong}, S. and {Pe'er}, A.},
        title = "{GRB110721A: An Extreme Peak Energy and Signatures of the Photosphere}",
      journal = {\apjl},
         year = 2012,
        month = oct,
       volume = {757},
       number = {2},
          eid = {L31},
        pages = {L31},
          doi = {10.1088/2041-8205/757/2/L31},
archivePrefix = {arXiv},
       eprint = {1207.6109},
 primaryClass = {astro-ph.HE},
       adsurl = {https://ui.adsabs.harvard.edu/abs/2012ApJ...757L..31A}
}

@ARTICLE{2013A&A...551A.124H,
       author = {{Hasco{\"e}t}, R. and {Daigne}, F. and {Mochkovitch}, R.},
        title = "{Prompt thermal emission in gamma-ray bursts}",
      journal = {\aap},
         year = 2013,
        month = mar,
       volume = {551},
          eid = {A124},
        pages = {A124},
          doi = {10.1051/0004-6361/201220023},
archivePrefix = {arXiv},
       eprint = {1302.0235},
 primaryClass = {astro-ph.HE},
       adsurl = {https://ui.adsabs.harvard.edu/abs/2013A&A...551A.124H}
}

@ARTICLE{2015ApJ...801..103G,
       author = {{Gao}, He and {Zhang}, Bing},
        title = "{Photosphere Emission from a Hybrid Relativistic Outflow with Arbitrary Dimensionless Entropy and Magnetization in GRBs}",
      journal = {\apj},
         year = 2015,
        month = mar,
       volume = {801},
       number = {2},
          eid = {103},
        pages = {103},
          doi = {10.1088/0004-637X/801/2/103},
archivePrefix = {arXiv},
       eprint = {1409.3584},
 primaryClass = {astro-ph.HE},
       adsurl = {https://ui.adsabs.harvard.edu/abs/2015ApJ...801..103G}
}

@ARTICLE{2020ApJ...894..100L,
       author = {{Li}, Liang},
        title = "{Thermal Components in Gamma-Ray Bursts. II. Constraining the Hybrid Jet Model}",
      journal = {\apj},
         year = 2020,
        month = may,
       volume = {894},
       number = {2},
          eid = {100},
        pages = {100},
          doi = {10.3847/1538-4357/ab8014},
archivePrefix = {arXiv},
       eprint = {1908.09240},
 primaryClass = {astro-ph.HE},
       adsurl = {https://ui.adsabs.harvard.edu/abs/2020ApJ...894..100L}
}

@ARTICLE{2009ApJ...706L.138A,
       author = {{Abdo}, A.~A. and {Ackermann}, M. and {Ajello}, M. and {Asano}, K. and {Atwood}, W.~B. and {Axelsson}, M. and {Baldini}, L. and {Ballet}, J. and {Barbiellini}, G. and {Baring}, M.~G. and {Bastieri}, D. and {Bechtol}, K. and {Bellazzini}, R. and {Berenji}, B. and {Bhat}, P.~N. and {Bissaldi}, E. and {Blandford}, R.~D. and {Bloom}, E.~D. and {Bonamente}, E. and {Borgland}, A.~W. and {Bouvier}, A. and {Bregeon}, J. and {Brez}, A. and {Briggs}, M.~S. and {Brigida}, M. and {Bruel}, P. and {Burgess}, J.~M. and {Burrows}, D.~N. and {Buson}, S. and {Caliandro}, G.~A. and {Cameron}, R.~A. and {Caraveo}, P.~A. and {Casandjian}, J.~M. and {Cecchi}, C. and {{\c{C}}elik}, {\"O}. and {Chekhtman}, A. and {Cheung}, C.~C. and {Chiang}, J. and {Ciprini}, S. and {Claus}, R. and {Cohen-Tanugi}, J. and {Cominsky}, L.~R. and {Connaughton}, V. and {Conrad}, J. and {Cutini}, S. and {d'Elia}, V. and {Dermer}, C.~D. and {de Angelis}, A. and {de Palma}, F. and {Digel}, S.~W. and {Dingus}, B.~L. and {Silva}, E. do Couto e. and {Drell}, P.~S. and {Dubois}, R. and {Dumora}, D. and {Farnier}, C. and {Favuzzi}, C. and {Fegan}, S.~J. and {Finke}, J. and {Fishman}, G. and {Focke}, W.~B. and {Fortin}, P. and {Frailis}, M. and {Fukazawa}, Y. and {Funk}, S. and {Fusco}, P. and {Gargano}, F. and {Gehrels}, N. and {Germani}, S. and {Giavitto}, G. and {Giebels}, B. and {Giglietto}, N. and {Giordano}, F. and {Glanzman}, T. and {Godfrey}, G. and {Goldstein}, A. and {Granot}, J. and {Greiner}, J. and {Grenier}, I.~A. and {Grove}, J.~E. and {Guillemot}, L. and {Guiriec}, S. and {Hanabata}, Y. and {Harding}, A.~K. and {Hayashida}, M. and {Hays}, E. and {Horan}, D. and {Hughes}, R.~E. and {Jackson}, M.~S. and {J{\'o}hannesson}, G. and {Johnson}, A.~S. and {Johnson}, R.~P. and {Johnson}, W.~N. and {Kamae}, T. and {Katagiri}, H. and {Kataoka}, J. and {Kawai}, N. and {Kerr}, M. and {Kippen}, R.~M. and {Kn{\"o}dlseder}, J. and {Kocevski}, D. and {Komin}, N. and {Kouveliotou}, C. and {Kuss}, M. and {Lande}, J. and {Latronico}, L. and {Lemoine-Goumard}, M. and {Longo}, F. and {Loparco}, F. and {Lott}, B. and {Lovellette}, M.~N. and {Lubrano}, P. and {Madejski}, G.~M. and {Makeev}, A. and {Mazziotta}, M.~N. and {McBreen}, S. and {McEnery}, J.~E. and {McGlynn}, S. and {Meegan}, C. and {M{\'e}sz{\'a}ros}, P. and {Meurer}, C. and {Michelson}, P.~F. and {Mitthumsiri}, W. and {Mizuno}, T. and {Moiseev}, A.~A. and {Monte}, C. and {Monzani}, M.~E. and {Moretti}, E. and {Morselli}, A. and {Moskalenko}, I.~V. and {Murgia}, S. and {Nakamori}, T. and {Nolan}, P.~L. and {Norris}, J.~P. and {Nuss}, E. and {Ohno}, M. and {Ohsugi}, T. and {Omodei}, N. and {Orlando}, E. and {Ormes}, J.~F. and {Paciesas}, W.~S. and {Paneque}, D. and {Panetta}, J.~H. and {Pelassa}, V. and {Pepe}, M. and {Pesce-Rollins}, M. and {Petrosian}, V. and {Piron}, F. and {Porter}, T.~A. and {Preece}, R. and {Rain{\`o}}, S. and {Rando}, R. and {Rau}, A. and {Razzano}, M. and {Razzaque}, S. and {Reimer}, A. and {Reimer}, O. and {Reposeur}, T. and {Ritz}, S. and {Rochester}, L.~S. and {Rodriguez}, A.~Y. and {Roming}, P.~W.~A. and {Roth}, M. and {Ryde}, F. and {Sadrozinski}, H.~F.-W. and {Sanchez}, D. and {Sander}, A. and {Saz Parkinson}, P.~M. and {Scargle}, J.~D. and {Schalk}, T.~L. and {Sgr{\`o}}, C. and {Siskind}, E.~J. and {Smith}, P.~D. and {Spinelli}, P. and {Stamatikos}, M. and {Stecker}, F.~W. and {Stratta}, G. and {Strickman}, M.~S. and {Suson}, D.~J. and {Swenson}, C.~A. and {Tajima}, H. and {Takahashi}, H. and {Tanaka}, T. and {Thayer}, J.~B. and {Thayer}, J.~G. and {Thompson}, D.~J. and {Tibaldo}, L. and {Torres}, D.~F. and {Tosti}, G. and {Tramacere}, A. and {Uchiyama}, Y. and {Uehara}, T. and {Usher}, T.~L. and {van der Horst}, A.~J. and {Vasileiou}, V. and {Vilchez}, N. and {Vitale}, V. and {von Kienlin}, A. and {Waite}, A.~P. and {Wang}, P.},
        title = "{Fermi Observations of GRB 090902B: A Distinct Spectral Component in the Prompt and Delayed Emission}",
      journal = {\apjl},
         year = 2009,
        month = nov,
       volume = {706},
       number = {1},
        pages = {L138-L144},
          doi = {10.1088/0004-637X/706/1/L138},
archivePrefix = {arXiv},
       eprint = {0909.2470},
 primaryClass = {astro-ph.HE},
       adsurl = {https://ui.adsabs.harvard.edu/abs/2009ApJ...706L.138A}
}

@ARTICLE{2012MNRAS.420..468P,
       author = {{Pe'Er}, Asaf and {Zhang}, Bin-Bin and {Ryde}, Felix and {McGlynn}, Sin{\'e}ad and {Zhang}, Bing and {Preece}, Robert D. and {Kouveliotou}, Chryssa},
        title = "{The connection between thermal and non-thermal emission in gamma-ray bursts: general considerations and GRB 090902B as a case study}",
      journal = {\mnras},
         year = 2012,
        month = feb,
       volume = {420},
       number = {1},
        pages = {468-482},
          doi = {10.1111/j.1365-2966.2011.20052.x},
archivePrefix = {arXiv},
       eprint = {1007.2228},
 primaryClass = {astro-ph.HE},
       adsurl = {https://ui.adsabs.harvard.edu/abs/2012MNRAS.420..468P}
}

@ARTICLE{2022ApJ...934L..22D,
       author = {{Deng}, Li-Tao and {Lin}, Da-Bin and {Zhou}, Li and {Wang}, Kai and {Yang}, Xing and {Hou}, Shu-Jin and {Li}, Jing and {Wang}, Xiang-Gao and {Lu}, Rui-Jing and {Liang}, En-Wei},
        title = "{Spectral Analysis of GRB 220426A: Another Case of a Thermally Dominated Burst}",
      journal = {\apjl},
         year = 2022,
        month = aug,
       volume = {934},
       number = {2},
          eid = {L22},
        pages = {L22},
          doi = {10.3847/2041-8213/ac8169},
       adsurl = {https://ui.adsabs.harvard.edu/abs/2022ApJ...934L..22D}
}

@ARTICLE{2022ApJ...940..142W,
       author = {{Wang}, Yun and {Zheng}, Tian-Ci and {Jin}, Zhi-Ping},
        title = "{GRB 220426A: A Thermal Radiation-Dominated Gamma-Ray Burst}",
      journal = {\apj},
         year = 2022,
        month = dec,
       volume = {940},
       number = {2},
          eid = {142},
        pages = {142},
          doi = {10.3847/1538-4357/aca017},
archivePrefix = {arXiv},
       eprint = {2205.08427},
 primaryClass = {astro-ph.HE},
       adsurl = {https://ui.adsabs.harvard.edu/abs/2022ApJ...940..142W}
}

@ARTICLE{2022MNRAS.517.2088S,
       author = {{Song}, Xin-Ying and {Zhang}, Shuang-Nan and {Ge}, Ming-Yu and {Zhang}, Shu},
        title = "{The origin of the photospheric emission of GRB 220426A}",
      journal = {\mnras},
         year = 2022,
        month = dec,
       volume = {517},
       number = {2},
        pages = {2088-2102},
          doi = {10.1093/mnras/stac2764},
archivePrefix = {arXiv},
       eprint = {2209.10832},
 primaryClass = {astro-ph.HE},
       adsurl = {https://ui.adsabs.harvard.edu/abs/2022MNRAS.517.2088S}
}

@ARTICLE{2024ApJ...961..137S,
       author = {{Song}, Xin-Ying and {Wang}, Ling-Jun and {Zhang}, Shu},
        title = "{Spectral and Jet Properties of the Quasi-thermal-dominated GRB 210121A, GRB 210610B, and GRB 221022B}",
      journal = {\apj},
         year = 2024,
        month = jan,
       volume = {961},
       number = {1},
          eid = {137},
        pages = {137},
          doi = {10.3847/1538-4357/ad0df3},
archivePrefix = {arXiv},
       eprint = {2306.02248},
 primaryClass = {astro-ph.HE},
       adsurl = {https://ui.adsabs.harvard.edu/abs/2024ApJ...961..137S}
}

@ARTICLE{2024ApJ...972..132C,
       author = {{Chen}, Jia-Ming and {Zhu}, Ke-Rui and {Peng}, Zhao-Yang and {Zhang}, Li},
        title = "{GRB 231129C: Another Thermal Emission Dominated Gamma-Ray Burst}",
      journal = {\apj},
         year = 2024,
        month = sep,
       volume = {972},
       number = {2},
          eid = {132},
        pages = {132},
          doi = {10.3847/1538-4357/ad5f93},
       adsurl = {https://ui.adsabs.harvard.edu/abs/2024ApJ...972..132C}
}

@ARTICLE{2009ApJ...702..791M,
       author = {{Meegan}, Charles and {Lichti}, Giselher and {Bhat}, P.~N. and {Bissaldi}, Elisabetta and {Briggs}, Michael S. and {Connaughton}, Valerie and {Diehl}, Roland and {Fishman}, Gerald and {Greiner}, Jochen and {Hoover}, Andrew S. and {van der Horst}, Alexander J. and {von Kienlin}, Andreas and {Kippen}, R. Marc and {Kouveliotou}, Chryssa and {McBreen}, Sheila and {Paciesas}, W.~S. and {Preece}, Robert and {Steinle}, Helmut and {Wallace}, Mark S. and {Wilson}, Robert B. and {Wilson-Hodge}, Colleen},
        title = "{The Fermi Gamma-ray Burst Monitor}",
      journal = {\apj},
         year = 2009,
        month = sep,
       volume = {702},
       number = {1},
        pages = {791-804},
          doi = {10.1088/0004-637X/702/1/791},
archivePrefix = {arXiv},
       eprint = {0908.0450},
 primaryClass = {astro-ph.IM},
       adsurl = {https://ui.adsabs.harvard.edu/abs/2009ApJ...702..791M}
}

@ARTICLE{2019ApJS..242...16L,
       author = {{Li}, Liang},
        title = "{Multipulse Fermi Gamma-Ray Bursts. I. Evidence of the Transition from Fireball to Poynting-flux-dominated Outflow}",
      journal = {\apjs},
         year = 2019,
        month = jun,
       volume = {242},
       number = {2},
          eid = {16},
        pages = {16},
          doi = {10.3847/1538-4365/ab1b78},
archivePrefix = {arXiv},
       eprint = {1810.03129},
 primaryClass = {astro-ph.HE},
       adsurl = {https://ui.adsabs.harvard.edu/abs/2019ApJS..242...16L}
}

@ARTICLE{2013ApJ...764..167S,
       author = {{Scargle}, Jeffrey D. and {Norris}, Jay P. and {Jackson}, Brad and {Chiang}, James},
        title = "{Studies in Astronomical Time Series Analysis. VI. Bayesian Block Representations}",
      journal = {\apj},
         year = 2013,
        month = feb,
       volume = {764},
       number = {2},
          eid = {167},
        pages = {167},
          doi = {10.1088/0004-637X/764/2/167},
archivePrefix = {arXiv},
       eprint = {1207.5578},
 primaryClass = {astro-ph.IM},
       adsurl = {https://ui.adsabs.harvard.edu/abs/2013ApJ...764..167S}
}

@ARTICLE{2018ApJS..236...17V,
       author = {{Vianello}, Giacomo},
        title = "{The Significance of an Excess in a Counting Experiment: Assessing the Impact of Systematic Uncertainties and the Case with a Gaussian Background}",
      journal = {\apjs},
         year = 2018,
        month = may,
       volume = {236},
       number = {1},
          eid = {17},
        pages = {17},
          doi = {10.3847/1538-4365/aab780},
archivePrefix = {arXiv},
       eprint = {1712.00118},
 primaryClass = {physics.data-an},
       adsurl = {https://ui.adsabs.harvard.edu/abs/2018ApJS..236...17V}
}

@ARTICLE{2008ApJ...682..463P,
       author = {{Pe'er}, Asaf},
        title = "{Temporal Evolution of Thermal Emission from Relativistically Expanding Plasma}",
      journal = {\apj},
         year = 2008,
        month = jul,
       volume = {682},
       number = {1},
        pages = {463-473},
          doi = {10.1086/588136},
archivePrefix = {arXiv},
       eprint = {0802.0725},
 primaryClass = {astro-ph},
       adsurl = {https://ui.adsabs.harvard.edu/abs/2008ApJ...682..463P}
}

@ARTICLE{2018ApJ...866...13H,
       author = {{Hou}, Shu-Jin and {Zhang}, Bin-Bin and {Meng}, Yan-Zhi and {Wu}, Xue-Feng and {Liang}, En-Wei and {L{\"u}}, Hou-Jun and {Liu}, Tong and {Liang}, Yun-Feng and {Lin}, Lin and {Lu}, Rui-jing and {Huang}, Jin-Shu and {Zhang}, Bing},
        title = "{Multicolor Blackbody Emission in GRB 081221}",
      journal = {\apj},
         year = 2018,
        month = oct,
       volume = {866},
       number = {1},
          eid = {13},
        pages = {13},
          doi = {10.3847/1538-4357/aadc07},
archivePrefix = {arXiv},
       eprint = {1808.03942},
 primaryClass = {astro-ph.HE},
       adsurl = {https://ui.adsabs.harvard.edu/abs/2018ApJ...866...13H}
}

@ARTICLE{2021ApJS..255...25I,
       author = {{Iyyani}, Shabnam and {Sharma}, Vidushi},
        title = "{Study of the Prompt Emission of Short Gamma-Ray Bursts Using a Multicolor Blackbody: A Clue to the Viewing Angle}",
      journal = {\apjs},
         year = 2021,
        month = aug,
       volume = {255},
       number = {2},
          eid = {25},
        pages = {25},
          doi = {10.3847/1538-4365/ac082f},
archivePrefix = {arXiv},
       eprint = {2007.00020},
 primaryClass = {astro-ph.HE},
       adsurl = {https://ui.adsabs.harvard.edu/abs/2021ApJS..255...25I}
}

@ARTICLE{2015ApJ...814...10G,
       author = {{Guiriec}, S. and {Mochkovitch}, R. and {Piran}, T. and {Daigne}, F. and {Kouveliotou}, C. and {Racusin}, J. and {Gehrels}, N. and {McEnery}, J.},
        title = "{GRB 131014A: A Laboratory for Studying the Thermal-like and Non-thermal Emissions in Gamma-Ray Bursts, and the New L$^{nTh}$$_{i}$-E$^{nTh,rest}$$_{peak,i}$ Relation}",
      journal = {\apj},
         year = 2015,
        month = nov,
       volume = {814},
       number = {1},
          eid = {10},
        pages = {10},
          doi = {10.1088/0004-637X/814/1/10},
archivePrefix = {arXiv},
       eprint = {1507.06976},
 primaryClass = {astro-ph.HE},
       adsurl = {https://ui.adsabs.harvard.edu/abs/2015ApJ...814...10G}
}

@ARTICLE{2016ApJ...833..139A,
       author = {{Arimoto}, Makoto and {Asano}, Katsuaki and {Ohno}, Masanori and {Veres}, P{\'e}ter and {Axelsson}, Magnus and {Bissaldi}, Elisabetta and {Tachibana}, Yutaro and {Kawai}, Nobuyuki},
        title = "{High-energy Non-thermal and Thermal Emission from GRB141207A Detected by Fermi}",
      journal = {\apj},
         year = 2016,
        month = dec,
       volume = {833},
       number = {2},
          eid = {139},
        pages = {139},
          doi = {10.3847/1538-4357/833/2/139},
archivePrefix = {arXiv},
       eprint = {1610.04867},
 primaryClass = {astro-ph.HE},
       adsurl = {https://ui.adsabs.harvard.edu/abs/2016ApJ...833..139A}
}

@ARTICLE{2022MNRAS.516.4180N,
       author = {{Nawaz}, Alina and {Sajjad}, Saeeda},
        title = "{The spectral analysis of GRB 150902A and the nature of its outflow}",
      journal = {\mnras},
         year = 2022,
        month = nov,
       volume = {516},
       number = {3},
        pages = {4180-4193},
          doi = {10.1093/mnras/stac2217},
       adsurl = {https://ui.adsabs.harvard.edu/abs/2022MNRAS.516.4180N}
}

@ARTICLE{2022ApJ...938..159S,
       author = {{Siddique}, Iqra and {Sajjad}, Saeeda and {Motiwala}, Khadeejah},
        title = "{The Prompt Emission of GRB 130518A and the Study of Its Outflow through Hybrid Jet Models}",
      journal = {\apj},
         year = 2022,
        month = oct,
       volume = {938},
       number = {2},
          eid = {159},
        pages = {159},
          doi = {10.3847/1538-4357/ac8d05},
       adsurl = {https://ui.adsabs.harvard.edu/abs/2022ApJ...938..159S}
}

@ARTICLE{2002MNRAS.336.1271D,
       author = {{Daigne}, Fr{\'e}d{\'e}ric and {Mochkovitch}, Robert},
        title = "{The expected thermal precursors of gamma-ray bursts in the internal shock model}",
      journal = {\mnras},
         year = 2002,
        month = nov,
       volume = {336},
       number = {4},
        pages = {1271-1280},
          doi = {10.1046/j.1365-8711.2002.05875.x},
archivePrefix = {arXiv},
       eprint = {astro-ph/0207456},
 primaryClass = {astro-ph},
       adsurl = {https://ui.adsabs.harvard.edu/abs/2002MNRAS.336.1271D}
}

@ARTICLE{2007ApJ...664L...1P,
       author = {{Pe'er}, Asaf and {Ryde}, Felix and {Wijers}, Ralph A.~M.~J. and {M{\'e}sz{\'a}ros}, Peter and {Rees}, Martin J.},
        title = "{A New Method of Determining the Initial Size and Lorentz Factor of Gamma-Ray Burst Fireballs Using a Thermal Emission Component}",
      journal = {\apjl},
         year = 2007,
        month = jul,
       volume = {664},
       number = {1},
        pages = {L1-L4},
          doi = {10.1086/520534},
archivePrefix = {arXiv},
       eprint = {astro-ph/0703734},
 primaryClass = {astro-ph},
       adsurl = {https://ui.adsabs.harvard.edu/abs/2007ApJ...664L...1P}
}

@ARTICLE{2011ApJ...733L..40M,
       author = {{M{\'e}sz{\'a}ros}, P. and {Rees}, M.~J.},
        title = "{GeV Emission from Collisional Magnetized Gamma-Ray Bursts}",
      journal = {\apjl},
         year = 2011,
        month = jun,
       volume = {733},
       number = {2},
          eid = {L40},
        pages = {L40},
          doi = {10.1088/2041-8205/733/2/L40},
archivePrefix = {arXiv},
       eprint = {1104.5025},
 primaryClass = {astro-ph.HE},
       adsurl = {https://ui.adsabs.harvard.edu/abs/2011ApJ...733L..40M}
}

@ARTICLE{2012ApJ...755...12V,
       author = {{Veres}, P. and {M{\'e}sz{\'a}ros}, P.},
        title = "{Single- and Two-component Gamma-Ray Burst Spectra in the Fermi GBM-LAT Energy Range}",
      journal = {\apj},
         year = 2012,
        month = aug,
       volume = {755},
       number = {1},
          eid = {12},
        pages = {12},
          doi = {10.1088/0004-637X/755/1/12},
archivePrefix = {arXiv},
       eprint = {1202.2821},
 primaryClass = {astro-ph.HE},
       adsurl = {https://ui.adsabs.harvard.edu/abs/2012ApJ...755...12V}
}

@ARTICLE{2019ApJS..245....7L,
       author = {{Li}, Liang},
        title = "{Thermal Components in Gamma-Ray Bursts. I. How Do They Affect Nonthermal Spectral Parameters?}",
      journal = {\apjs},
         year = 2019,
        month = nov,
       volume = {245},
       number = {1},
          eid = {7},
        pages = {7},
          doi = {10.3847/1538-4365/ab42de},
archivePrefix = {arXiv},
       eprint = {1905.02340},
 primaryClass = {astro-ph.HE},
       adsurl = {https://ui.adsabs.harvard.edu/abs/2019ApJS..245....7L}
}

@ARTICLE{2015arXiv150708343V,
       author = {{Vianello}, Giacomo and {Lauer}, Robert J. and {Younk}, Patrick and {Tibaldo}, Luigi and {Burgess}, James M. and {Ayala}, Hugo and {Harding}, Patrick and {Hui}, Michelle and {Omodei}, Nicola and {Zhou}, Hao},
        title = "{The Multi-Mission Maximum Likelihood framework (3ML)}",
      journal = {arXiv e-prints},
         year = 2015,
        month = jul,
          eid = {arXiv:1507.08343},
        pages = {arXiv:1507.08343},
          doi = {10.48550/arXiv.1507.08343},
archivePrefix = {arXiv},
       eprint = {1507.08343},
 primaryClass = {astro-ph.HE},
       adsurl = {https://ui.adsabs.harvard.edu/abs/2015arXiv150708343V}
}

@ARTICLE{2019MNRAS.490..927B,
       author = {{Burgess}, J. Michael and {Greiner}, Jochen and {B{\'e}gu{\'e}}, Damien and {Berlato}, Franceso},
        title = "{A Bayesian Fermi-GBM short GRB spectral catalogue}",
      journal = {\mnras},
         year = 2019,
        month = nov,
       volume = {490},
       number = {1},
        pages = {927-946},
          doi = {10.1093/mnras/stz2589},
       adsurl = {https://ui.adsabs.harvard.edu/abs/2019MNRAS.490..927B}
}

@ARTICLE{2021ApJS..254...35L,
       author = {{Li}, Liang and {Ryde}, Felix and {Pe'er}, Asaf and {Yu}, Hoi-Fung and {Acuner}, Zeynep},
        title = "{Bayesian Time-resolved Spectroscopy of Multipulse GRBs: Variations of Emission Properties among Pulses}",
      journal = {\apjs},
         year = 2021,
        month = jun,
       volume = {254},
       number = {2},
          eid = {35},
        pages = {35},
          doi = {10.3847/1538-4365/abee2a},
archivePrefix = {arXiv},
       eprint = {2012.03038},
 primaryClass = {astro-ph.HE},
       adsurl = {https://ui.adsabs.harvard.edu/abs/2021ApJS..254...35L}
}

@ARTICLE{2021ApJ...920...53C,
       author = {{Chen}, Jia-Ming and {Peng}, Zhao-Yang and {Du}, Tan-Tan and {Yin}, Yue and {Wu}, Hui},
        title = "{GRB 180720B: A GRB with Interesting Spectral Characteristics}",
      journal = {\apj},
         year = 2021,
        month = oct,
       volume = {920},
       number = {1},
          eid = {53},
        pages = {53},
          doi = {10.3847/1538-4357/ac14b8},
       adsurl = {https://ui.adsabs.harvard.edu/abs/2021ApJ...920...53C}
}

@ARTICLE{2022ApJ...941...27L,
       author = {{Li}, Liang},
        title = "{Standard GRB Spectral Models ``Misused''?}",
      journal = {\apj},
         year = 2022,
        month = dec,
       volume = {941},
       number = {1},
          eid = {27},
        pages = {27},
          doi = {10.3847/1538-4357/ac3d89},
archivePrefix = {arXiv},
       eprint = {2103.11091},
 primaryClass = {hep-ph},
       adsurl = {https://ui.adsabs.harvard.edu/abs/2022ApJ...941...27L}
}

@ARTICLE{2023ApJS..266...31L,
       author = {{Li}, Liang},
        title = "{Revisiting the Spectral Energy Correlations of GRBs with Fermi Data. I. Model-wise Properties}",
      journal = {\apjs},
         year = 2023,
        month = jun,
       volume = {266},
       number = {2},
          eid = {31},
        pages = {31},
          doi = {10.3847/1538-4365/acc867},
archivePrefix = {arXiv},
       eprint = {2211.12187},
 primaryClass = {astro-ph.HE},
       adsurl = {https://ui.adsabs.harvard.edu/abs/2023ApJS..266...31L}
}

@ARTICLE{2023ApJ...944L..57L,
       author = {{Li}, Liang and {Wang}, Yu and {Ryde}, Felix and {Pe'er}, Asaf and {Zhang}, Bing and {Guiriec}, Sylvain and {Castro-Tirado}, Alberto J. and {Kann}, D. Alexander and {Axelsson}, Magnus and {Page}, Kim and {Veres}, P{\'e}ter and {Bhat}, P.~N.},
        title = "{A Cosmological Fireball with 16\% Gamma-Ray Radiative Efficiency}",
      journal = {\apjl},
         year = 2023,
        month = feb,
       volume = {944},
       number = {2},
          eid = {L57},
        pages = {L57},
          doi = {10.3847/2041-8213/acb99d},
archivePrefix = {arXiv},
       eprint = {2212.02141},
 primaryClass = {astro-ph.HE},
       adsurl = {https://ui.adsabs.harvard.edu/abs/2023ApJ...944L..57L}
}

@ARTICLE{2024ApJ...972..195L,
       author = {{Li}, Liang and {Wang}, Yu},
        title = "{GRB 190114C: Fireball Energy Budget and Radiative Efficiency Revisited}",
      journal = {\apj},
         year = 2024,
        month = sep,
       volume = {972},
       number = {2},
          eid = {195},
        pages = {195},
          doi = {10.3847/1538-4357/ad2511},
archivePrefix = {arXiv},
       eprint = {2302.06116},
 primaryClass = {astro-ph.HE},
       adsurl = {https://ui.adsabs.harvard.edu/abs/2024ApJ...972..195L}
}

@ARTICLE{2018NatAs...2...69Z,
       author = {{Zhang}, B.-B. and {Zhang}, B. and {Castro-Tirado}, A.~J. and {Dai}, Z.~G. and {Tam}, P.-H.~T. and {Wang}, X.-Y. and {Hu}, Y.-D. and {Karpov}, S. and {Pozanenko}, A. and {Zhang}, F.-W. and {Mazaeva}, E. and {Minaev}, P. and {Volnova}, A. and {Oates}, S. and {Gao}, H. and {Wu}, X.-F. and {Shao}, L. and {Tang}, Q.-W. and {Beskin}, G. and {Biryukov}, A. and {Bondar}, S. and {Ivanov}, E. and {Katkova}, E. and {Orekhova}, N. and {Perkov}, A. and {Sasyuk}, V. and {Mankiewicz}, L. and {{\.Z}arnecki}, A.~F. and {Cwiek}, A. and {Opiela}, R. and {Zadro{\.Z}ny}, A. and {Aptekar}, R. and {Frederiks}, D. and {Svinkin}, D. and {Kusakin}, A. and {Inasaridze}, R. and {Burhonov}, O. and {Rumyantsev}, V. and {Klunko}, E. and {Moskvitin}, A. and {Fatkhullin}, T. and {Sokolov}, V.~V. and {Valeev}, A.~F. and {Jeong}, S. and {Park}, I.~H. and {Caballero-Garc{\'\i}a}, M.~D. and {Cunniffe}, R. and {Tello}, J.~C. and {Ferrero}, P. and {Pandey}, S.~B. and {Jel{\'\i}nek}, M. and {Peng}, F.~K. and {S{\'a}nchez-Ram{\'\i}rez}, R. and {Castell{\'o}n}, A.},
        title = "{Transition from fireball to Poynting-flux-dominated outflow in the three-episode GRB 160625B}",
      journal = {Nature Astronomy},
         year = 2018,
        month = nov,
       volume = {2},
        pages = {69-75},
          doi = {10.1038/s41550-017-0309-8},
archivePrefix = {arXiv},
       eprint = {1612.03089},
 primaryClass = {astro-ph.HE},
       adsurl = {https://ui.adsabs.harvard.edu/abs/2018NatAs...2...69Z}
}

@misc{GCN14687,
  author       = {Cucchiara, A. and Cenko, S. B.},
  title        = {GRB 130518A: Gemini Redshift Confirmation},
  year         = {2013},
  howpublished = {GCN Circular 14687},
  url          = {https://gcn.nasa.gov/circulars/14687}
}

@misc{GCN30194,
  author       = {de Ugarte Postigo, A. and others},
  title        = {GRB 210610B: Redshift Confirmation from GTC},
  year         = {2021},
  howpublished = {GCN Circular 30194},
  url          = {https://gcn.nasa.gov/circulars/30194}
}

@misc{GCN30272,
  author       = {de Ugarte Postigo, A. and others},
  title        = {GRB 210619B: Redshift from OSIRIS/GTC},
  year         = {2021},
  howpublished = {GCN Circular 30272},
  url          = {https://gcn.nasa.gov/circulars/30272}
}

@misc{GCN31221,
  author       = {Malesani, D. B. and others},
  title        = {GRB 211211A: Optical Spectroscopy and the Candidate Associated Galaxy},
  year         = {2021},
  howpublished = {GCN Circular 31221},
  url          = {https://gcn.nasa.gov/circulars/31221}
}

@misc{GCN37293,
  author       = {Martin-Carrillo, A. and others},
  title        = {GRB 240825A: VLT/X-shooter Redshift},
  year         = {2024},
  howpublished = {GCN Circular 37293},
  url          = {https://gcn.nasa.gov/circulars/37293}
}

@misc{GCN38704,
  author       = {An, J. and others},
  title        = {GRB 241228B: VLT/X-shooter Redshift of $z=2.674$},
  year         = {2024},
  howpublished = {GCN Circular 38704},
  url          = {https://gcn.nasa.gov/circulars/38704}
}

@misc{GCN40175,
  author       = {Malesani, D. B. and others},
  title        = {GRB 250407A: GTC/OSIRIS+ Spectroscopic Redshift $z=1.36$},
  year         = {2025},
  howpublished = {GCN Circular 40175},
  url          = {https://gcn.nasa.gov/circulars/40175}
}

\end{document}